\documentclass[12pt]{iopart}

\usepackage[perpage]{footmisc}
\usepackage{verbatim}
\usepackage{epsfig,multicol,bbm,physics,lscape,multirow}
\usepackage{amsmath}
\usepackage{amssymb}
\usepackage{subfigure}
\usepackage{threeparttable}
\usepackage{graphicx}
\usepackage{url}
\usepackage{xspace}
\usepackage{hyperref}
\hypersetup{
  colorlinks=true,
  linkcolor=red,
  citecolor=red,
  urlcolor=blue
}

\usepackage{color}
\definecolor{comment}{rgb}{0,0.3,0}
\definecolor{identifier}{rgb}{0.0,0,0.3}
\usepackage{listings}
\def\boost{{\sc boost}}

\def\pythia{{\sc pythia}}

\def\herwig{{\sc herwig}}
\def\herwigpp{{\sc herwig++}}
\def\alpgen{{\sc alpgen}}
\def\powheg{{\sc powheg}}
\def\jimmy{{\sc jimmy}}

\def\ambt{{\sc ambt1}}
\def\sherpa{{\sc sherpa}}
\def\comix{{\sc comix}}

\def\boostauthor[#1]#2{{#2}$^{\, #1}$}
\def\specialboostauthortwo[#1][#2]#3{{#3}$^{\, #1, #2}$}
\def\specialboostauthorthree[#1][#2][#3]#4{{#4}$^{\, #1, #2, #3}$}
\def\boosteditor[#1]#2{{#2}$^{\, #1, *}$}
\def\boostaffiliation[#1]#2{$^{#1\,}$ {#2}\\}

\newcommand{\Npv}           {\ensuremath{N_{\rm PV}}\xspace}
\renewcommand{\pt}            {\ensuremath{p_{T}}\xspace}
\newcommand{\kt}            {\ensuremath{k_{T}}\xspace}
\newcommand{\antikt}        {anti-\kt}
\newcommand{\ca}            {Cambridge-Aachen\xspace}

\newcommand{\ptjet}{p_{ T}}

\newcommand{\rapjet}{y}

\newcommand{\akt}{\antikt}
\newcommand{\njet}{N^{\rm jet}}

\newcommand{\invfb}{\ensuremath{{\rm fb}^{-1}}\xspace}

\newcommand{\figrefcap}[1]{Figure \ref{fig:#1}}
\newcommand{\erefcap}[1]{Eq.~\ref{eq:#1}}
\newcommand{\tabrefcap}[1]{Table \ref{tab:#1}}

\newcommand{\secrefcap}[1]{Section \ref{sec:#1}}

\newcommand{\hep}{HEP top tagger\xspace}

\newcommand{\inst}[1]{$^{#1}$}
\newcommand{\insttwo}[2]{$^{#1,#2}$}
\newcommand{\instthr}[3]{$^{#1,#2,#3}$}

\graphicspath{{figures/}}

\begin{document}


\title[Jet Substructure at the Tevatron and LHC]{Jet Substructure at the Tevatron and LHC:\\New results, new tools, new benchmarks\footnote{Report prepared by the participants of the \boost\ 2011 workshop at Princeton University, May 22--26, 2011.  L.~Asquith (lasquith@hep.anl.gov), S.~Rappoccio (rappocc@fnal.gov), and C.~K.~Vermilion (verm@uw.edu), editors.}}

\author{A Altheimer\inst{1}, 
S Arora\inst{2},
L Asquith\inst{3},
G Brooijmans\inst{1},
J Butterworth\inst{4},
M Campanelli\inst{4},
B Chapleau\inst{5},
A E Cholakian\insttwo{1}{6},
J P Chou\inst{7},
M Dasgupta\inst{8},
A Davison\inst{4},
J Dolen\inst{9},
S D Ellis\inst{10},
R Essig\instthr{11}{12}{13},
J J Fan\inst{14},
R Field\inst{15},
A Fregoso\inst{8},
J Gallicchio\inst{6},
Y Gershtein\inst{2},
A Gomes\inst{16},
A Haas\inst{11},
E Halkiadakis\inst{2},
V Halyo\inst{14},
S Hoeche\inst{11},
A Hook\insttwo{11}{17},
A Hornig\inst{10},
P Huang\inst{18},
E Izaguirre\insttwo{11}{17},
M Jankowiak\insttwo{11}{17},
G Kribs\insttwo{19}{20},
D Krohn\inst{6},
A J Larkoski\inst{11},
A Lath\inst{2},
C Lee\inst{21},
S J Lee\inst{22},
P Loch\inst{23},
P Maksimovic\inst{24},
M Martinez\inst{25},
D W Miller\insttwo{11}{17},
T Plehn\inst{26},
K Prokofiev\inst{27},
R Rahmat\inst{28},
S Rappoccio\inst{24},
A Safonov\inst{29},
G P Salam\instthr{30}{14}{31},
S Schumann\inst{32},
M D Schwartz\inst{6},
A Schwartzman\inst{11},
M Seymour\inst{8},
J Shao\inst{33},
P Sinervo\inst{34},
M Son\inst{35},
D E Soper\inst{20},
M Spannowsky\inst{36},
I W Stewart\inst{21},
M Strassler\inst{2},
E Strauss\inst{11},
M Takeuchi\inst{26},
J Thaler\inst{21},
S Thomas\inst{2},
B Tweedie\inst{37},
R Vasquez Sierra\inst{9},
C K Vermilion\insttwo{38}{39},
M Villaplana\inst{40},
M Vos\inst{40},
J Wacker\inst{11},
D Walker\inst{6},
J R Walsh\insttwo{38}{41},
L-T Wang\inst{14},
S Wilbur\inst{42} and
W Zhu\inst{14}
}

\address{$^1$ Columbia University, Nevis Laboratory, 136 S Broadway, Irvington, NY 10533, USA} 
\address{$^2$ Rutgers University, Department of Physics and Astronomy, 136 Frelinghuysen Road, Piscataway, NJ 08854, USA}
\address{$^3$ Argonne National Laboratory, 9700 S. Cass Avenue  Argonne, IL 60439, USA}
\address{$^4$ Department of Physics and Astronomy, University College London, WC1E 6BT, UK }
\address{$^5$ McGill University, High Energy Physics Group, 3600 University Street, Montr\'{e}al, Qu\'{e}bec H3A 2T8, Canada }
\address{$^6$ Department of Physics, Harvard University, Cambridge, MA 02138, USA}
\address{$^7$ Department of Physics, Brown University, Box 1843, Providence, RI 02912, USA}
\address{$^8$ School of Physics and Astronomy, University of Manchester, Manchester, M13 9PL, UK }
\address{$^9$ University of California, Davis, Davis, CA 95616, USA }
\address{$^{10}$ Department of Physics, University of Washington, Box 351560, Seattle, WA 98195, USA }
\address{$^{11}$ SLAC National Accelerator Laboratory, Menlo Park, CA 94025, USA}
\address{$^{12}$ C.N.~Yang Institute for Theoretical Physics, Stony Brook University, Stony Brook, NY 11794, USA}
\address{$^{13}$ School of Natural Sciences, Institute for Advanced Study, Einstein Drive, Princeton, NJ 08544, USA}
\address{$^{14}$ Department of Physics, Princeton University, Princeton, NJ 08544, USA }
\address{$^{15}$ Department of Physics, University of Florida, Gainesville, FL 32611, USA}
\address{$^{16}$ Laborat\'orio de Instrumenta\c{c}ao e F\'isica Experimental de Part\'iculas, 1000-149 Lisboa, Portugal }
\address{$^{17}$ Department of Physics, Stanford University, 382 Via Pueblo Mall, Stanford, CA 94305, USA}
\address{$^{18}$ Department of Physics, University of Wisconsin--Madison, 1150 University Ave, Madison, WI 53706, USA}
\address{$^{19}$ Fermi National Accelerator Laboratory, Batavia, IL, 60510, USA}
\address{$^{20}$ Institute of Theoretical Science, University of Oregon, Eugene, OR 97403, USA}
\address{$^{21}$ Center for Theoretical Physics, Massachusetts Institute of Technology, Cambridge, MA 02139}
\address{$^{22}$ Department of Physics, KAIST, Daejeon 305-701, Korea}
\address{$^{23}$ Department of Physics, University of Arizona, Tucson, AZ 85719, USA}
\address{$^{24}$ Department of Physics and Astronomy, Johns Hopkins University, 3400 N. Charles St., Baltimore, MD 21218, USA}
\address{$^{25}$ ICREA and Institut de F'sica d'Altes Energies, UAB Campus Bellaterra, 08193 Barcelona, Spain}
\address{$^{26}$ Institute for Theoretical Physics, Uni Heidelberg, Philosophenweg 16,  D-69120 Heidelberg, Germany}
\address{$^{27}$ Department of Physics, New York University, 4 Washington Pl, New York, NY 10003, USA}
\address{$^{28}$ Department of Physics and Astronomy, The University of Mississippi, University, MS 38677, USA}
\address{$^{29}$ Department of Physics and Astronomy, Texas A\&M University, 4242 TAMU, College Station, TX 77843, USA}
\address{$^{30}$ Department of Physics, Theory Unit, CERN, CH-1211 Geneva 23, Switzerland }
\address{$^{31}$ LPTHE, UPMC Univ.~Paris 6 and CNRS UMR 7589, Paris, France}
\address{$^{32}$ II. Physikalisches Institut, Universit\"at G\"ottingen, 37077 G\"ottingen, Germany}
\address{$^{33}$ Department of Physics, Syracuse University, Syracuse, NY 13244, USA}
\address{$^{34}$ Department of Physics, University of Toronto, 60 Saint George Street, Toronto, M5S 1A7, Ontario, Canada          	}
\address{$^{35}$ Department of Physics, Yale University, New Haven, CT 06511, USA    }
\address{$^{36}$ IPPP, Department of Physics, Durham University, UK}
\address{$^{37}$ Physics Department, Boston University, Boston, MA 02215, USA}
\address{$^{38}$ Ernest Orlando Lawrence Berkeley National Laboratory, University of California, Berkeley, CA 94720    }
\address{$^{39}$ Department of Physics \& Astronomy, University of Louisville, Louisville, KY 40292, USA}
\address{$^{40}$ Instituto de F\'isica Corpuscular, IFIC/CSIC-UVEG, PO Box 22085, 46071 Valencia, Spain}
\address{$^{41}$ Center for Theoretical Physics, University of California, Berkeley, CA 94720, USA}
\address{$^{42}$ Department of Physics, University of Chicago, 5720 S Ellis Ave, Chicago, IL 60637, USA}

\maketitle

\begin{abstract}
In this report we review recent theoretical progress and
the latest experimental results in jet substructure from the Tevatron and the LHC.
We review the status of and outlook for calculation and simulation tools for studying
jet substructure.
Following up on the report of the Boost 2010 workshop, we present a
new set of benchmark comparisons of substructure techniques,
focusing on the set of variables and grooming methods that are
collectively known as ``top taggers''.  To facilitate further
exploration, we have attempted to collect, harmonise, and publish software implementations of these techniques.
\end{abstract}




\section{Introduction}
\label{sec:introduction}

At the time of the first \boost\ meeting, at SLAC in July 2009, several groups had begun to argue that jet substructure --- the internal characteristics of hadronic jets --- could be useful in identifying the decays of heavy particles at the Large Hadron Collider.  By the following year, a trickle had become a flood.  Some techniques had received detailed attention from experimental groups, and the increasing quantity of data available meant that background studies were beginning to be possible.  Another year has passed, and the stream of theoretical advances shows no signs of abating.  Many new substructure measurements and techniques have been proposed, and significant progress has been made in developing the theoretical tools to calculate distributions in substructure observables.  Meanwhile, sufficient data now exist to study the boosted hadronic decays of heavy Standard Model particles such as the $W$ boson and top quark.

This report, an outgrowth of the \boost\ 2011 workshop held at Princeton University in May 2011, aims to summarise recent theoretical and experimental progress, outline goals for the near future, and provide benchmark comparisons and tools to help achieve these goals.

In \secrefcap{results}, we review recent substructure results at the Tevatron and LHC.  In \secrefcap{tools}, we survey new proposals for substructure techniques.  In \secrefcap{software}, we describe the new software tools for studying jet substructure in FastJet 3.  In \secrefcap{compare}, we extend the benchmark top-tagging comparisons found in last year's report \cite{Boost2010} with new methods, new Monte Carlo samples, and detector simulation.  The samples, as well as software to implement the techniques compared and our detector model, are publicly available, either as part of the FastJet 3.0.0 package or as FastJet-based tools.  Finally, in \secrefcap{theory}, we survey the status of substructure predictions and discuss goals for new calculations and measurements in the coming year.

\section{New results from the Tevatron and LHC}
\label{sec:results}
Several recent experimental results from the Fermilab Tevatron were presented at the \boost\ 2011 workshop, exploring different aspects of boosted physics.  These could be categorised into two broad classes:  studies that elucidate the behaviour of Standard Model physics processes when subjected to various degrees of boost, and searches for new physics using boosted object signatures. All the analyses were performed on samples of $\rts=1.96$~\tev proton-antiproton collisions produced over the last eight years at the Tevatron. Results from the Tevatron's two detectors are presented here in \secrefcap{results:cdf} for CDF and \secrefcap{results:dzero} for D0.

The LHC started colliding protons at a center of mass energy $\sqrt{s} = $7~\tev on March 30, 2010. Just over 18 months and 5 fb$^{-1}$ later, the ATLAS and CMS experiments have collected sufficient data to have a realistic chance of using substructure techniques to uncover massive boosted particles such as $W$/$Z$ bosons, top quarks, and whatever else may lie in wait. The experimental results presented in these proceedings focus on the results made public prior to the \boost\ workshop in May 2011. For both ATLAS and CMS this means two analyses: a paper from each on jet shapes \cite{jshapes}, \cite{cms-pas-qcd-10-014} and a conference note on jet substructure \cite{ATLAS-CONF-2011-073}, \cite{cms-pas-jme-10-013}. 

The most important and interesting results of these studies are summarised in \secrefcap{results:atlas} for ATLAS and \secrefcap{results:cms} for CMS.  Sensitivities to pile-up and detector effects, seen by the experimental community as the most pressing issues in jet and jet substructure studies, feature prominently.

All of these experiments must in varying degrees grapple with the presence of multiple simultaneous proton-(anti-)proton interactions, or pile-up, within every bunch crossing. These additional collisions are uncorrelated with the hard-scattering process that triggers the event and create a background of soft diffuse radiation that offsets the energy measurement of jets and impacts jet shape and substructure measurements. It is essential that measurements of jet substructure be able to disentangle or correct for the influence of pile-up.

Observables designed to be sensitive to the internal structure of jets are expected to also be  sensitive to pile-up \cite{Boost2010}. Large-radius jets, such as those used in the measurements of jet substructure, are naturally more susceptible to pile-up due to their larger catchment area \cite{Cacciari:2008gn}; the invariant mass of these large jets is particularly affected \cite{massareaspaper}. Techniques for correcting for these effects --- as in the CDF analyses --- or mitigating their impact --- such as the splitting and filtering procedure pioneered in ATLAS --- are essential in producing precision measurements for several of the analyses presented by the Tevatron and LHC experiments.  A more thorough review of these issues at ATLAS can be found in \cite{MillerDWThesis}.



%
%
%
\subsection{Results from CDF}\label{sec:results:cdf}

\subsubsection{Pile-up at CDF}

\begin{figure}
    \subfigure[Mass shift.]{
      \includegraphics[width=0.32\linewidth]{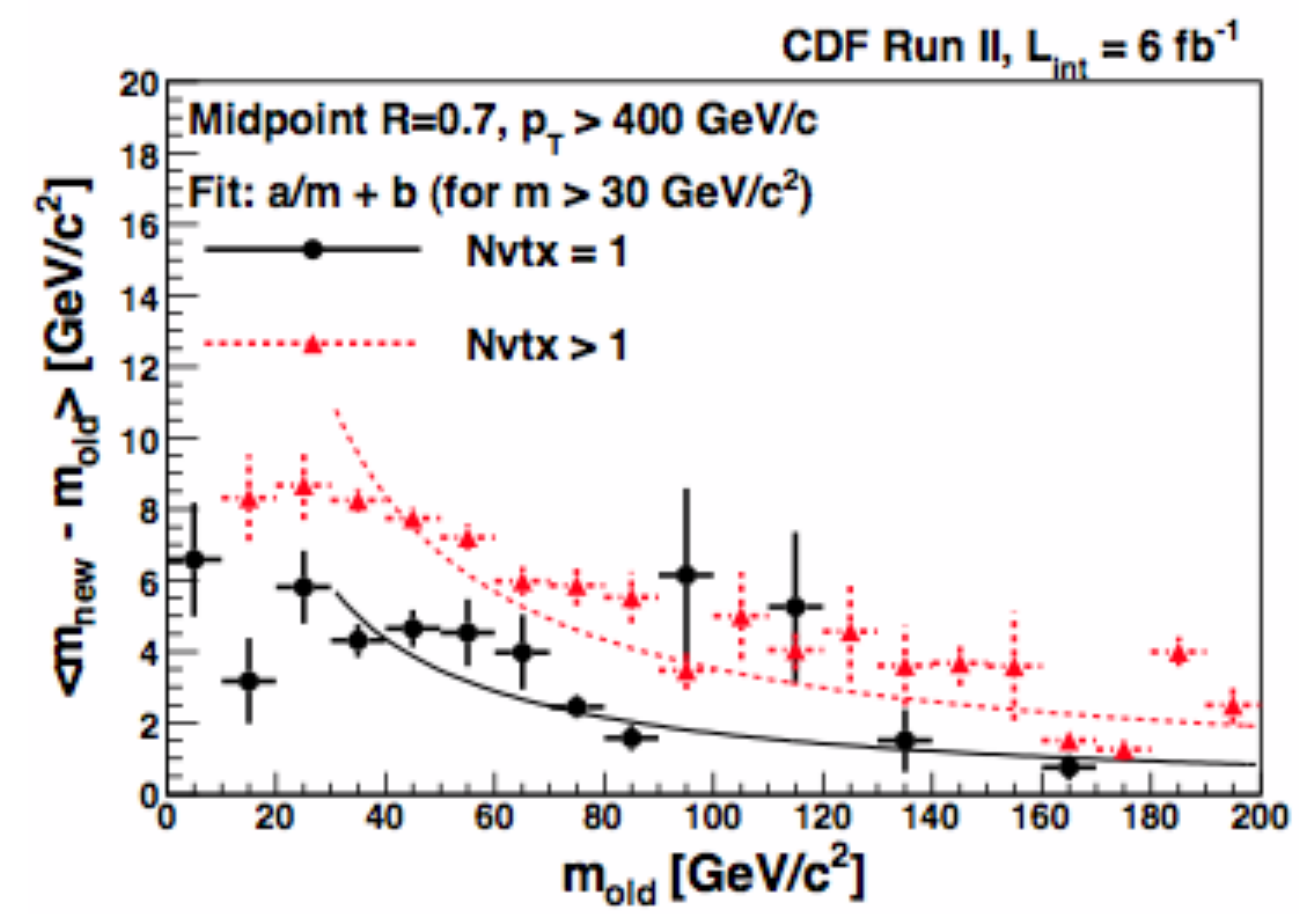}
    }
    \subfigure[Planar flow shift.]{
      \includegraphics[width=0.32\linewidth]{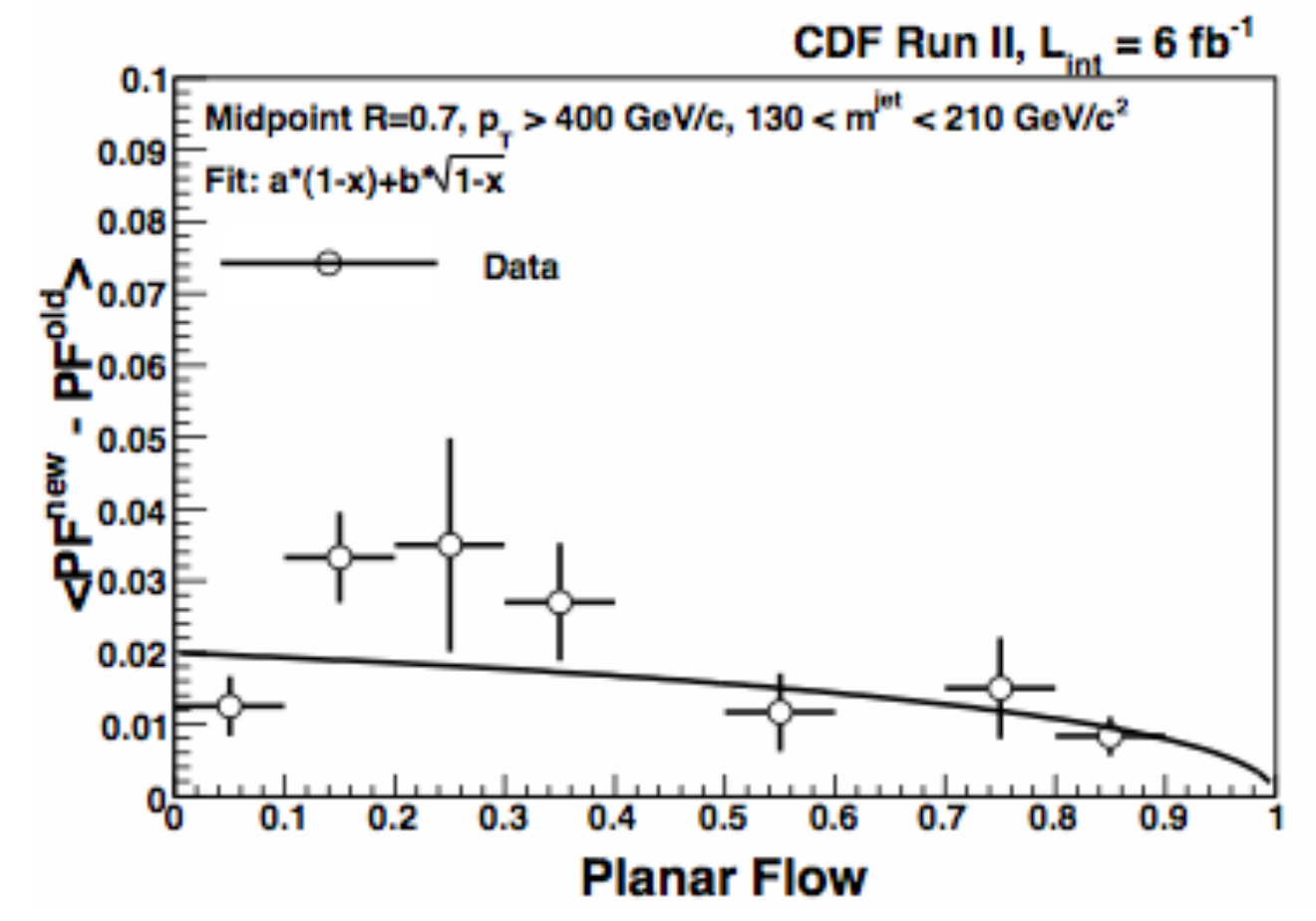}
    }
    \subfigure[Angularity shift.]{
      \includegraphics[width=0.32\linewidth]{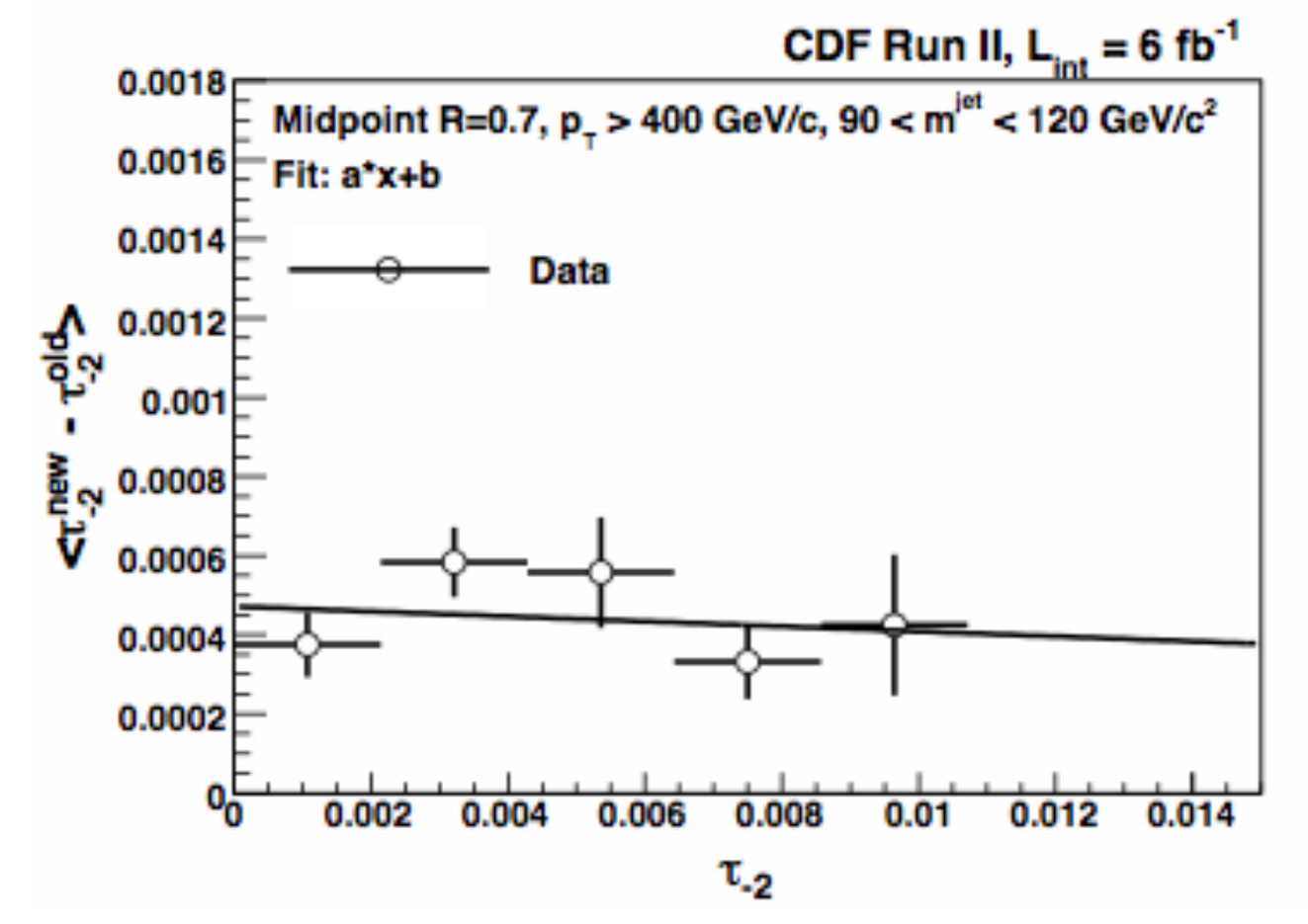}
    }
  \caption{\small
    The correction to the jet mass (a) from additional energy deposition due to MI+UE (Nvtx $>$ 1 events) compared with the jet mass corrections for UE alone (Nvtx = 1 events) for jets with a cone size of $R$ = 0.7. The estimated shift from the combination of UE and MI in planar flow (b) and angularity (c) as measured in data. The average number of collisions per bunch crossing is $\sim$ 3 for this data sample.
  }
    \label{fig:cdfpileup}
\end{figure}

In the context of measurements of the mass, angularity and planar flow of high-\pT jets \cite{CDF:subjet2010a}, a new method of correcting the jet mass for pile-up was developed. The use of a complementary cone at right angles to the jet in azimuthal angle $\phi$ and at approximately the same pseudorapidity allows the energy density from both underlying event (UE) and multiple interactions (MI) to be measured.  The underlying event refers to the interactions and hadronisation of  partons in the colliding protons other than the partons in the hard process.  Multiple interactions refer to both ``in-time'' and ``out-of-time'' pile-up: the former describing multiple proton-proton collisions in a single bunch crossing, the latter describing the delayed instrumental effects of previous crossings.  The incoherent contributions to the shift in jet mass from MI were isolated from the partially coherent contributions due to UE by examining the mass shift as a function of the number of good vertices in the event. The average shift in the jet mass when adding the towers from the complementary cone into the jet are shown in \figrefcap{cdfpileup} along with the MI+UE corrections measured for angularity and planar flow. The high mass selections made as part of the angularity and planar flow measurements resulted in too few events to separate the UE (single-vertex) and MI (single-  and multiple-vertex) components.

\subsubsection{New physics searches in multijet events at CDF}

The results of a search for pair production of a supersymmetric particle that decays strongly and violates $R$-parity were described in \cite{Aaltonen:2011sg}.
The final state of interest was at least six quarks, most observed as separate jets,  and no missing transverse energy in the event.  
The challenge for this search was to reduce the large backgrounds from QCD multijet production, which was done by placing specific kinematic requirements on three-jet triplets in the final state.

The analysis sought events in a CDF sample that satisfied a trigger requiring at least four jets with $\pT> 15$ \gev and the sum of the calorimeter transverse energy greater than 175 \gev.
Events were furthermore required to have at least six jets with $\pT > 15$ \gev and $|\eta|<2.5$, and the scalar sum of the six highest-energy jets was required to be greater than 250 \gev.
The missing transverse energy, $\met$, was required to be less than 50 \gev in order to reduce contributions from $W$ boson final states and mis-measured QCD events.
 
All twenty combinations (or more) of three-jet triplets that could be produced from jets with $\pT > 15$ \gev were considered and the invariant mass of each combination, $M_{jjj}$, and the sum of the magnitudes of the transverse momenta of the three jets, $\sum_{jjj} |\pT|$, were formed.  
Monte Carlo studies have shown that requiring 
\begin{eqnarray}
\sum_{jjj} |\pT| - M_{jjj} > \Delta
\end{eqnarray}
is an efficient way of separating potential signal combinations (where $M_{jjj}$ would be a constant reflecting the mass of the supersymmetric parent) from the QCD and combinatorial backgrounds. $\Delta$  is a constant optimized for each assumed parent mass.
This is in effect a boosted three-jet final state.
The shape of the backgrounds in $M_{jjj}$ were estimated by using a five-jet final state, showing that the background is expected to peak around 100 \gev.  

The $M_{jjj}$ distribution was then fit to a combination of signal and background terms, where the signal was defined by a \pythia\ Monte Carlo calculation for $R$-parity violating (RPV) gluino pair production.
Although the acceptance for the gluino final state is quite low (roughly $5\times 10^{-5}$), CDF was able to set significant limits on the RPV gluino cross section, which were then converted into lower limits on the gluino mass.
Lower mass limits ranging from 144 to 154 \gev at 95\% C.L. were set on the gluino mass, depending on the assumptions about the spectrum of intermediate supersymmetric final states.
What is perhaps as interesting is that CDF observed evidence for a boosted top quark signature.  In the top quark mass region of $M_{jjj}\sim175$ GeV, CDF observed $11\pm5$ jet triplets. 
Although one expects on average only one top quark event in this kinematic region, the shape of the mass distribution is consistent with what one expects from MC simulations.  

\subsubsection{Boosted top quark search at CDF}

CDF presented updated results on the measurements of jet mass, angularity and planar flow for jets with $p_T > 400$ \gev from a sample of 5.95 \invfb \cite{CDF:boostedtop}.  
The measured distributions were compared with analytical expressions from NLO QCD calculations, as well as \pythia\ 6.1.4 predictions incorporating full detector simulation.  
The theory predictions for jet mass were in good agreement with the data, whereas the angularity and planar flow predictions by \pythia\ showed disagreement in detail (primarily at low angularity and low planar flow).

\begin{figure}
\hspace{-.25cm}
    \subfigure[{{\sc pythia}} $t\bar{t}$ signal.]{
      \includegraphics[width=0.31\linewidth]{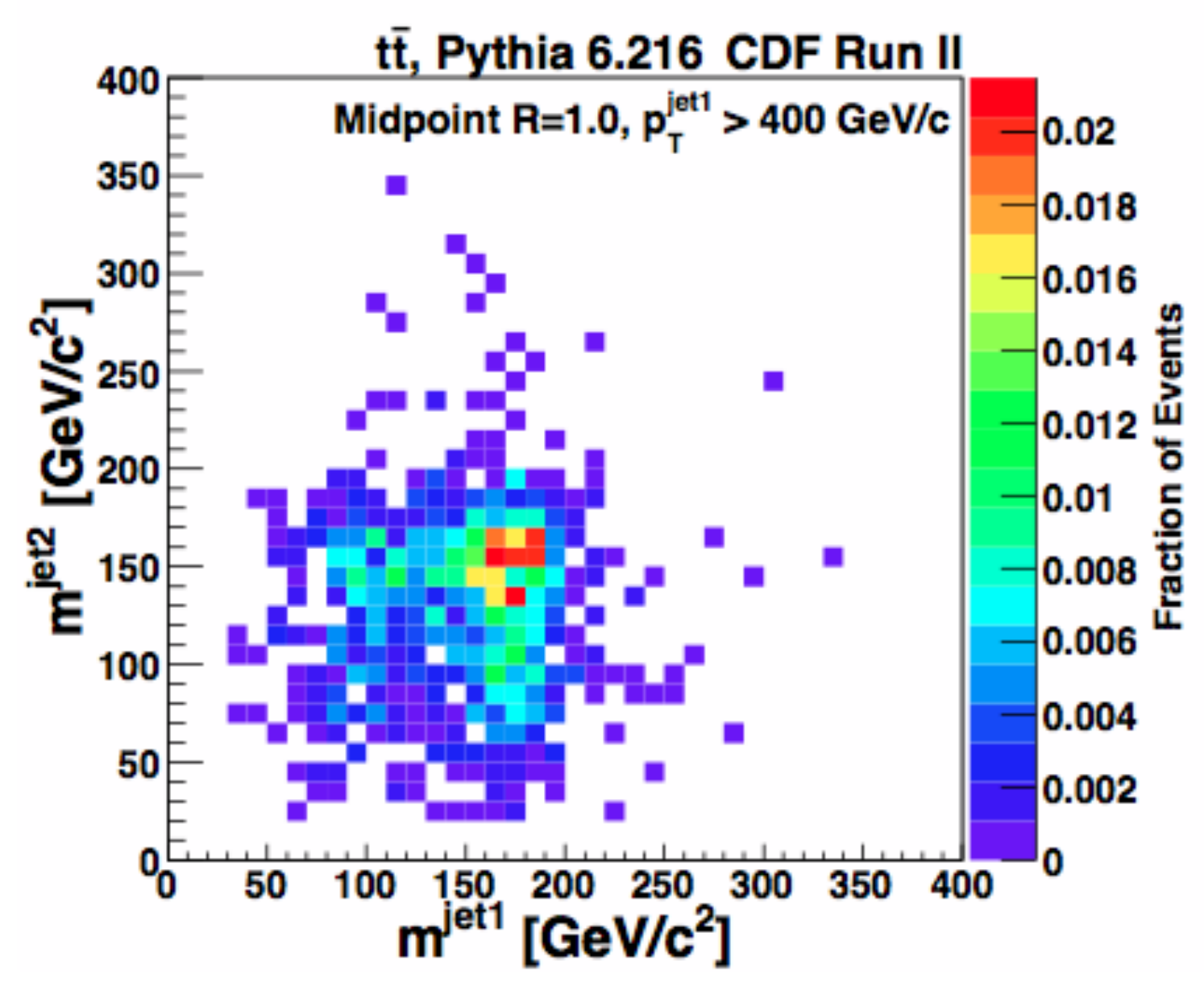}
      \label{fig:m1m2signal}
    }
    \subfigure[{{\sc pythia}} QCD dijets.]{
      \includegraphics[width=0.31\linewidth]{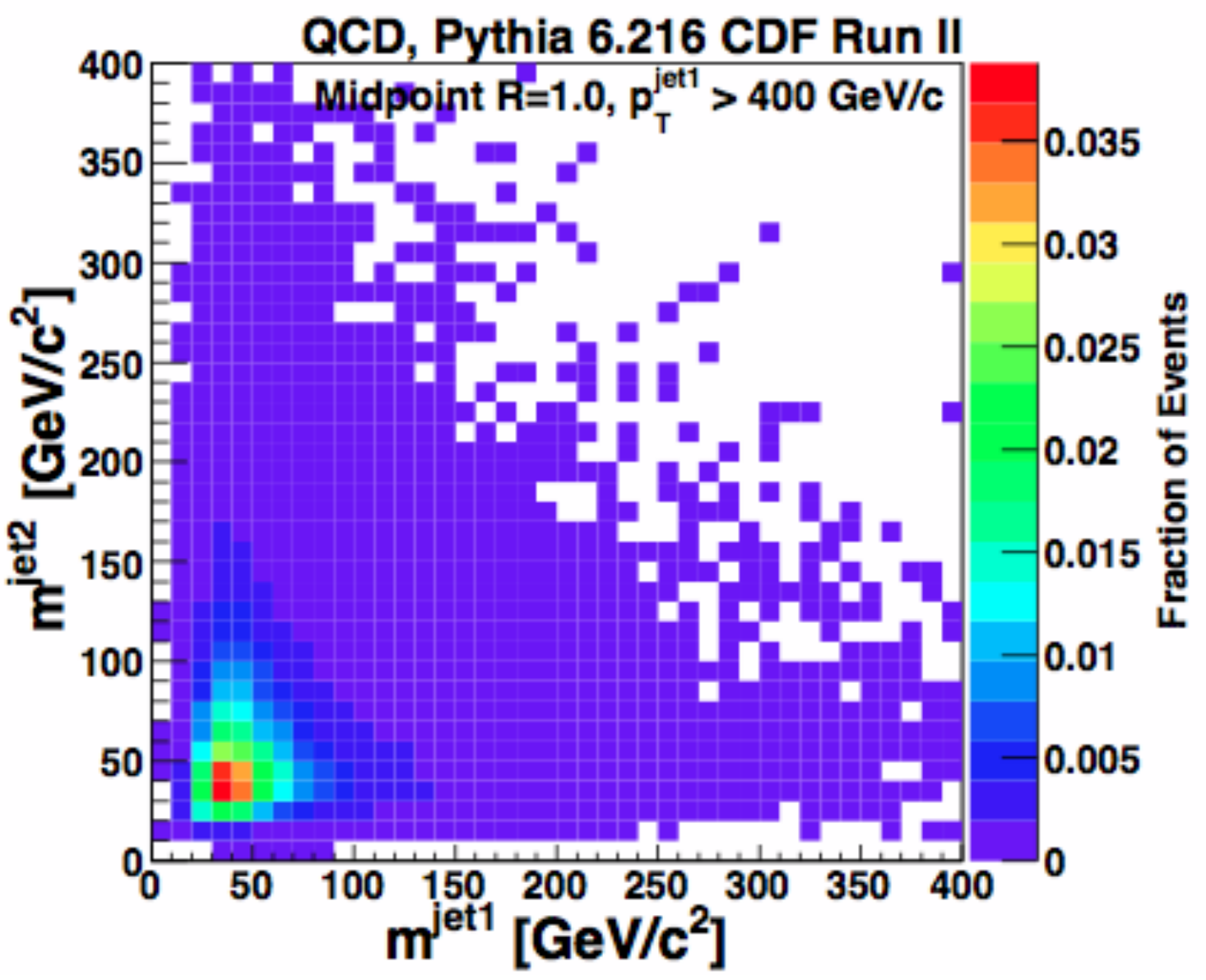}
      \label{fig:m1m2bg}
    }
    \subfigure[CDF Run II data, 6 fb$^{-1}$.]{
      \includegraphics[width=0.31\linewidth]{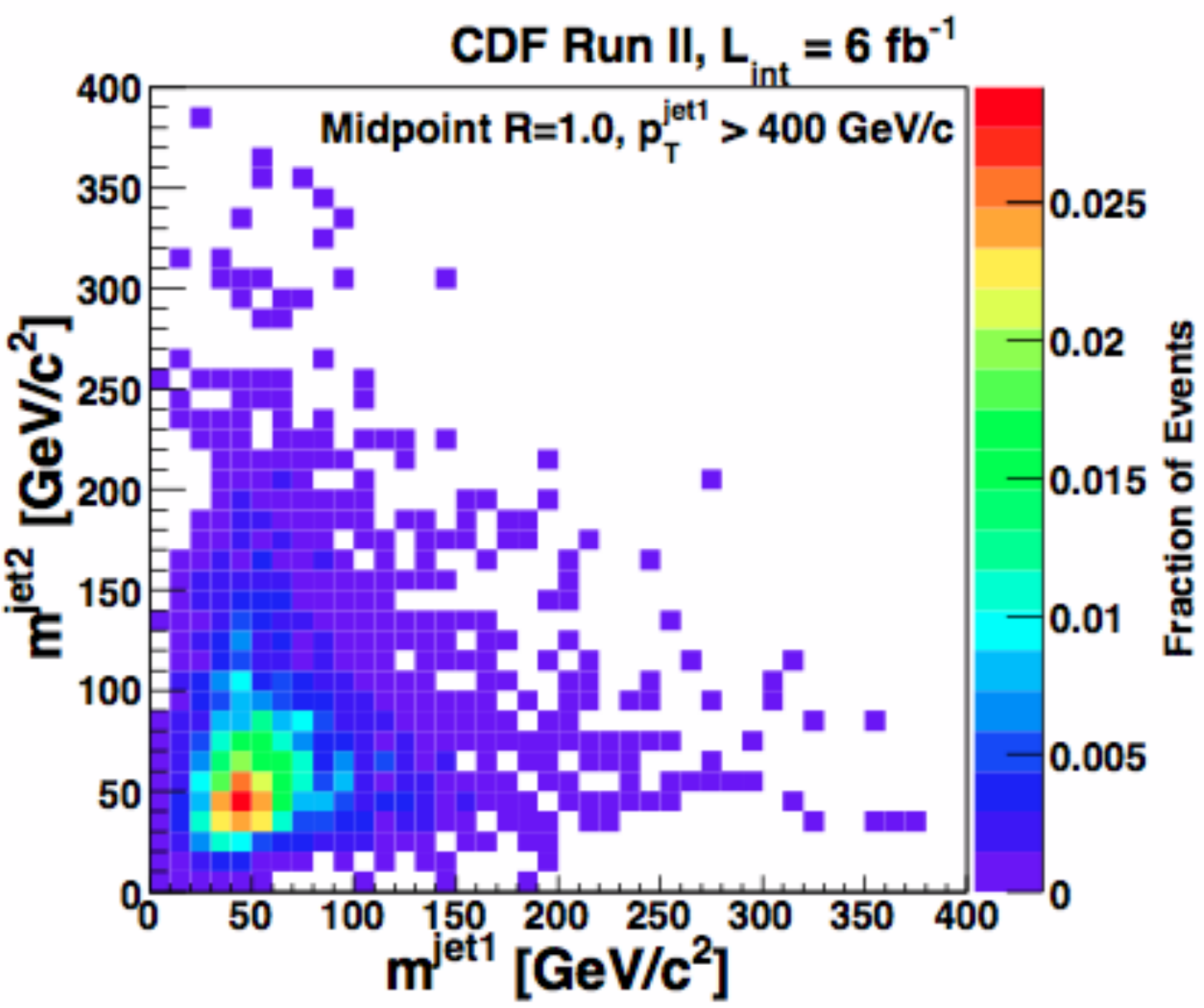}
      \label{fig:m1m2data}
    }
  \caption{\small
    The $m^\text{jet2}$ versus $m^\text{jet1}$ distribution for all events with at least one jet with $\pT > 400$ GeV and $\eta < 0.7$, using $R=1.0$ Midpoint cones.  MI corrections have been performed and all events are required to have $S_{MET} < 4$.
  }
    \label{fig:cdfm1m2}
\end{figure}

CDF used these data to also search for a signal of boosted top quark production.
Candidate events were selected in two channels:  the fully hadronic decays where both top quarks produce massive high-$\pT$ jets (the ``1+1'' channel), and the decays where one top quark decays semi-leptonically resulting in one massive high-$\pT$ jet recoiling against a second lower-mass jet and significant missing transverse energy (the ``SL'' final state).  
A signal region in the 1+1 mode was defined by requiring both leading jets to have a mass between 130 and 210 \gev (see \figrefcap{cdfm1m2}), while the signal region in the SL mode requires the leading jet to have $130 < m^\text{jet1} < 210$ \gev and the event to have missing transverse energy significance
\begin{eqnarray}
S_{MET} \equiv \met / \sqrt{\sum E_T} \in (4,10).
\end{eqnarray}
This cut rejects primarily QCD dijet events.
The remaining QCD backgrounds are estimated by looking at the event rates in sideband regions of jet mass and $S_{MET}$.  
The analysis observes 31 and 26 candidate events in the 1+1 and SL channels, respectively, and the backgrounds are estimated to be $14.6\pm2.7$ and $31.3\pm8.1$ events, respectively.  
The total number of top quark events expected in the two channels is $4.9\pm 2.1$ candidates.  

Although the data are consistent with a boosted top quark signature, they are not statistically strong enough to claim observation.
Rather, they are used to set an upper limit of 38 fb on top quarks produced with $\pT>400$ \gev.  
The SM expectation for this cross section is 4.5 fb.  

\subsubsection{Search for lepton jets at CDF}

There are several theories beyond the Standard Model that predict the production of cascades of particles that appear in the final state as a ``lepton jet''.  
The CDF collaboration reported on a search for such objects using 5.1 \invfb of $\rts = 1.96$ \tev proton-antiproton collisions at the Tevatron.
The search looked for events with a large number of low-energy leptons produced in association with a $W$ or $Z$ boson.  

Events were selected by performing the standard selection for $W$ or $Z$ boson candidates, requiring at least one well-identified electron or muon candidate and then either significant \met or a second well-identified charged lepton of the same flavour but opposite charge.  
A ``soft lepton'' algorithm was then employed to identify additional electron or muon candidates down to a transverse momentum of 1 \gev for electrons and 3 \gev for muons.
The numbers of events in the zero or one additional lepton bins were used to scale the expected backgrounds for the signal region defined by two or more soft lepton candidates.  
The analysis found that the dominant backgrounds came from inclusive $W$+jets production where one or more of the jets were mis-identified, or Drell-Yan production where the additional leptons were mis-identified jets.
Other sources of background, such was $W$+$c$-quark, $W$+$b$-quark, \ttbar and di-vector boson production were also evaluated.

The potential signal was modelled on a benchmark process defined by a neutralino model with a ``hidden'' Higgs boson coupled to a dark sector, where the dark sector particles decay to pairs of charged leptons \cite{Falkowski:2010cm}.  
The channels with the best signal-to-background for this model were those with a $W$ or $Z$ boson with at least three additional muons, either with none or one additional electron; for example, in the channel with a $W+3$ additional muons, one expected $1.5\pm1.2$ background events and nine signal events (only two events were observed).
There was no signal observed above background, and a 95\% C.L. upper limit on the production cross section of a $W$ or $Z$ boson produced in association with a Higgs boson with the expected couplings of 27 fb was set.
This allowed the collaboration to rule out the benchmark model.

This analysis showed the effectiveness of using the lepton jet signatures to search for evidence of new physics.

\subsection{Results from D0}\label{sec:results:dzero}

\subsubsection{Colour Flow in D0 \ttbar Events}
The D0 experiment presented recent results showing how the colour flow that is expected to arise between two jets can be used as a discriminant to identify $W$ boson hadronic decays in a 5.3~\invfb sample of \ttbar events \cite{PhysRevD.83.092002}.
Since the $W$ decay products form a colour singlet, they produce an antenna radiation pattern, with most soft particles emitted between the two jet directions.
The ``pull'' of the jets' radiation toward each other can be used to more effectively identify dijet systems arising from a specific colour state.

\begin{figure}
  \subfigure[]{
  \includegraphics[width=0.5\linewidth]{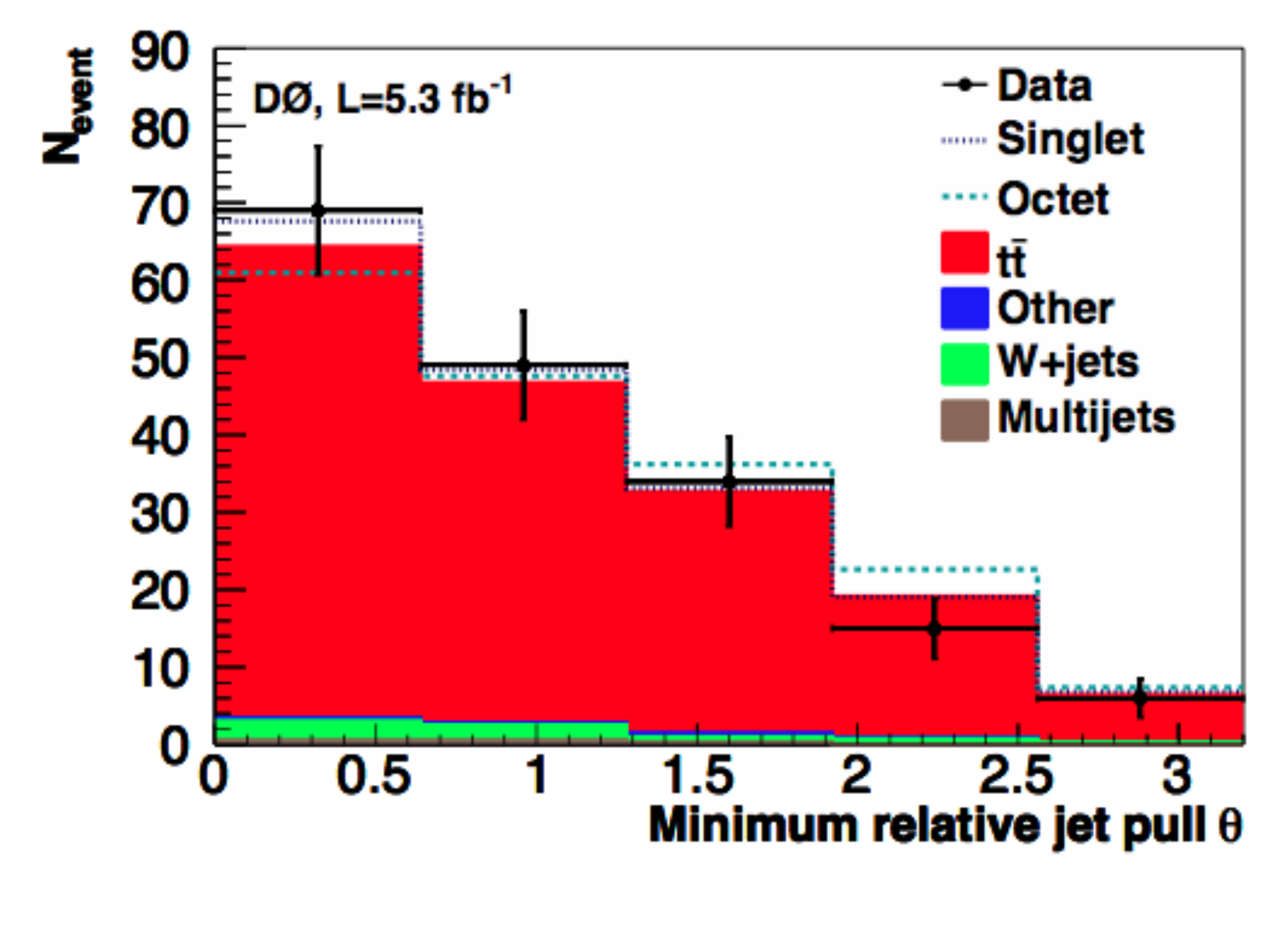}
  \label{fig:dzeropullw}
  }
  \subfigure[]{
    \includegraphics[width=0.45\linewidth]{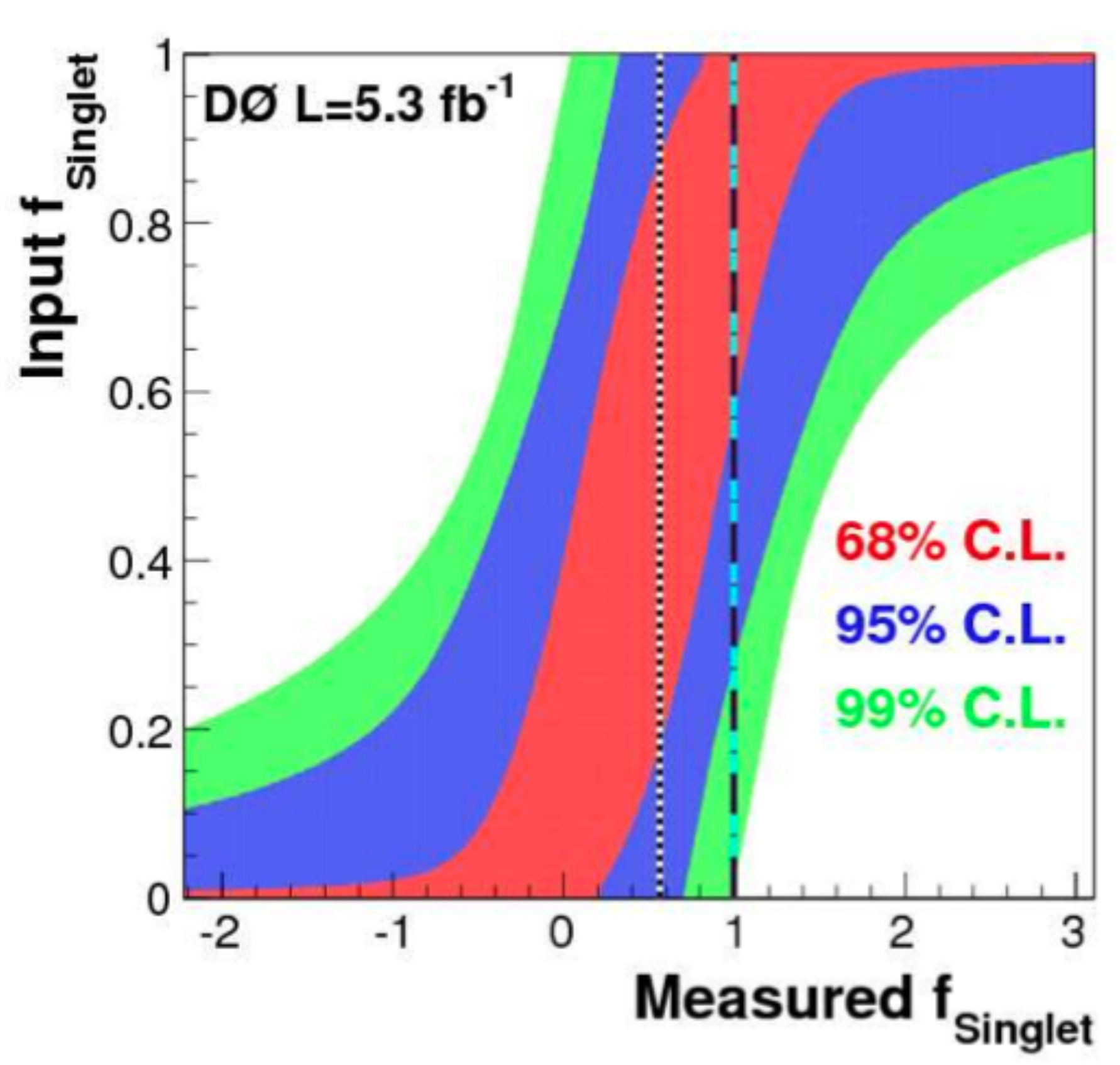}
    \label{fig:dzerofsinglet}
  }    
  \caption{\small
    The minimum relative pull angle for $W$ jets \subref{fig:dzeropullw} is measured for jets satisfying $|m_{W} - m_{j1j2}| < 30$ \gev, $\Delta R_{j1,j2} < 2$ and $|\eta| < 1$. 
    The expected C.L. bands for $f_\text{singlet}$ are shown in \subref{fig:dzerofsinglet}; the dashed line shows the expected value and the dotted line indicates the measured value of $f_\text{singlet}$.
  }
\end{figure}

Events were selected by requiring the traditional lepton+jets final state, with a charged lepton, missing transverse energy and four or more jets.
At least two of the jets had to be tagged as b-quark candidates, resulting in 728 candidate events with an estimated background rate of $82\pm9$ events in the sample not arising from \ttbar production.
The jet pairs that had an invariant mass within 30 \gev of the $W$ boson mass were then selected and the shape of the energy depositions studied to look for a signal consistent with the colour singlet. The minimum relative pull angle for jet pairs satisfying $|m_{W} - m_{j1j2}| < 30$ \gev, $\Delta R_{j1,j2} < 2$ and $|\eta| < 1$ is shown in \figrefcap{dzeropullw}.
The expected colour effect in the relative orientations of the jet pulls was observed when comparing the daughter jets from $W$ boson decays and the $b$-quark jets, though the statistical power of the measurement was modest and the kinematics of the pairs of jets were not identical.  
A more direct comparison was made by producing Monte Carlo \ttbar decays with a $W$ boson produced as a colour octet compared with the expected colour singlet state, and a series of detailed studies were performed to evaluate the systematic uncertainties on this measurement.

The actual quantity measured was the fraction of singlet $W$ boson decays compared with the total number, giving
\begin{eqnarray}
f_\text{singlet} = 0.65^{+0.37}_{-0.38}\, {\rm (stat.)} \pm 0.22\, {\rm (syst.)}.
\end{eqnarray}
This could be turned into a lower limit on the fraction of singlet decays of $f_\text{singlet}>0.277$ at 95\% C.L.
Although this is somewhat higher than the limit that might have been expected given the data sample, it is consistent with the expectation within statistical uncertainties. The expected C.L. bands for this limit are shown in \figrefcap{dzerofsinglet}.

This means of measuring the colour flow provides an additional discriminant that can be used to search for decays of other boosted objects (such as $H \rightarrow b\bar{b}$) and gives us confidence that the modelling of colour flow is being done well in current Monte Carlo calculations.  

\subsection{Results from ATLAS}\label{sec:results:atlas}

\subsubsection{ATLAS and pile-up}\label{sec:pileupatlas}

\begin{figure}
    \subfigure[Effect from jet area.]{
      \includegraphics[width=0.5\linewidth]{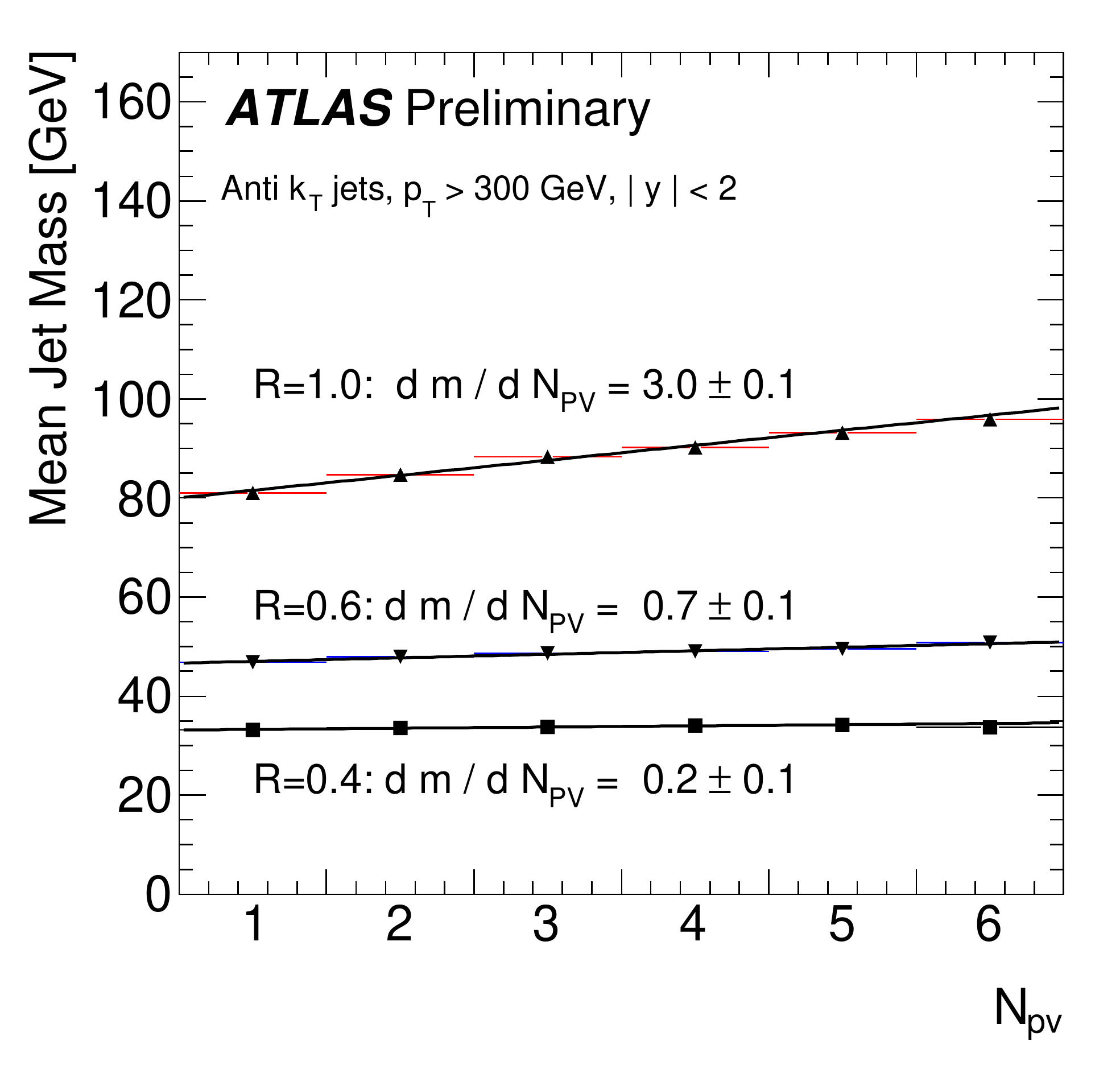}
      \label{fig:pileupeffect:radius}
    }
    \subfigure[Effect from filtering.]{
      \includegraphics[width=0.5\linewidth]{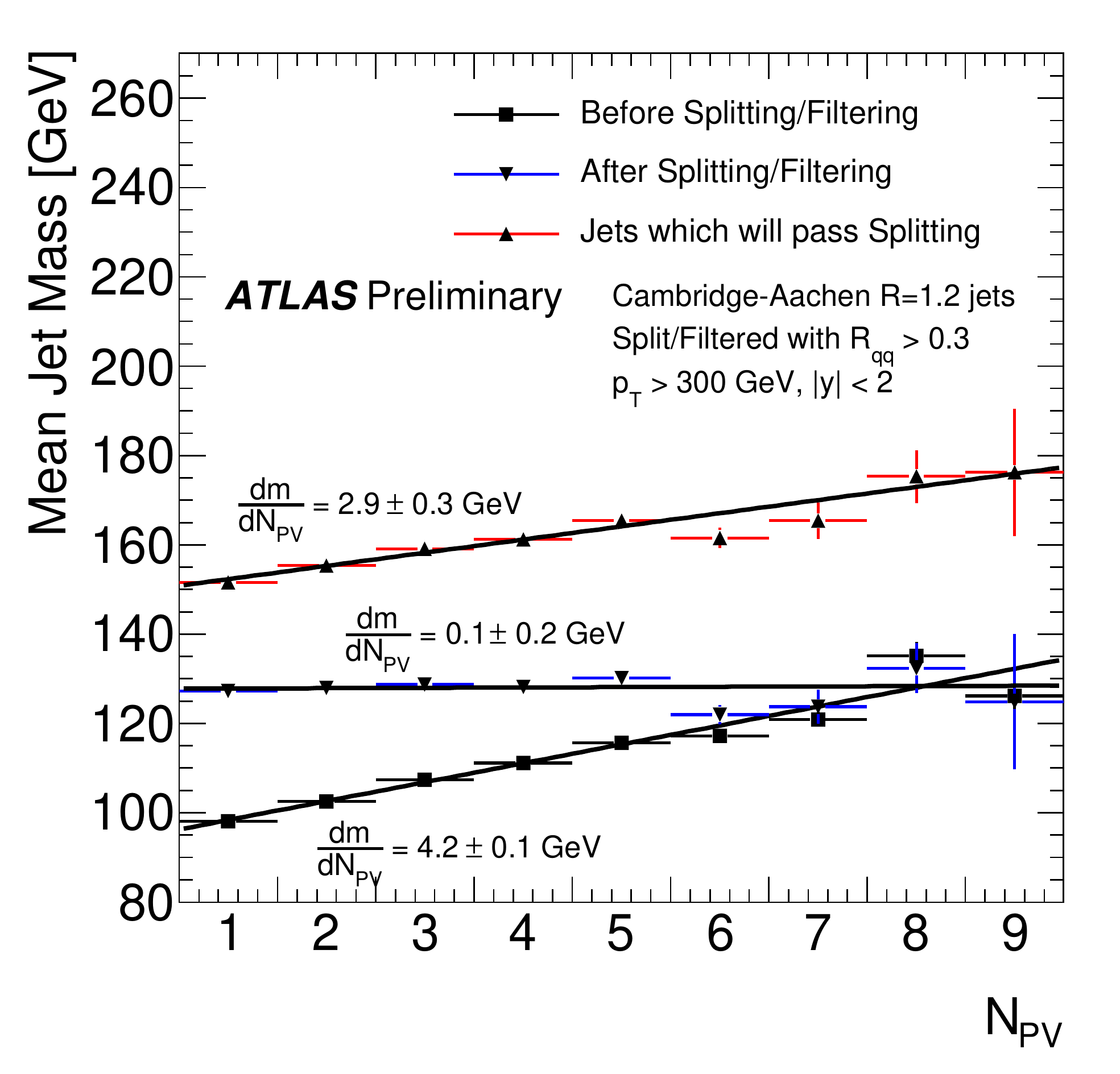}
      \label{fig:pileupeffect:filtering}
    }
  \caption{\small
The mean mass for jets with $\pt > 300$  \gev as a function of
    the number of primary vertices identified in the event.  Comparisons
    show the effect for \subref{fig:pileupeffect:radius} \antikt jets with different resolution
    parameters  and \subref{fig:pileupeffect:filtering} \ca $R=1.2$ jets with and without a splitting and
    filtering procedure. Each set of points is fitted with a straight line. }
    \label{fig:pileupeffect}
\end{figure}

In the recent ATLAS analysis of 35 pb$^{-1}$ of data \cite{ATLAS-CONF-2011-073} , the sensitivity of the individual jet mass to pile-up is directly tested. The increase in the mean jet mass as a function of the number of reconstructed good vertices (good vertices being defined as those with at least 5 tracks with $p_{T}>$ 150 MeV), \Npv, is measured for jets with a \pt of at least 300  \gev and rapidity $|y|<2$. The results are shown in \figrefcap{pileupeffect} for jets of various radii and built using two different jet algorithms.  The mean jet mass is observed to increase  linearly with \Npv.  For events with only a single primary vertex, the mean jet mass increases linearly with $R$. The slope of the mean jet mass as a function of the number of primary vertices, though, exhibits a dependence on $R$ that is consistent with the ratio of the third power of the jet resolution parameter.  This follows the prediction of pQCD calculations \cite{Dasgupta:2007wa}: the extra mass from an additional particle $k$ scales like $\delta m^2 \approx 2 \pt \cdot k \sim \pt \pt^k \Delta R_k^2$, so the net change from a uniform pile-up \pt density $\rho$ is $\delta m^2 \sim \pt \rho \int d A \Delta R^2 \sim \pt \rho R^4$, and hence $\delta m \sim R^4 \rho \pt/m$.  Since $m \sim R$, the derivative $dm/d\rho$ then scales like $R^3$ for fixed \pt.

\figrefcap{pileupeffect} also shows the impact of pile-up on each of the three jet algorithms used in the ATLAS study. Specifically, \figrefcap{pileupeffect:filtering} shows the dependence on \Npv for the mean jet mass before and after the splitting and filtering procedure. The filtering procedure significantly reduces the effect of pile-up on jet mass. In fact, the slope of a straight line fit is consistent with zero. This result demonstrates that certain well-designed jet grooming procedures can significantly reduce the impact of pile-up and that the efficacy of the method can be tested \textit{in-situ}. Methods to correct for these effects and to provide more general pile-up mitigation procedures using jet grooming are underway in ATLAS.

\subsubsection{Jet shapes with ATLAS}\label{sec:results:atlasjetshapes}

The study of jet shapes in proton-proton collisions provides 
information about the details of the parton-to-jet fragmentation process \cite{jetshape}. The internal structure of sufficiently energetic jets is 
primarily dictated by the emission of  multiple gluons from the primary parton, 
calculable in perturbative QCD (pQCD). The shape of the jet depends on the type of parton 
(quark or gluon) and is also sensitive to non-perturbative fragmentation 
effects and underlying event (UE) contributions from the interaction between
proton remnants. A proper modelling of the soft contributions is crucial 
for the understanding of  jet production in hadron-hadron collisions 
and for comparison of the jet cross section measurements with  pQCD  
theoretical predictions.

The ATLAS collaboration recently published results on differential and integrated jet shapes 
for jets constructed with the \akt 
algorithm \cite{Cacciari_antikt} 
with distance parameter $R=0.6$ \cite{jshapes}. 
The jet shapes were calculated using the constituents associated with the jets, which were 3D topological calorimeter clusters \cite{Lampl:1099735}.

The analysis follows similar measurements undertaken by CDF for $p\bar{p}$ 
collisions \cite{cdf_dijet,Abe:1992wv,Abachi:1995zw}, by ZEUS for $e^{\pm}p$ \cite{ZEUS:subjet2004,Breitweg:1998gf,Adloff:1998ni,Breitweg:1997gg},
and OPAL for $e^+ e^-$ \cite{Akers:1994wj,Ackerstaff:1997xg} collisions. 

The ATLAS analysis uses data corresponding to $3$ pb${}^{-1}$ of total integrated luminosity, collected during 2010. 
This period of data collection can safely be referred to as ATLAS's early days, when pile-up was not the burning issue that it quickly became in 2011, after about 35 pb$^{-1}$ had been collected.
In 2010 the average number of good vertices in a single hard scatter was  approximately 2.
In 2011 data this average is closer to 10, and increasing with luminosity.

The measurements are carried out for jets with $\ptjet > 30$  \gev and $|\rapjet|<2.8$.

The measurements of integrated and differential jet shapes, $\Psi$ (\erefcap{psi}) and $\rho$ (\erefcap{rho}), are corrected to hadron level and compared to several Monte Carlo predictions.
Different phenomenological models are employed to describe the fragmentation processes and UE contributions.

The comparisons presented in \cite{jshapes} were complemented with additional results in a second 
publication \cite{pubnote} including the predictions from new generators and up-to-date MC tunes, and are reported in this contribution to the proceedings.  

The differential jet shape $\rho(r)$ as a function of
the distance 
 $r=\sqrt{\Delta y^2 + \Delta \phi^2}$ to the jet axis is
defined as the average fraction of the 
jet $\ptjet$ that lies inside an annulus of inner radius
${r - \Delta r/2}$ and outer radius ${r + \Delta r/2}$ around the jet axis:

\begin{equation}\label{eq:rho}
\rho (r) = \frac{1}{\Delta r}\frac{1}{\njet}\sum_{\rm jets}\frac{p_{T}(r-\Delta r/2,r+\Delta r/2)}{p_{T}
(0,R)},\ \Delta r/2 \le r \le R-\Delta r/2, 
\end{equation}

\noindent
where $  p_{T}(r_1,r_2)$ denotes the summed $p_{T}$ of the clusters in the annulus between 
radius $r_1$ and $r_2$, $\njet$ is the number of jets, and $R=0.6$ and $\Delta r = 0.1$ are used. 

 The points from the differential jet shape 
at different $r$ values are correlated since, by definition, $\sum_0^R \rho (r) \ \Delta r = 1$. 
Alternatively, the integrated jet shape $\Psi(r)$ is defined as the 
average fraction of the jet $\ptjet$ that lies inside a cone of radius $r$ 
concentric with the jet cone:

\begin{equation}\label{eq:psi}
\Psi (r) = \frac{1}{\njet}\sum_{\rm jets}\frac{p_{T}(0,r)}{p_{T}(0,R)},\ 0 \le r \le R, 
\end{equation}

\noindent
where, by definition, $\Psi (r = R) = 1$, and the points at different $r$ values are correlated. 

\begin{figure}[tbh!]
\subfigure[{\sc pythia} tunes.]{\includegraphics[width=0.5\textwidth]{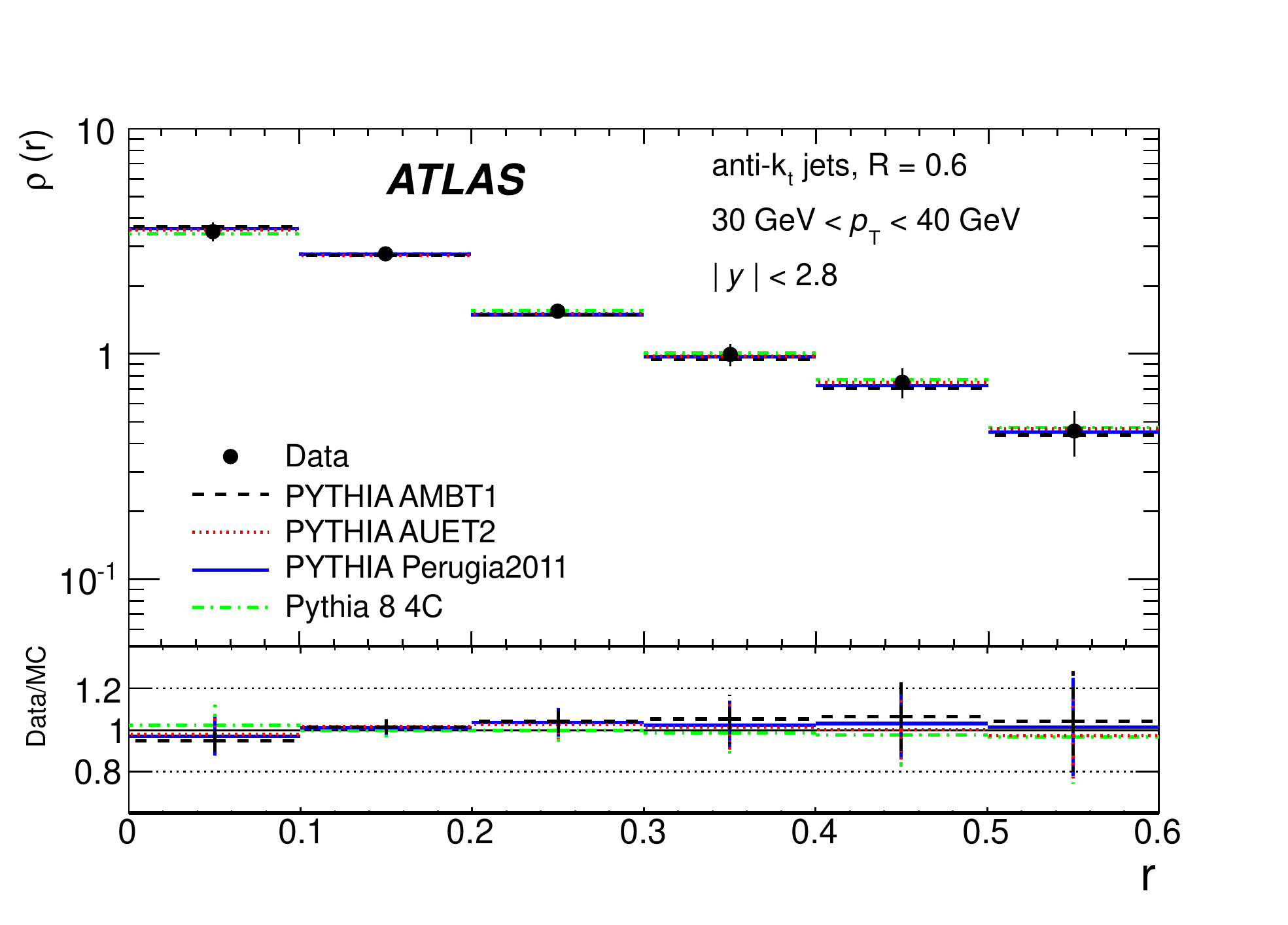}
\label{fig:diffshapes_pythia}}
\subfigure[{\sc herwig++} tunes.]{\includegraphics[width=0.5\textwidth]{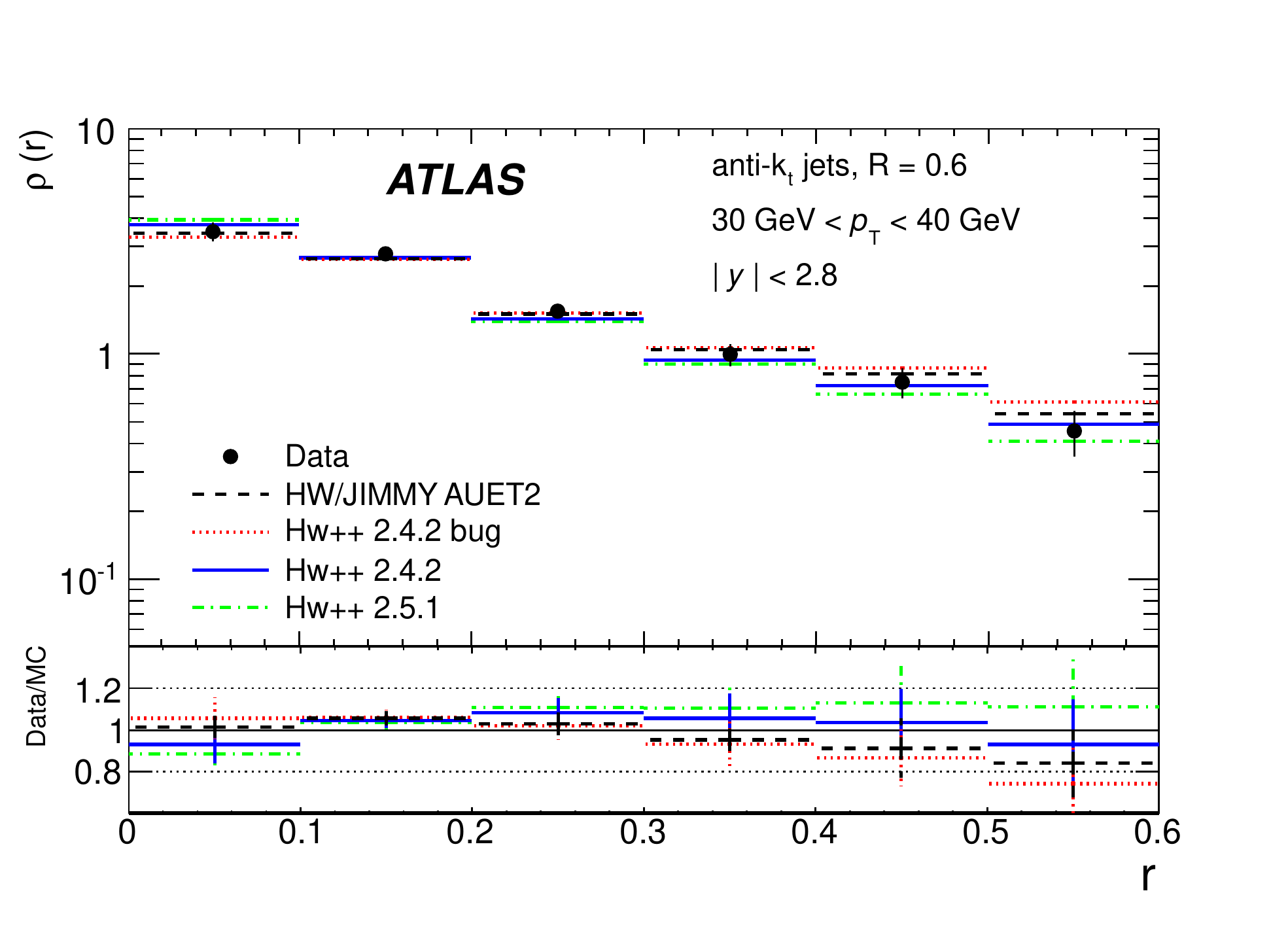}
\label{fig:diffshapes_herwig}}

\subfigure[{\sc sherpa} and {\sc alpgen}.]{\includegraphics[width=0.5\textwidth]{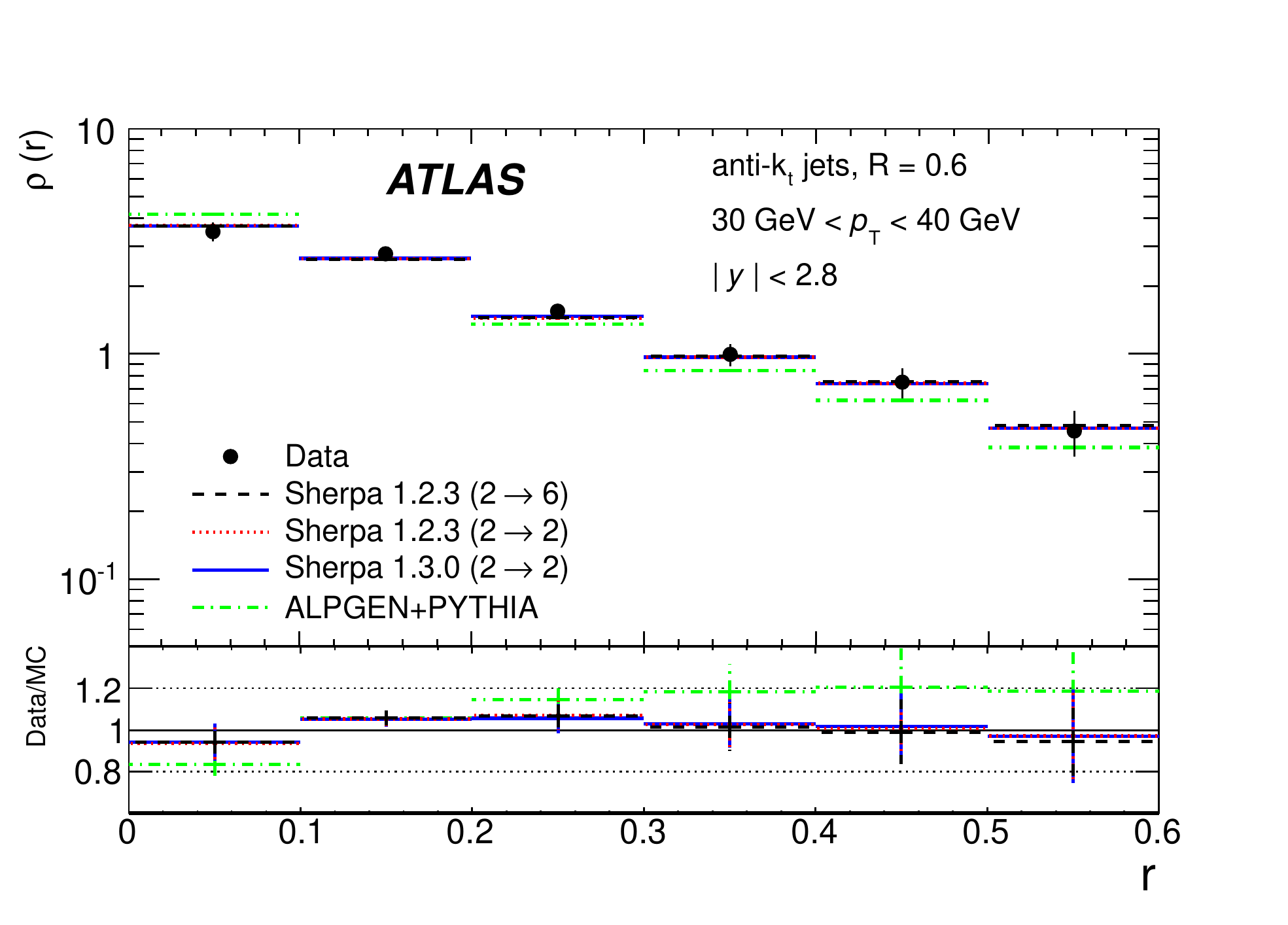}
\label{fig:diffshapes_sherpa}}
\subfigure[{\sc powheg} comparison.]{\includegraphics[width=0.5\textwidth]{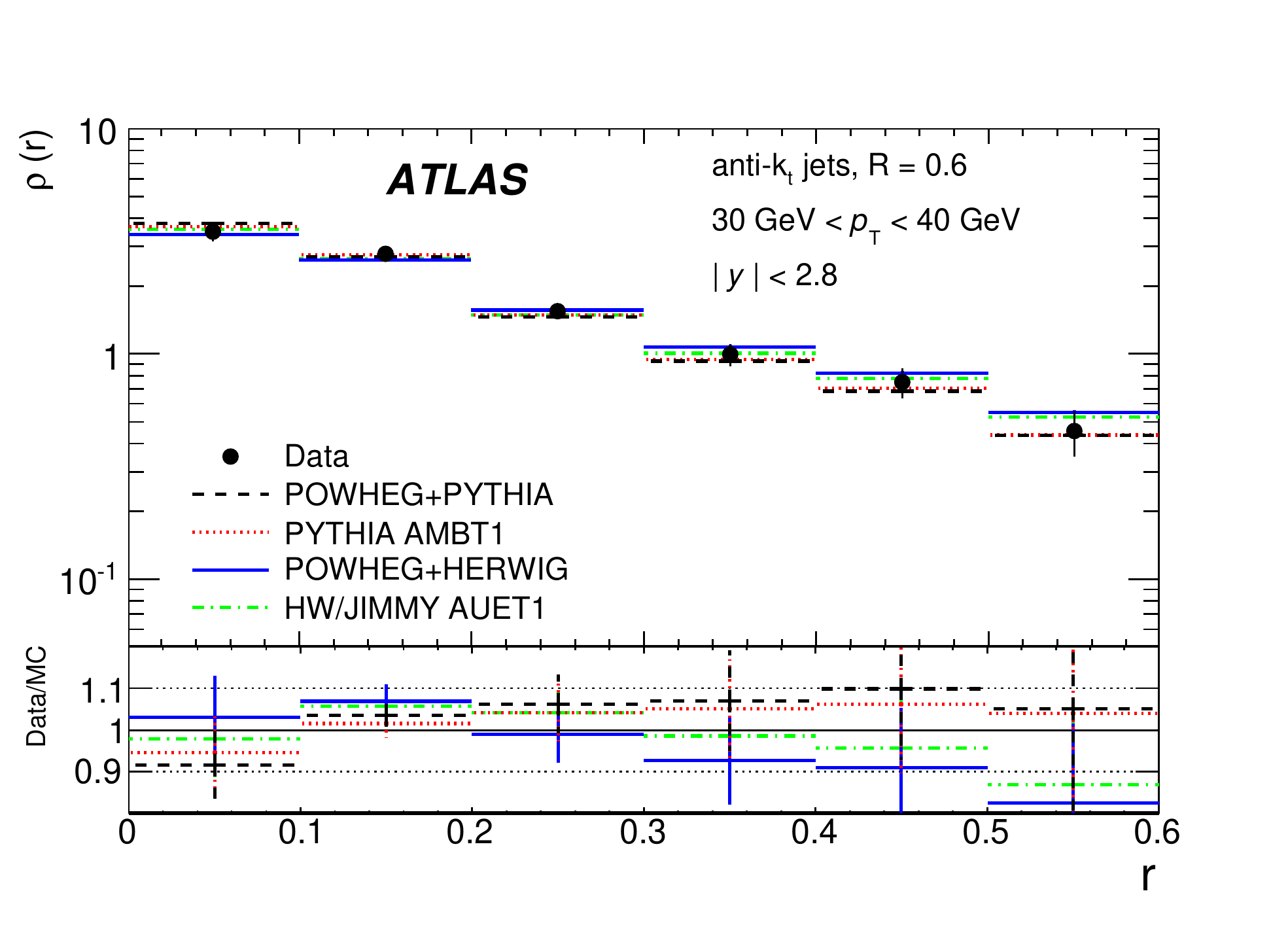}
\label{fig:diffshapes_powheg}}
\caption{\small
The measured differential jet shape, $\rho(r)$, in inclusive jet production for jets
with $|\rapjet| < 2.8$ and $30 \gev < \ptjet < 40  \gev$. Error bars indicate the statistical and systematic uncertainties added in quadrature. The measurements are compared to different MC predictions.}
\label{fig:rho}
\end{figure}

\begin{figure}[tbh!]
\subfigure[{\sc pythia} tunes.]{\includegraphics[width=0.5\textwidth]{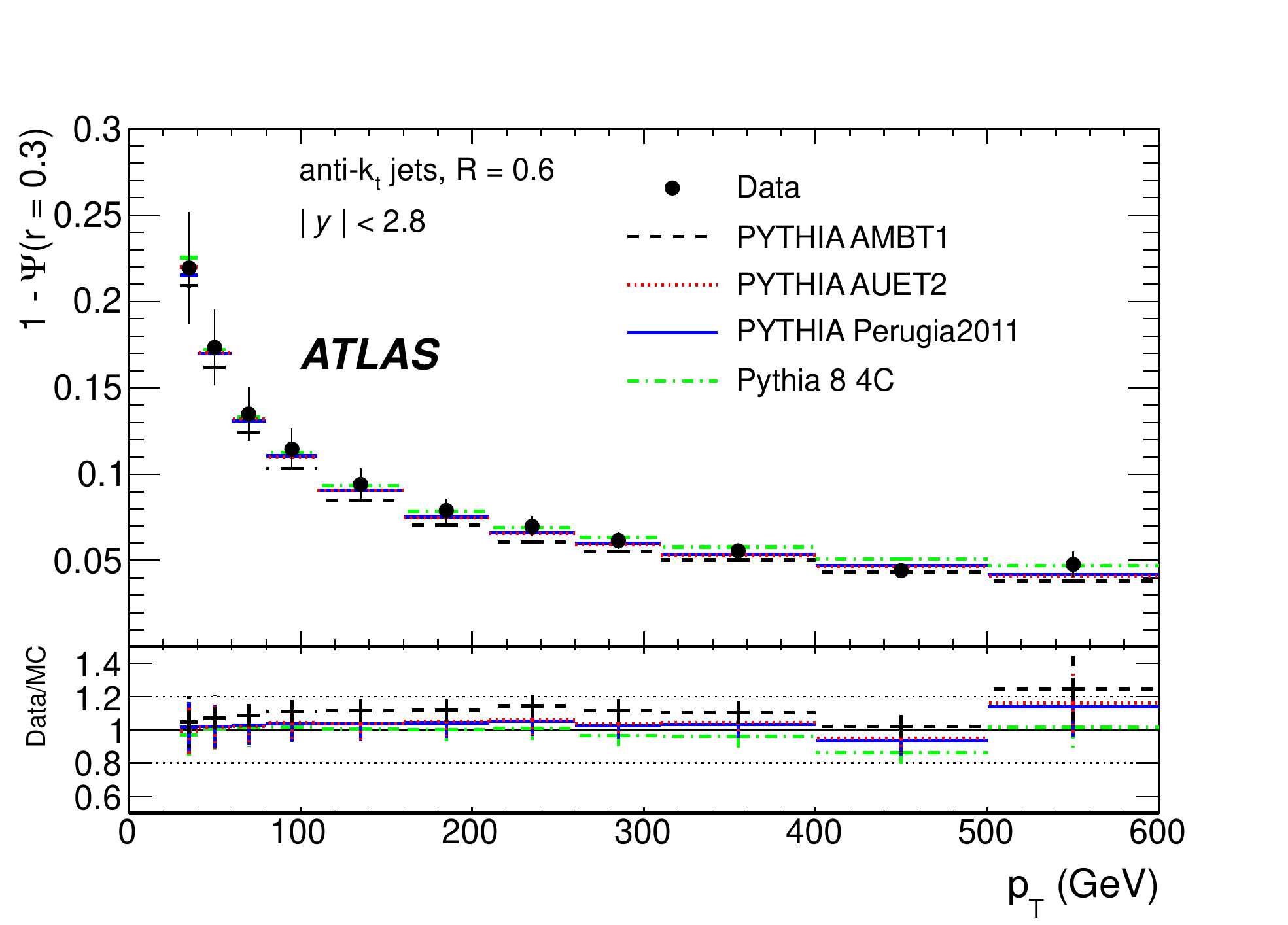}
\label{fig:shapes_pythia}}
\subfigure[{\sc herwig++} tunes.]{\includegraphics[width=0.5\textwidth]{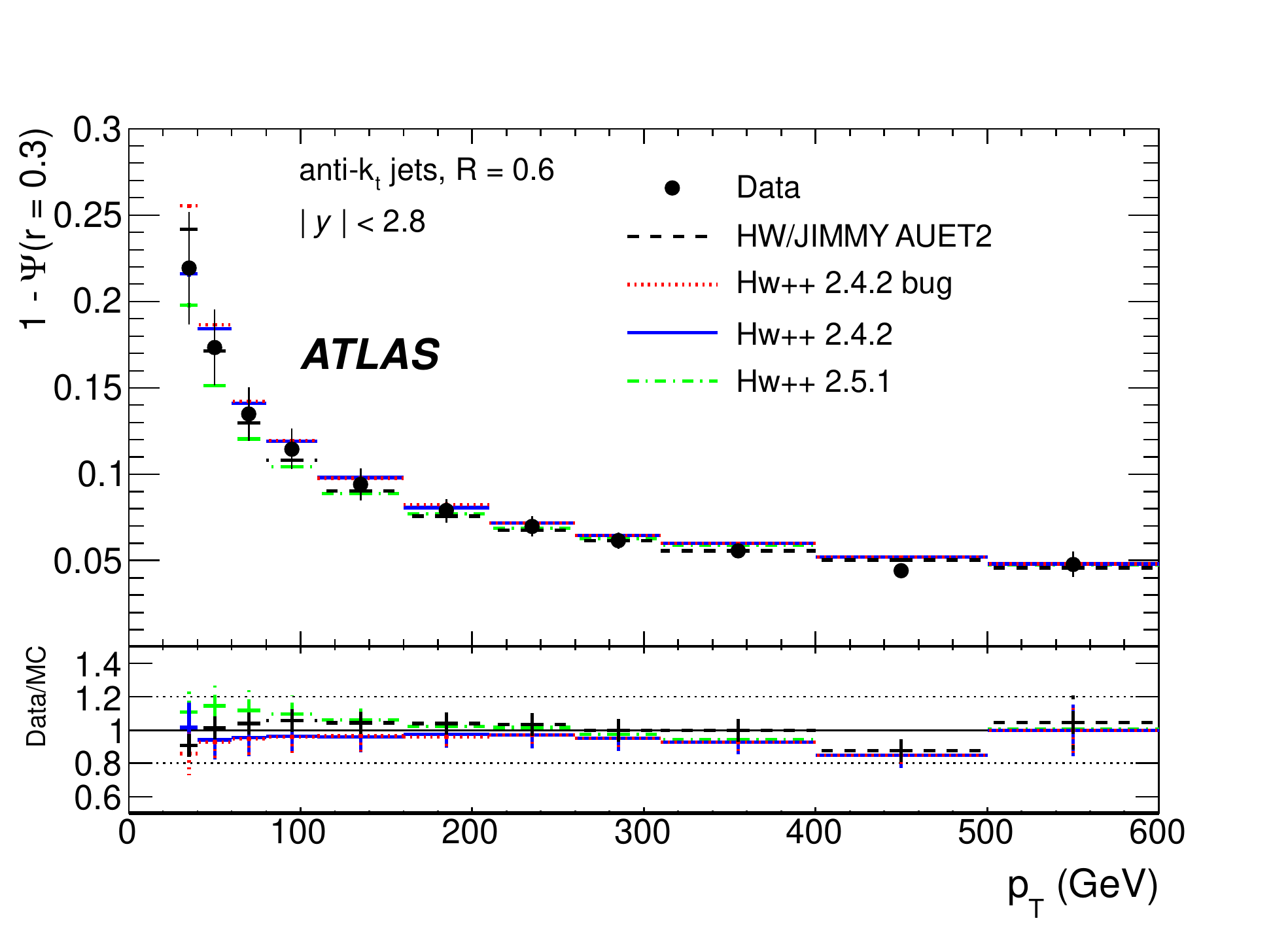}
\label{fig:shapes_herwig}}

\subfigure[{\sc sherpa} and {\sc alpgen}.]{\includegraphics[width=0.5\textwidth]{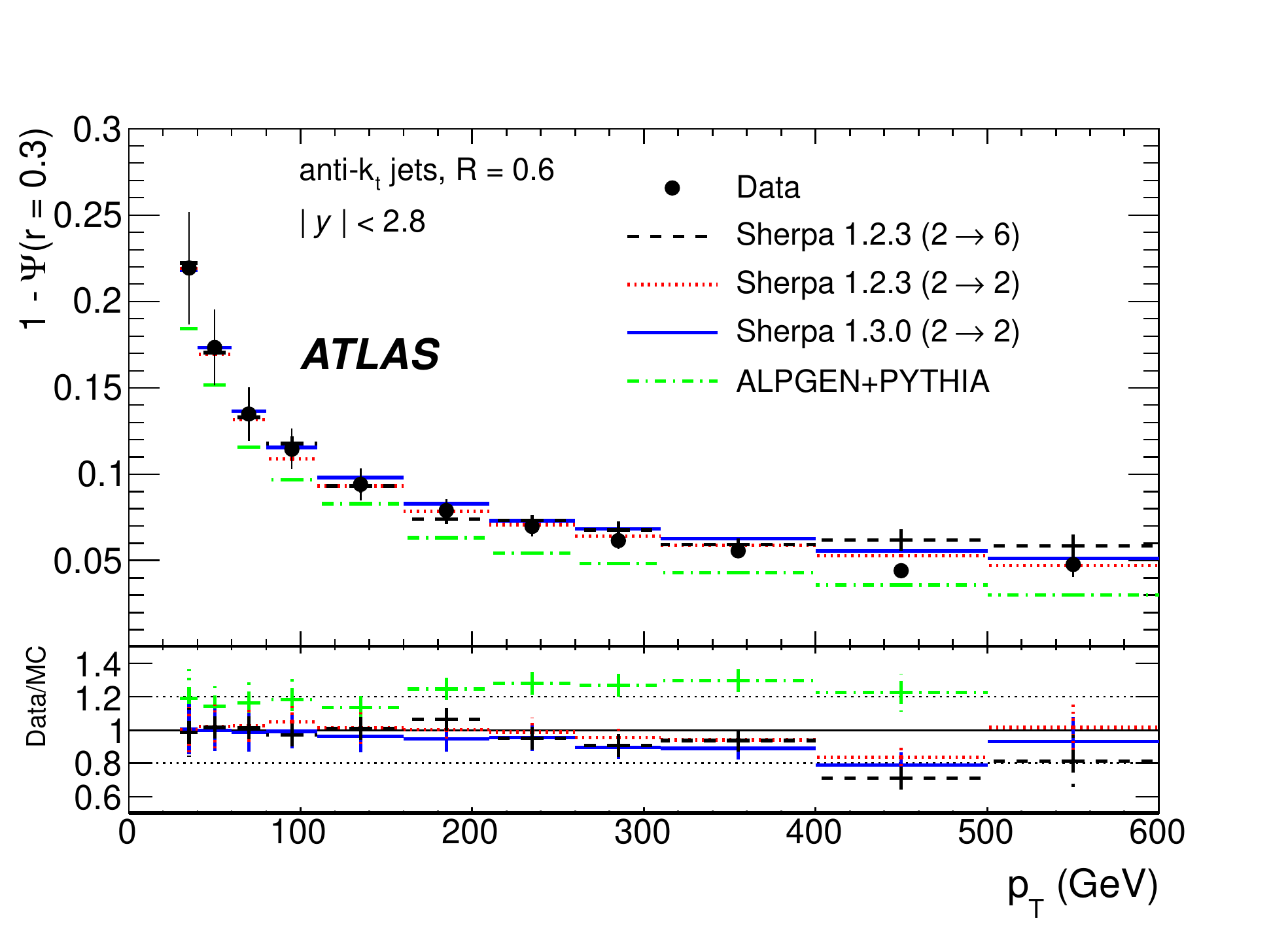}
\label{fig:shapes_sherpa}}
\subfigure[{\sc powheg} comparison.]{\includegraphics[width=0.5\textwidth]{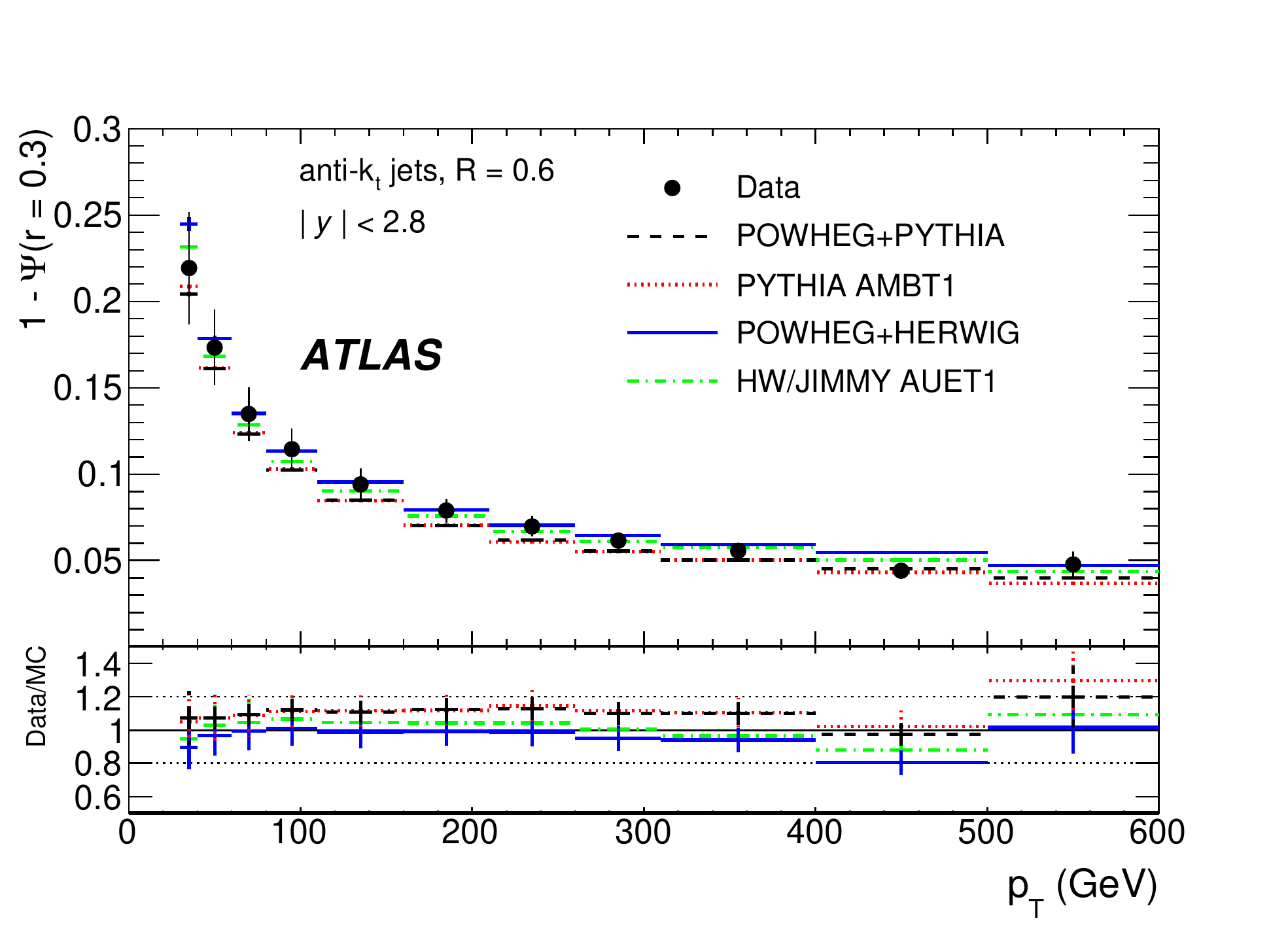}
\label{fig:shapes_powheg}}
\caption{\small
The measured integrated jet shape, $1 - \Psi(r=0.3)$, as a function of $\ptjet$ for jets with $|\rapjet| < 2.8$ and $30 \gev < \ptjet < 600 \gev$.
Error bars indicate the statistical and systematic uncertainties added in quadrature.
The measurements are compared to the different MC predictions considered.}
\label{fig:sum}
\end{figure}

A summary of the results presented in \cite{jshapes} are given in \figrefcap{rho} for the differential jet shape in the bin $30 \gev < \ptjet < 40 \gev$ and in \figrefcap{sum} for the integrated jet shape at $r=0.3$ in jets with $30 \gev < \ptjet < 600 \gev$.

The measured jets become narrower as the jet transverse
momentum and rapidity increase, although with a 
rather mild rapidity dependence.
The results indicate that the predicted jet shapes are mainly dictated by the details of the implementation of  
the parton showers and the modelling of the underlying event 
in the Monte Carlo generators.  The presence of contributions from higher-order pQCD matrix elements 
in the final state does not affect the predicted jet shapes significantly. 

The \pythia, \herwigpp, and \herwig/\jimmy\ 
Monte Carlo generators, using the latest tunes  to describe minimum bias and 
underlying-event--related observables in data, provide a good description of the measured jet shapes, although
the latest version of \herwigpp\ tends to produce jets narrower than the data.
As expected, \pythia\ 6 \ambt\ tends to produce jets slightly narrower than the data as it underestimates the 
UE activity in dijet events \cite{mc10} and lacks a tuned final-state parton shower from the initial-state 
radiation. The effect of the wrong setting in ``\herwigpp\ 2.4.2 bug''\footnote{The ATLAS Monte Carlo generation of \herwigpp\ 2.4.2 had the wrong settings for parameters controlling the multiple-parton interactions.} is only visible for jets with $\ptjet < 40$ \gev.

 The different \sherpa\ predictions are similar and provide a 
reasonable description of the data. This indicates that the presence of additional partons from 
higher-order matrix elements contributions do not affect the predicted jet shapes, mainly dictated by the 
soft radiation in the parton shower. The comparison between \sherpa\ 1.2.3 and \sherpa\ 1.3.0 shows that the 
NLL-inspired corrections included in the latter for the  parton shower  do not impact significantly 
the predicted jet shapes. \alpgen\ interfaced with \pythia\ predicts too-narrow jets and does not describe the data. This was already the case for \alpgen\ interfaced with \herwig/\jimmy\ \cite{jshapes} and requires further 
investigation to determine whether the disagreement observed with the data 
can be completely attributed to the  UE modelling in the MC samples or is 
also related to the prescription followed by \alpgen\ in merging the partons from the 
matrix elements with the parton showers in the final state.

Finally, \powheg\ interfaced with \herwig/\jimmy\ provides a reasonable description of the data while the 
interface with \pythia\ predicts too-narrow jets, which is mainly attributed to the details of the underlying event modelling.

 These results confirm the sensitivity jet shape measurements, and their potential to 
constrain different Monte Carlo models.

\subsubsection{Boosted top quarks with ATLAS}



First steps have been made this year by ATLAS in the quest to reconstruct the decay of a  boosted top quark in a single jet. The first publication 
has been delivered in the context of the search for massive exotic resonances decaying to $t\bar{t}$ \cite{ATLAS-CONF-2011-070}. This analysis was able
to isolate a clean sample of boosted top quarks.

\begin{figure}[t]
\begin{center}
\begin{tabular}{ l | l}
\multicolumn{2}{c}{ \includegraphics[width=0.9\textwidth]{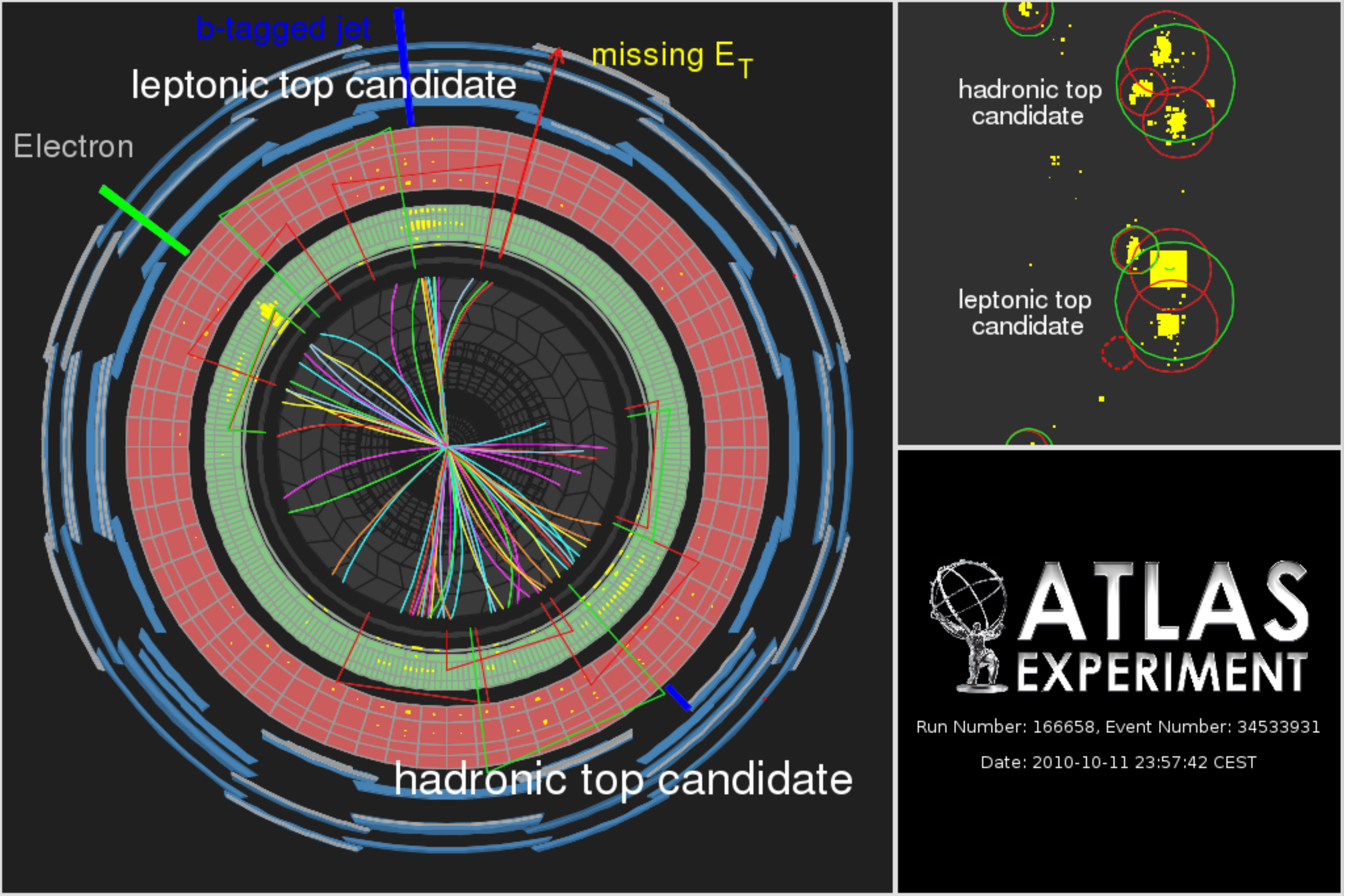}} \\ 
\hline
\scriptsize
Leptonic top & $E_T^\text{miss}$:  $E_T =  36$ \gev, $\phi = -1.5$    \\ 
                     & electron:    $p_T = 145$  \gev, $\eta = 1.1$, $\phi=2.5$ \\
                      & jet:  $E_T = 194$ \gev, $\eta = 1.2$, $\phi= 1.7$, $m_j=  16.6$ \gev \\ \hline
Hadronic top  & jet 2, $ E_T = 155$ \gev, $\eta = 1.1$, $ \phi= -0.7$, $m_j= 22.7$ \gev    \\
  ($R=0.4$ clustering)  & jet 3, $ E_T = 113$   \gev, $\eta =  1.3$, $ \phi= -1.7$, $m_j= 14.0$ \gev  \\
                        & jet 4, $ E_T = 54$   \gev, $\eta =   0.6$, $ \phi= -1.7$, $m_j=  8.1$ \gev  \\ \hline
Hadronic top             & jet 1, $ E_T = 355.5$ \gev, $\eta = 1.3$, $ \phi= -1.1$   \\          
($R=1.0$ clustering)     & \hskip 1cm $\sqrt{d_{12}} = 110$, $\sqrt{d_{23}} = 40$, $m_j= 197.1$ \gev  \\ \hline
\end{tabular}
\caption{\small
Summary table for event 34533931 of run 166658. The leptonic top candidate is formed by high-$p_T$ electron (145 \gev, 10 o'clock), $E_T^\text{miss}$ (1 o'clock), and the $b$-tagged jet at 12 o'clock. The three \antikt\ $R=$ 0.4 jets between 4 and 6 o'clock are identified as the hadronic top candidate. When reclustered with $R = $ 1.0 the three jets merge into a single jet with $m_j =$ 198 \gev and subjet splitting scales $\sqrt{d_{12}}= $ 110 and $\sqrt{d_{23}}=$ 40 \gev.} 
\label{fig:event34533931}
\end{center}
\end{figure}

\figrefcap{event34533931} is an event display showing a candidate boosted top quark event. The \antikt jets with $R = 0.4$ used in the 
standard ATLAS $t\bar{t}$ selection are indicated in red. The result of reclustering the constituents of these jets 
(topological calorimeter clusters) with $R =1.0$ is shown in green on the same figure. For these events the three $R =0.4$ jets that 
are combined to form the hadronic top candidate merge into a single jet when clustered with $R =1.0$. These events are thus the first candidates 
for boosted top quarks reconstructed as single jets at ATLAS.

Although the statistics with the 2010 dataset were too low to provide a testing ground for the many substructure techniques and variables we are interested 
in testing on such a clean sample of boosted top quarks, the groundwork was laid down to make the prospects for boosted top analyses on 2011 data very exciting.

\subsubsection{Jet substructure with ATLAS}

One of the most important tasks for the early ATLAS boosted object analyses was to begin subjecting 
the various grooming methods and variables, initially implemented in Monte Carlo, to the tests that come with the conditions of a real detector.
The recent ATLAS analysis on jet mass and substructure has done just that \cite{ATLAS-CONF-2011-073}. The analysis was based on 35 pb$^{-1}$ of data 
taken in 2010 and presented the mass and splitting scale of high $p_{T}$ ($>$ 300 \gev) jets reconstructed with the \antikt ($R=1.0$) algorithm and 
with the \ca ($R=1.2$) algorithm at different stages of grooming. The grooming method employed was a mass-drop filtering 
method which looks for 
hard subjets within the very large parent jet, and discards any radiation not falling within the subjets designated to the hard splitting. As already 
discussed in \secrefcap{pileupatlas}, the behaviour of the mass of large-area jets as a function of the number of good vertices in an event was studied and 
the process of filtering away the soft radiation in such jets was found to effectively remove this dependence. This is perhaps unsurprising, as the radiation
 present in an event due to pile-up or underlying event is expected to be a fairly soft continuum. The analysis discarded all events with more than one good 
vertex, focusing on the 28$\%$ of the 2010 dataset that is regarded as being free from pile-up.\footnote{The authors acknowledge that selecting on events with a single primary vertex is a limited opportunity for ignoring pile-up, as making such a demand in the 2011 data would result in approximately zero events.}

\begin{figure}
    \subfigure[C/A R=1.2 jet mass.]{
      \includegraphics[width=0.5\linewidth]{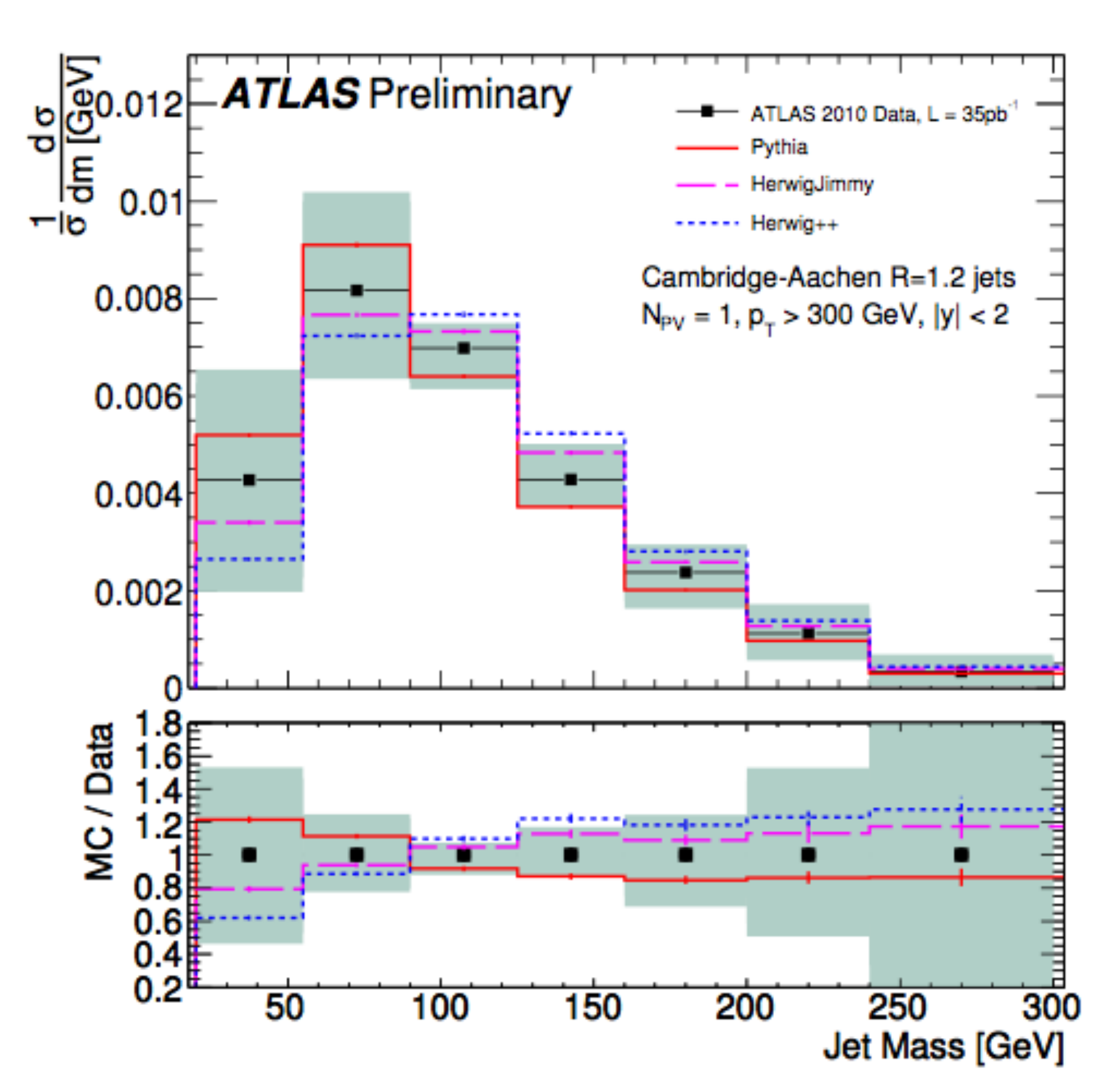}
      \label{fig:camass}
    }
    \subfigure[Filtered C/A R=1.2 jet mass.]{
      \includegraphics[width=0.5\linewidth]{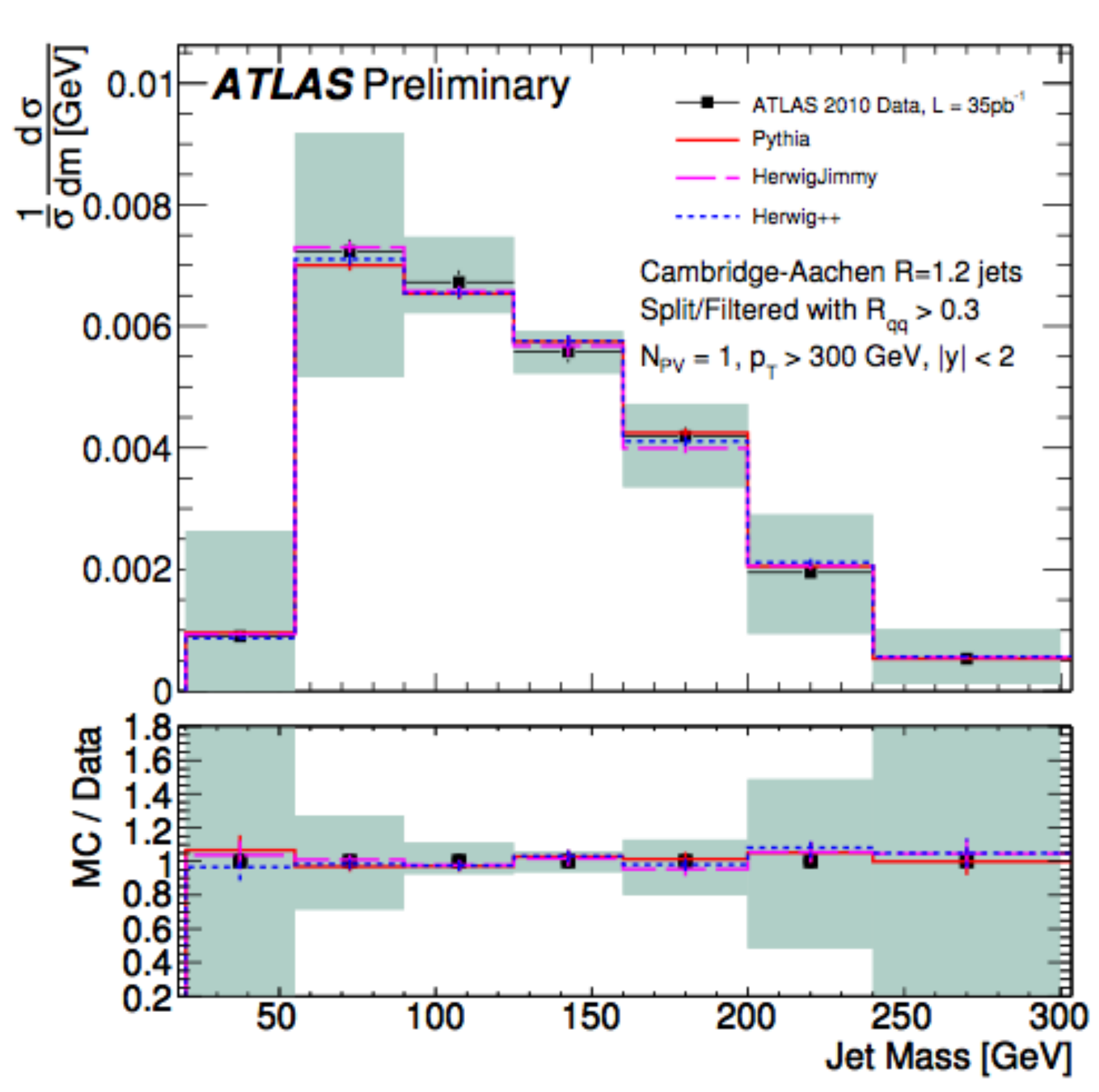}
      \label{fig:cafilt}
    }
  \caption{\small
The invariant mass distribution of \ca $R = 1.2$ jets (left) and the mass distribution after filtering (right). Both distributions are unfolded to stable-particle level and compared to \pythia, \herwig, and \herwigpp. The systematic uncertainties are shown by the shaded band.}
    \label{fig:conf_mass_ca}
\end{figure}

\begin{figure}
    \subfigure[Anti-$k_{t}$ R=1.0 jet mass.]{
      \includegraphics[width=0.5\linewidth]{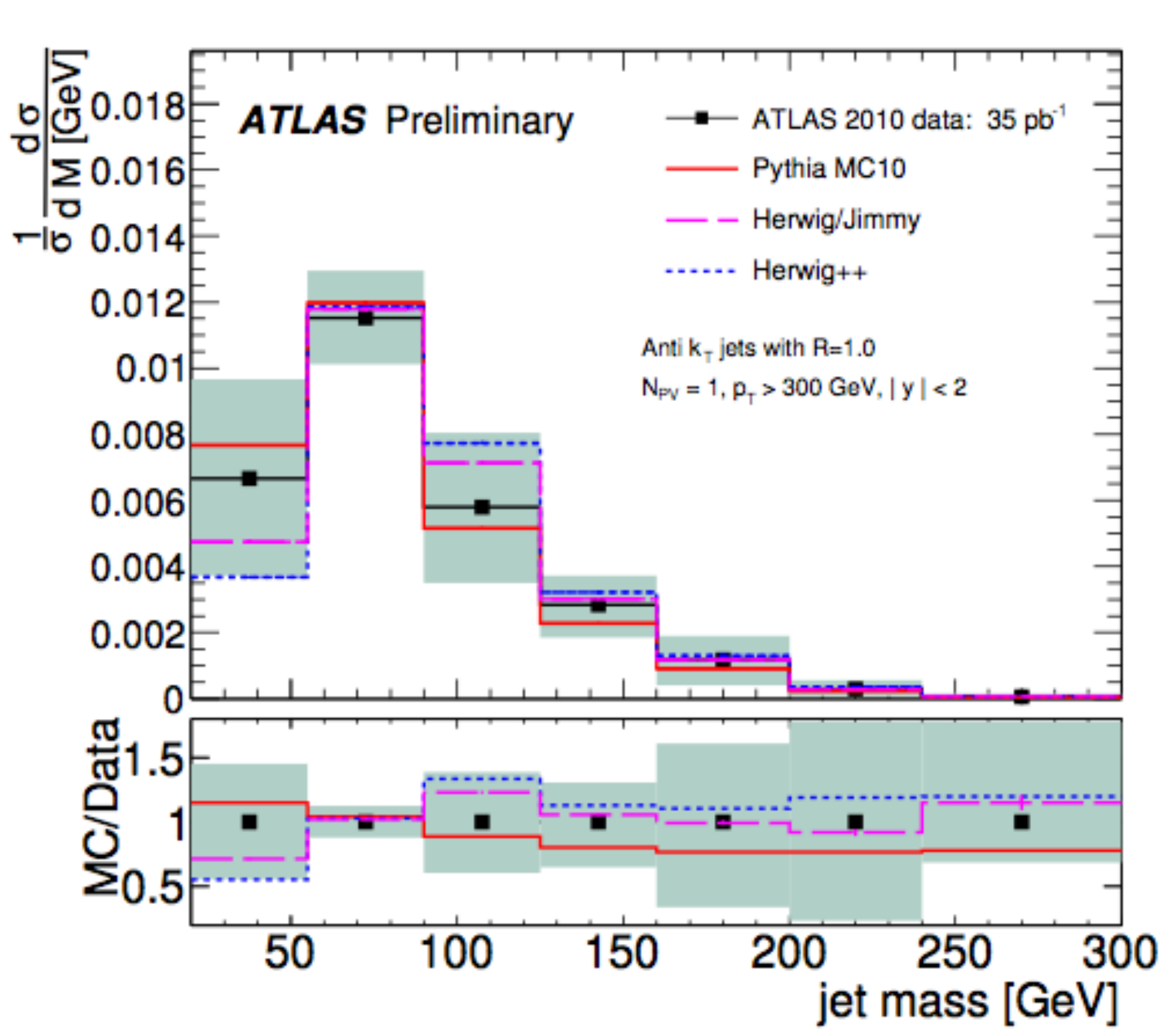}
      \label{fig:aktmass}
    }
    \subfigure[Anti-$k_{t}$ R=1.0 splitting scale.]{
      \includegraphics[width=0.5\linewidth]{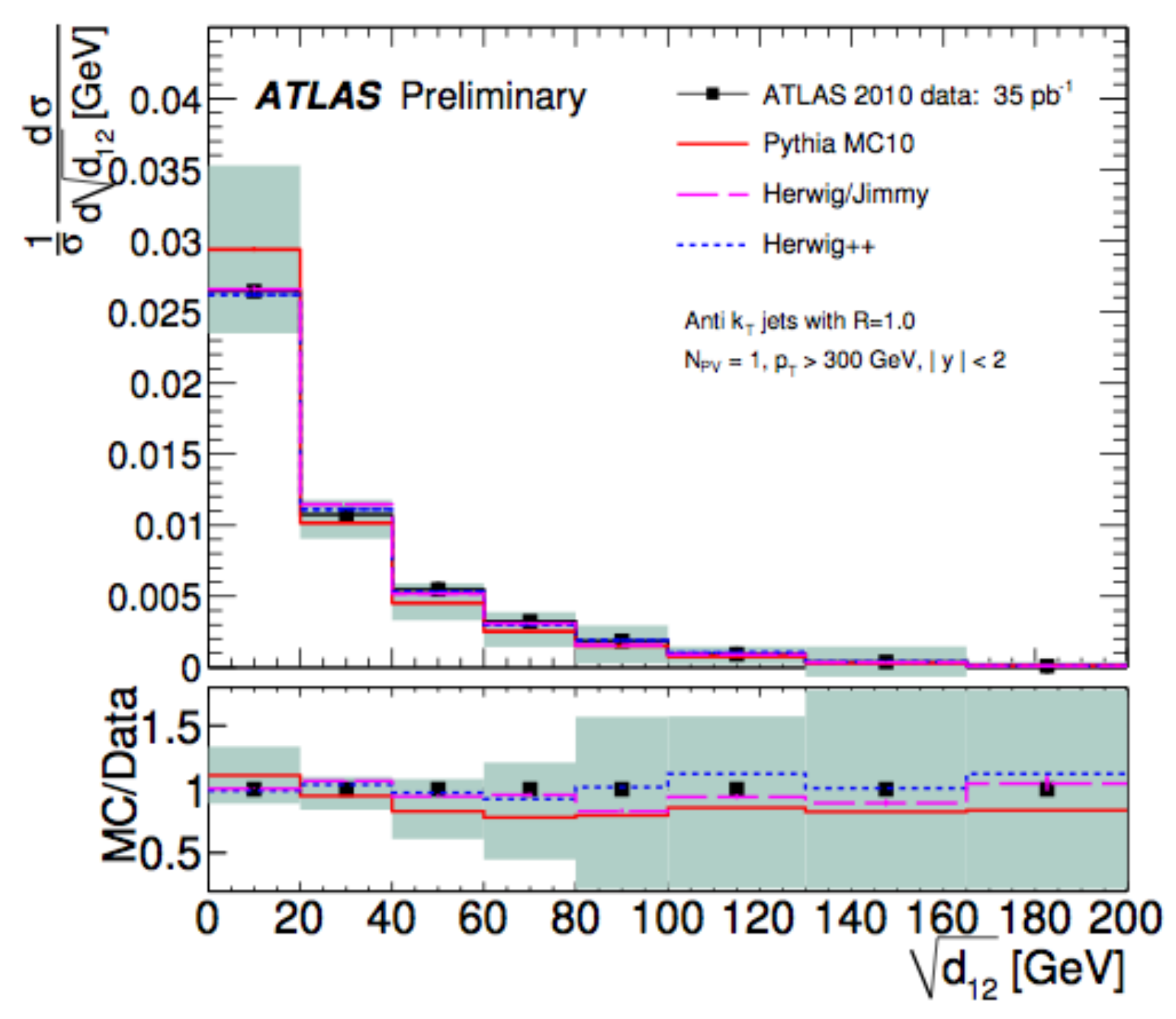}
      \label{fig:aktsplit}
    }
  \caption{\small The invariant mass distribution of \akt $R = 1.0$ jets (left) and the splitting scale $\sqrt{d_{12}}$ of the same jets (right). Both distributions are unfolded to stable-particle level and compared to \pythia, \herwig, and \herwigpp. The systematic uncertainties are shown by the shaded band.}
    \label{fig:conf_mass_antikt}
\end{figure}

The final distributions resulting from the analysis were unfolded to hadron level. The mass distributions of \ca jets with $R =$ 1.2 are shown in \figrefcap{conf_mass_ca} before and after employing mass-drop filtering with the requirement that the heavier subjet carry less than $\mu=67\%$ of the mass of the parent jet and the splitting between subjets is fairly symmetric, $y_2 \equiv \frac{\min\left(p^2_{T,a}, \,\, p^2_{T,b}\right)}{m^2_\text{jet}} \Delta R^2_{ab} < y_2^\text{cut}=0.09$. \figrefcap{conf_mass_antikt} shows the mass and splitting scale ($\sqrt{d_{12}}= \min(p_{T,a}, p_{T,b})\times \Delta R_{ab}$) of \antikt jets with $R =$ 1.0.

In all cases the distributions are found to be well modelled by a range 
of leading-order parton shower Monte Carlos, to within the systematic uncertainties. 

\subsection{Results from CMS}\label{sec:results:cms}

\subsubsection{CMS and pile-up}

One advantage of the CMS detector and reconstruction pipeline is the so-called ``particle flow'' algorithm for a more holistic approach to jet reconstruction. Because of this approach, it is possible to remove around 60\% of the pile-up from jets directly as it corresponds to charged tracks which can be associated to subleading primary vertices. This aids in the reduction of pile-up in jets with substructure at CMS, and studies show that there is only moderate dependence of the substructure variables on pile-up for the luminosities considered \cite{cms-pas-jme-10-013}.

The CMS collaboration plans to retain a data-driven approach to measuring mis-tag rates, ameliorating pile-up issues there.  Further studies are ongoing to utilise other advanced jet reconstruction techniques to further reduce pile-up dependence. 

\subsubsection{Jet Shapes with CMS}

Several jet shape measurements have been performed at CMS similar to those at ATLAS described above in \secrefcap{results:atlasjetshapes}. The first is the examination of  the classic jet shapes, and the second is the examination of the jet charged component structure, both in \cite{cms-pas-qcd-10-014}.

Several differences exist between the ATLAS and CMS results. While both collaborations use the \antikt jet algorithm,  CMS uses different values of the radius parameter, $R$. 
For the inclusive shape measurements, two types of jets were used: calorimeter-only jets
and jets where the calorimeter response was corrected using associated tracks (``jet-plus-tracks'' jets) \cite{JetPlusTracks}.
To reduce out-of-cone radiation, a large $R$ of 0.7 was used. The jets were required to have
(track-corrected) transverse momenta greater than 15  \gev. 
For the charged multiplicity measurement, only the jet-plus-tracks jets were used, with $R =  0.5$. The (track-corrected)  transverse momenta for the first two jets were required to 
be larger than 20 and 10  \gev, respectively. 

\begin{figure}[tbh]
\subfigure[$20<\pT<30$ \gev]{\includegraphics[width=0.5\textwidth]{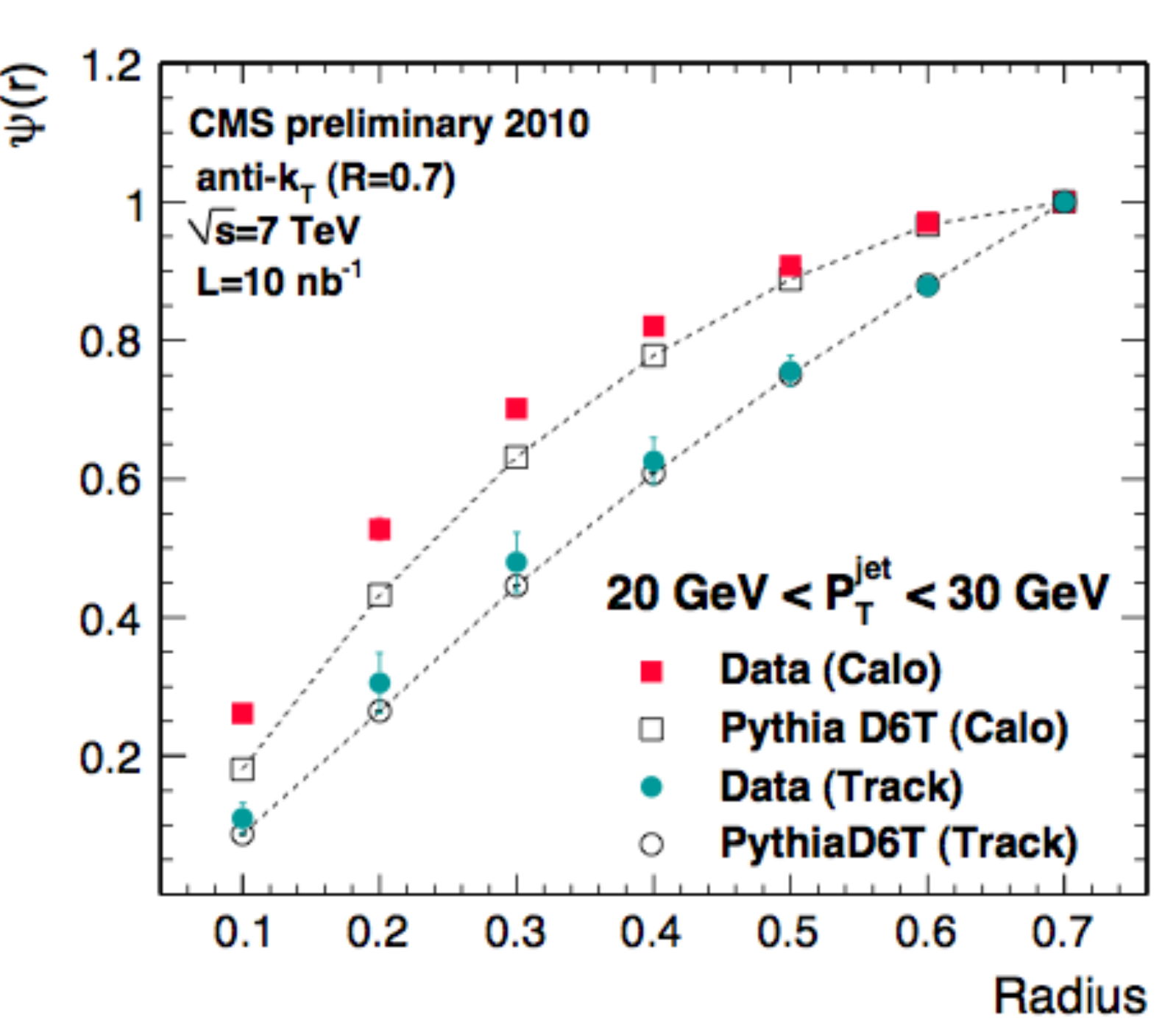}
\label{fig:shape1cms}}
\subfigure[$40<\pT<50$ \gev]{\includegraphics[width=0.5\textwidth]{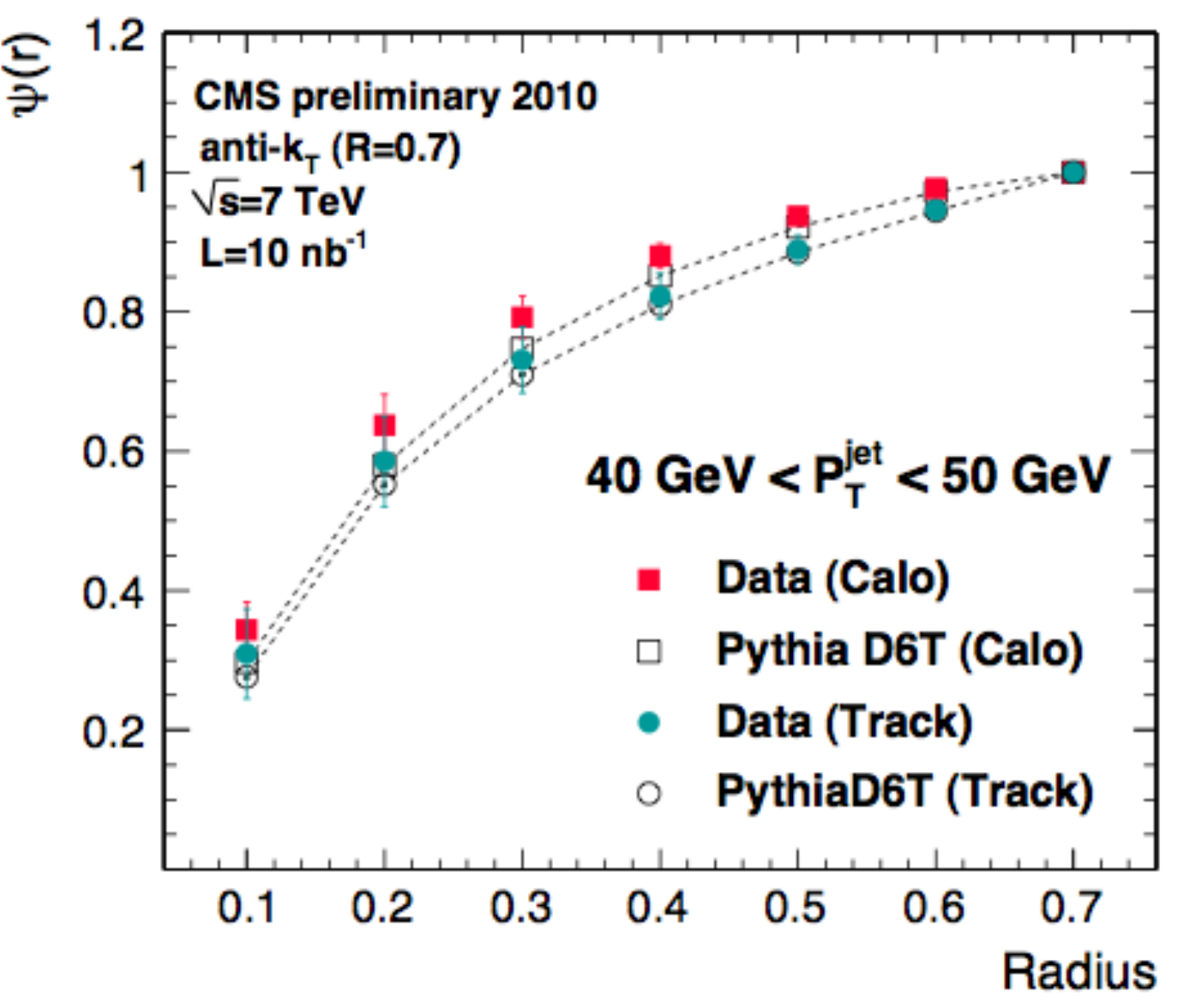}
\label{fig:shape2cms}}

\subfigure[$60<\pT<70$ \gev]{\includegraphics[width=0.5\textwidth]{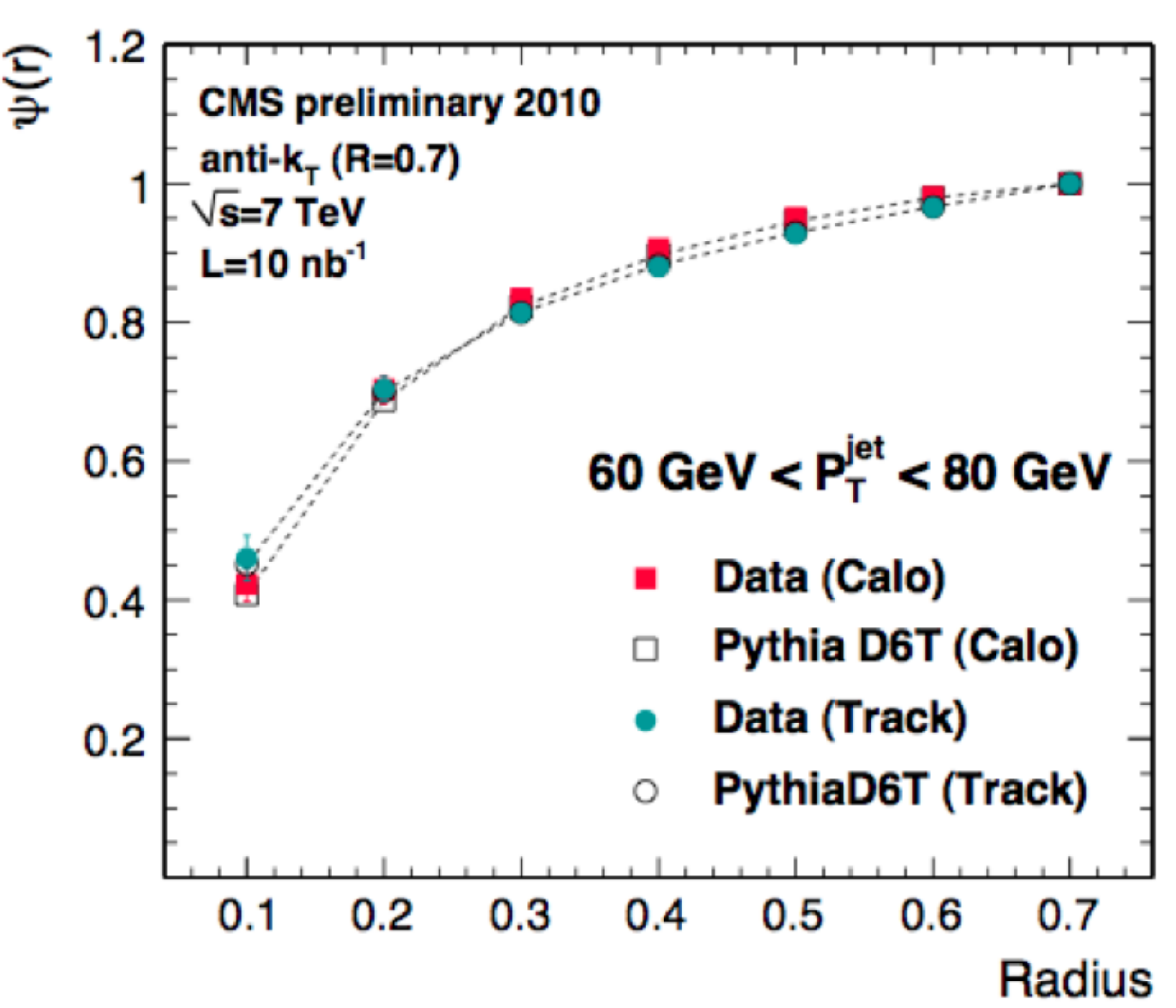}
\label{fig:shape3cms}}
\subfigure[$80<\pT<100$ \gev]{\includegraphics[width=0.5\textwidth]{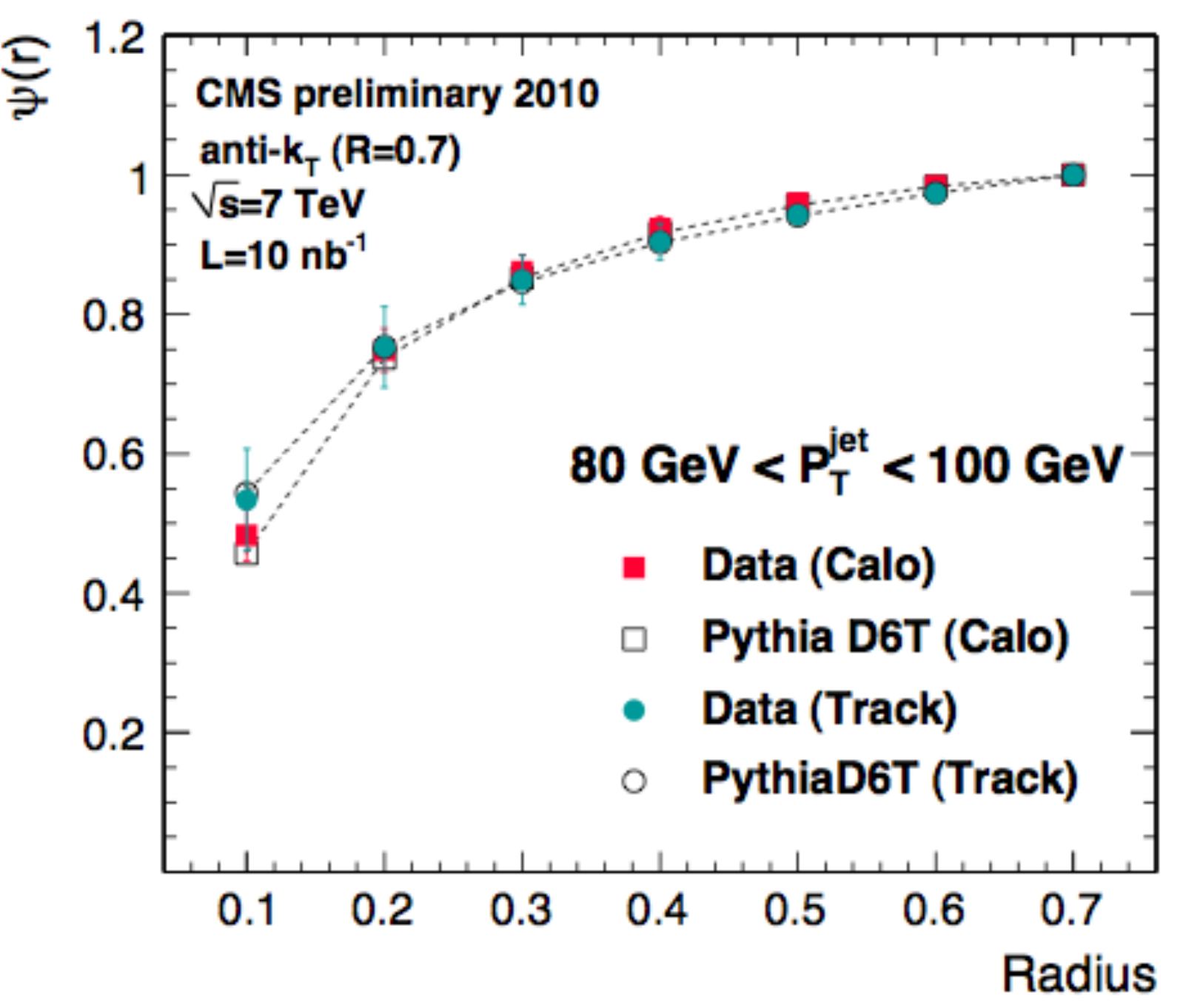}
\label{fig:shape4cms}}
\caption{\small
The measured differential jet shape at CMS, $\rho(r)$, in inclusive jet production for jets
with $\ptjet > 15$  \gev. Error bars indicate the statistical and systematic uncertainties added in quadrature. 
The measurements are compared to MC predictions in different $\ptjet$ bins.}
\label{fig:rho_cms}
\end{figure}

The distribution of $\Psi(r)$ (\erefcap{psi}) is shown in \figrefcap{rho_cms}.
The Monte Carlo simulation is doing a very reasonable job at predicting the data, and the
agreement improves with larger transverse momentum. 

\subsubsection{Boosted top quarks and W bosons with CMS}

CMS has developed algorithms to identify the hadronic decays of boosted top quarks and $W$ bosons. The top-tagging algorithm is based on the Johns Hopkins tagger \cite{Kaplan:2008ie}, with small modifications as described in ~\cite{Rappoccio:2009cr,Boost2010}. The $W$-tagging algorithm is based on jet pruning \cite{Ellis:2009su,Ellis:2009me}, adding a requirement on the mass drop of the subjet splitting, which was motivated by the BDRS subjet/filter algorithm \cite{Butterworth:2008iy}.

The algorithms were characterised as of the \boost\ 2011 conference \cite{cms-pas-jme-10-013}, and subsequent
studies have advanced this characterisation \cite{cms-pas-exo-11-006}. The strategy for the
studies here is to have data-driven mis-tag predictions which can accurately represent non-top
and non-$W$ backgrounds. Sidebands in the data are selected that have little to no 
true top or $W$ contributions, and the tagging rate is derived as a function of the 
jet transverse momentum. These per-jet parameterisations are then applied to each jet
in a signal selection, which is used to form a data-driven background estimate. 
This has several advantages, primarily that modelling effects are not relevant.  Moreover,
pile-up effects are handled directly, because the same data sample is used 
in the sidebands and the signal region, so the pile-up contribution is the same.

The efficiency for the tagging algorithms is derived from
Monte Carlo, and corrected for a data-to-Monte-Carlo scale factor using
Standard Model $t\overline{t}$ events in the semi-leptonic channel in \cite{cms-pas-exo-11-006}. 
In that public result, limits are set on the possible cross section for narrow resonances
decaying to $t\overline{t}$ pairs. 

\begin{figure}[tbh]
\subfigure[]{\includegraphics[width=0.5\textwidth]{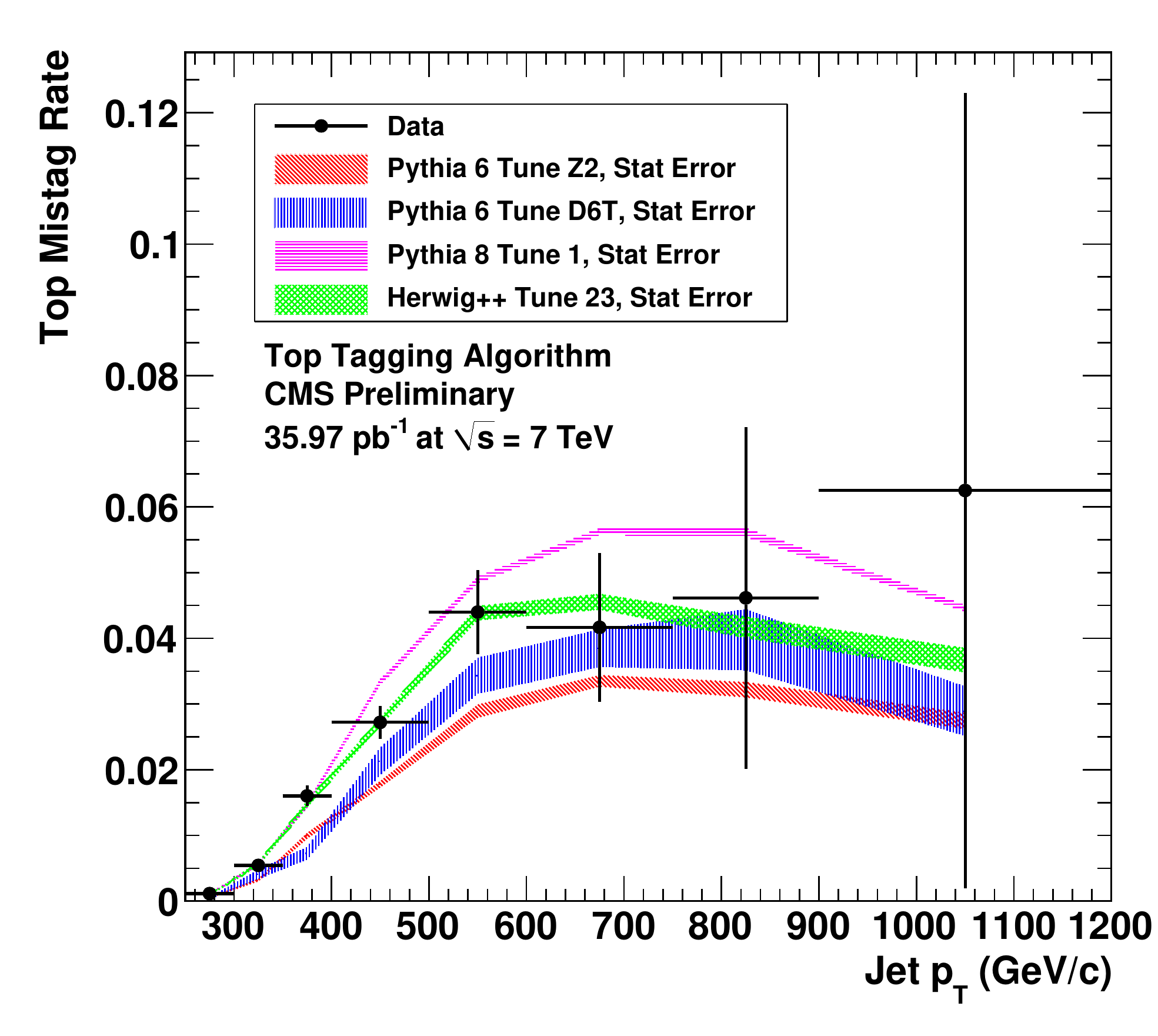}
}
\subfigure[]{\includegraphics[width=0.5\textwidth]{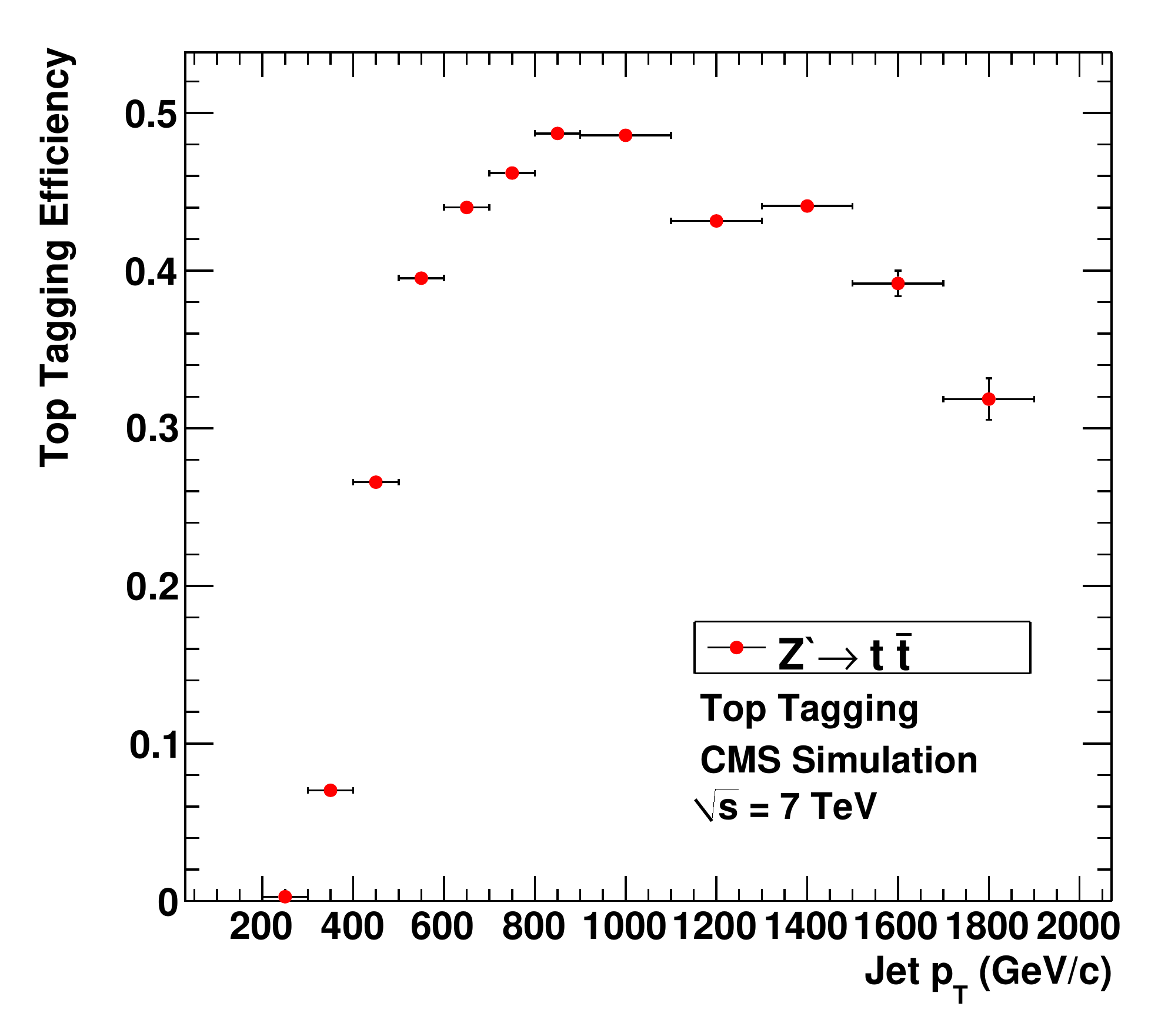}
}

\subfigure[]{\includegraphics[height=0.5\textwidth,angle=90]{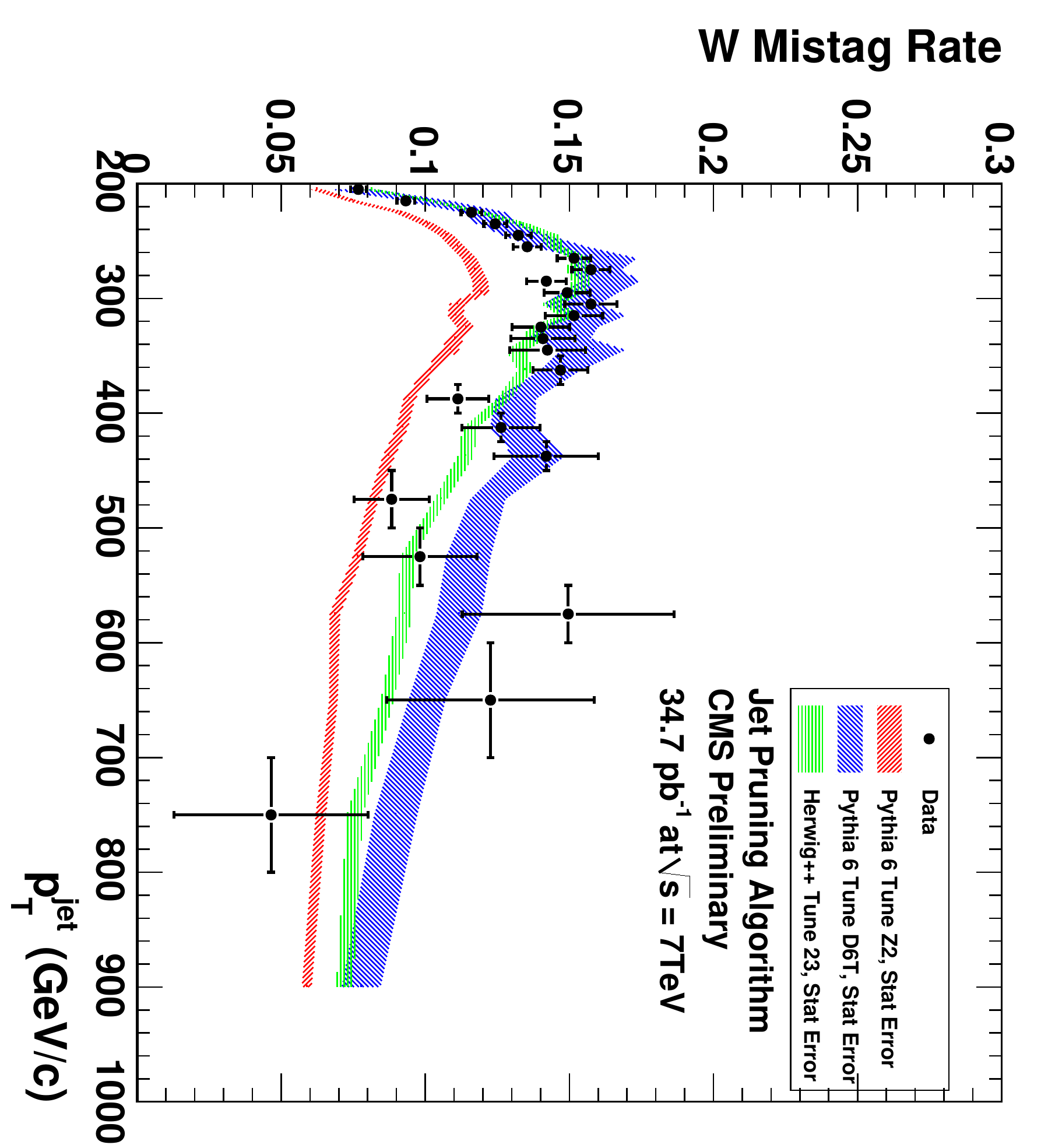}
}
\subfigure[]{\includegraphics[height=0.5\textwidth, angle=90]{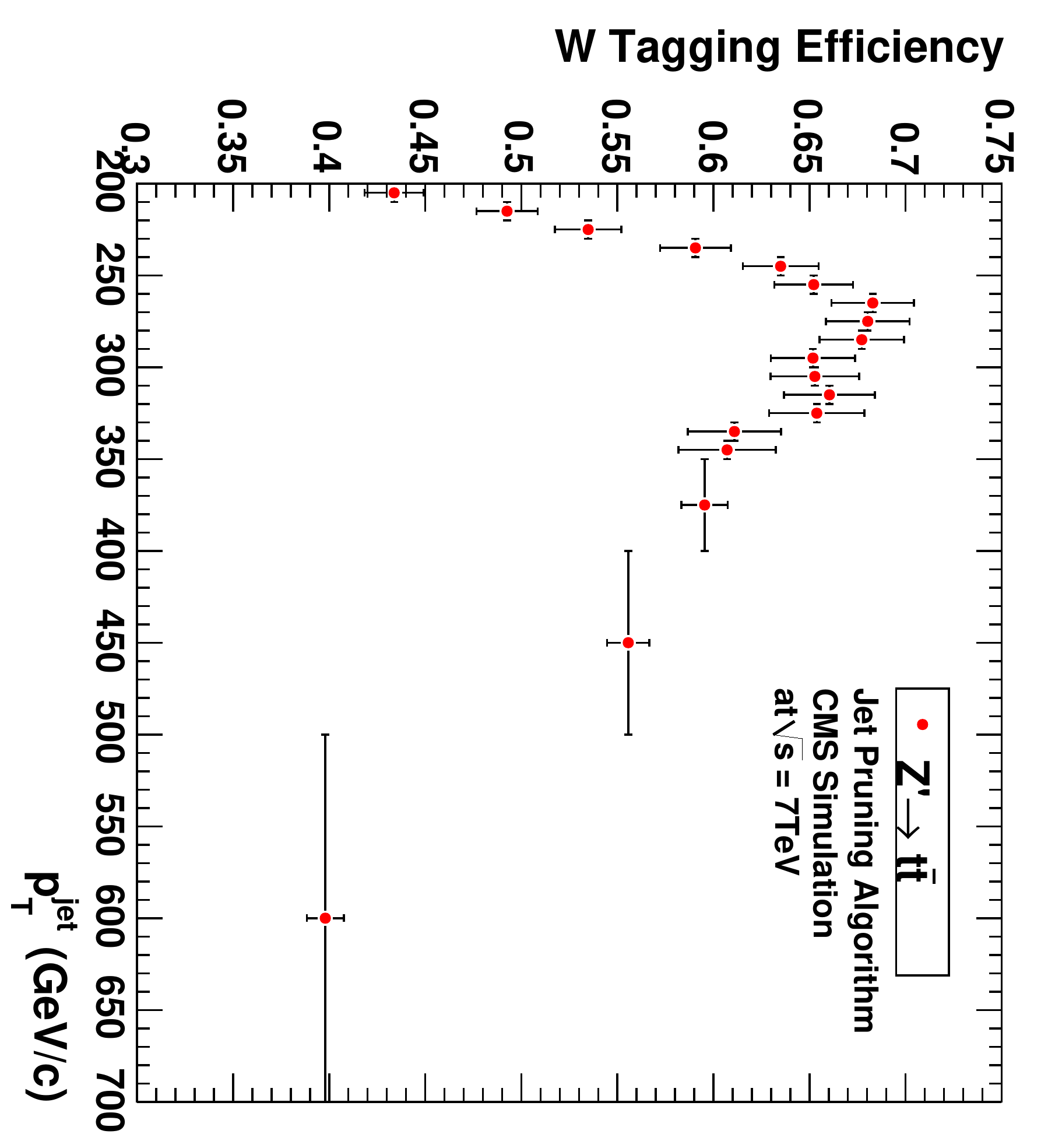}
}
\caption{\small
Mis-tag rates and efficiencies for the top tagging (top) and $W$ tagging (bottom) algorithms. The mis-tag rates are measured in a signal-depleted
control sample and compared to \pythia\ and \herwigpp\ predictions. The efficiencies for tagging top quarks and Ws are estimated from \pythia\ simulation. }
\label{fig:tags_cms}
\end{figure}

\figrefcap{tags_cms} shows the mis-tag rates and efficiencies for the
top and $W$ tagging algorithms from CMS.

\subsubsection{Jet substructure at CMS}

The study of the performance of jet substructure techniques in generic QCD samples is
critically important to understand these algorithms, and to derive appropriate
sideband regions for data-driven background estimates at CMS. As such, CMS has performed 
detailed comparisons of jet substructure observables in \cite{cms-pas-jme-10-013}.

\begin{figure}[tbh]
\subfigure[Top-tagged jet mass.]{
\includegraphics[width=0.51\textwidth]{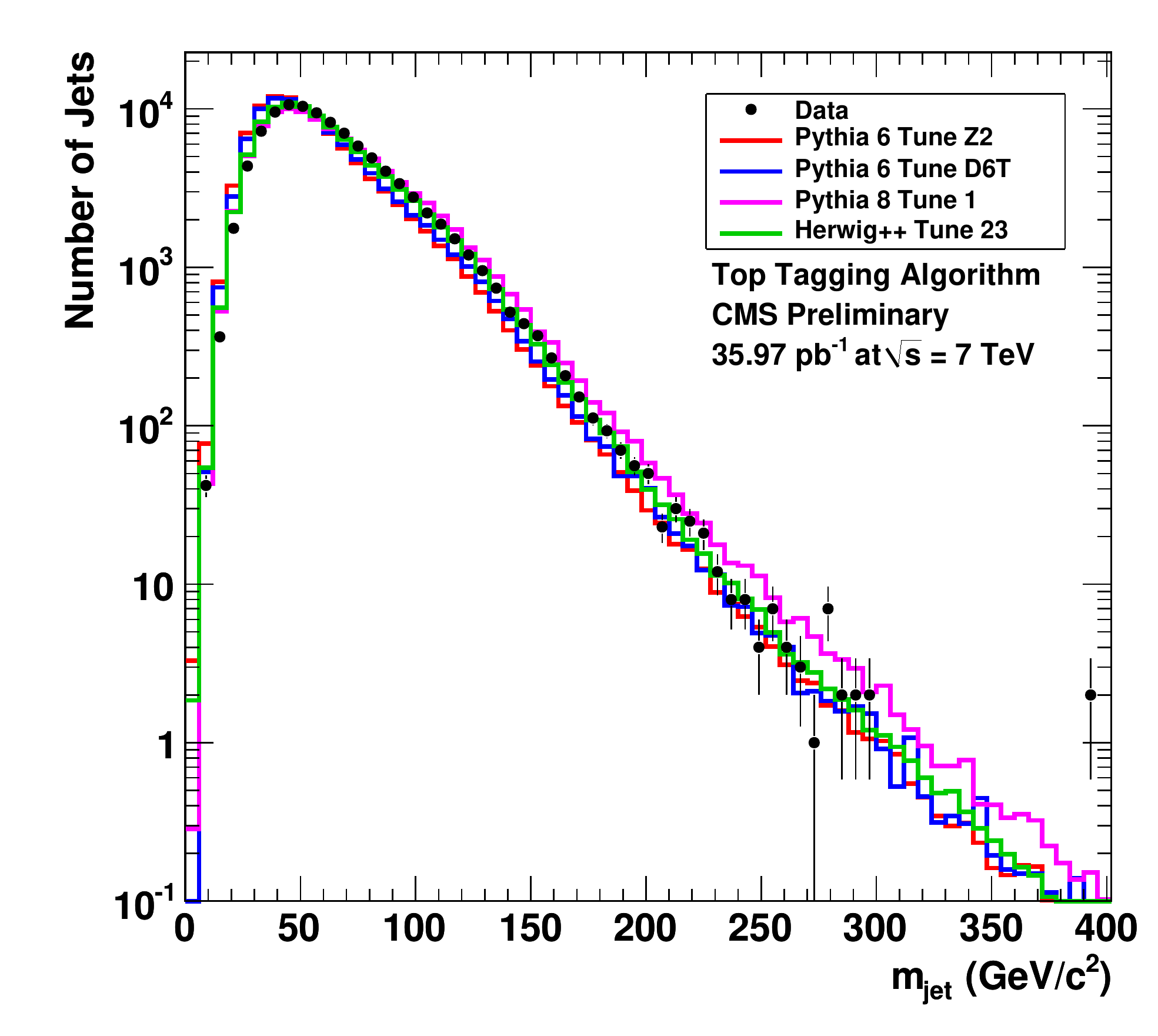}
}
\subfigure[$W$-tagged jet mass.]{
\includegraphics[height=0.49\textwidth,angle=90]{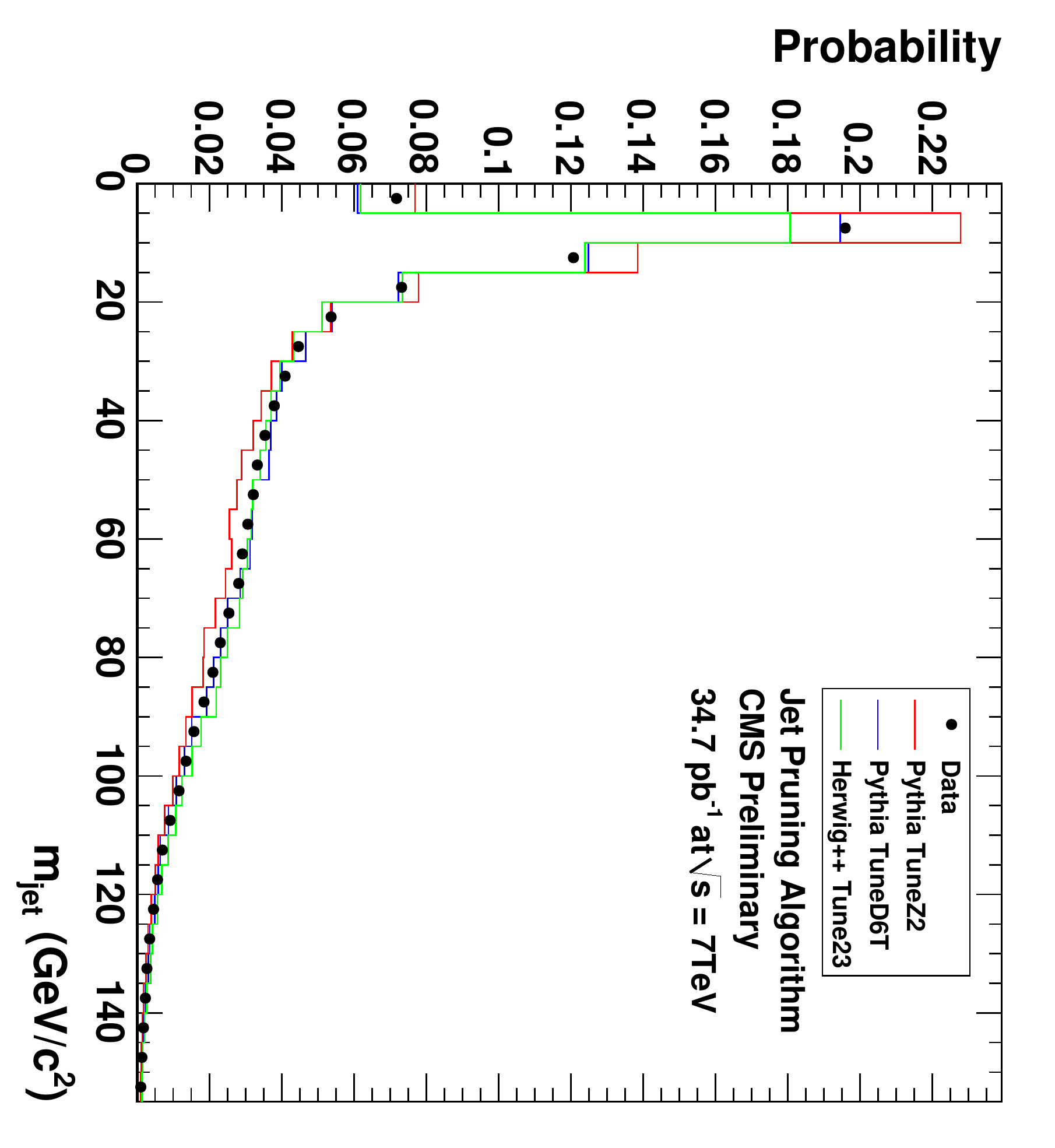}
}
\caption{\small
The jet mass distribution obtained with the top-tagging algorithm (left) and the $W$-tagging algorithm using jet pruning (right). Data are compared to \pythia\ and \herwigpp.}
\label{fig:subjets_cms}
\end{figure}

For example, \figrefcap{subjets_cms} shows the jet mass for the top-tagging
and $W$-tagging algorithms, compared to several different predictions from Monte Carlo. 
The simulation seems to be predicting the data quite nicely for all of the generators
examined, with slight variations based on shower model and underlying event tunes.
This seems to have a much larger effect on the substructure than the pile-up contributions.

\section{New tools and techniques}
\label{sec:tools}

\newcommand{\jdt}[1]{\textbf{(#1 --jdt)}}

Jet substructure techniques attempt to extract additional information about a jet from a detailed study of its constituents.   These methods have been mainly aimed at distinguishing boosted hadronic objects like top quarks and $W/Z$/Higgs bosons from the background of jets initiated by light quarks and gluons.
Many such techniques exist, and to help make sense of them, the \boost\ 2010 report \cite{Boost2010} proposed that they be classified according to three broad categories:  (1) methods that directly identify subjets within a jet, (2) methods which ``groom'' away uncorrelated radiation within a jet, and (3) methods based on energy flow within a jet.  

Since the \boost\ 2010 workshop, a variety of new techniques have been introduced which will be described in more detail below.  A recent and thorough review of substructure techniques applied to top tagging can be found in \cite{Plehn:2011tg}; here we focus on the developments since \boost\ 2010.  $N$-subjettiness \cite{Thaler:2010tr,Kim:2010uj} and dipolarity \cite{Hook:2011cq} are examples of hybrid jet shapes, which describe the energy flow of a jet (as in method (3) above) with respect to candidate subjet axes (determined using e.g. method (1) above).  Jet substructure through angular correlation functions \cite{Jankowiak:2011qa} is a complementary technique to energy flow observables.  The template overlap method \cite{Almeida:2010pa} and the shower deconstruction method \cite{Soper:2011cr} classify jets with the help of approximations to hard matrix elements and the parton shower.  Beyond the highly boosted regime, the HEP (Heidelberg-Eugene-Paris) top tagger \cite{Plehn:2009rk} is appropriate for identifying moderately boosted top quarks.  Meanwhile, substructure techniques have been been used in a variety of interesting applications, including separating quark jets from gluon jets \cite{Gallicchio:2011xq,Gallicchio:2011xc}, tagging jets from initial state radiation \cite{Krohn:2011zp}, and identifying boosted decay products of new physics \cite{Kribs:2009yh,Kribs:2010hp}.
 
\subsection{N-subjettiness}

In \cite{Thaler:2010tr}, Thaler and van Tilburg introduced a new jet shape ``$N$-subjettiness'' (denoted $\tau_N$), designed to identify boosted $N$-prong hadronic decays.  $N$-subjettiness quantifies the degree to which jet radiation is aligned along specified subjet axes, such that small values of $\tau_N$ correspond to $N$ or fewer subjets, while large values of $\tau_N$ indicate more than $N$ subjets.  This jet shape was adapted from the event shape $N$-jettiness introduced in \cite{Stewart:2010tn} to define exclusive jet cross sections, and similar ideas were pursued by Kim in \cite{Kim:2010uj}.\footnote{\cite{Kim:2010uj} focused on boosted Higgs identification, using a Lorentz-invariant version of $N$-subjettiness defined in the jet rest frame.}

Given candidate subjet directions determined by an external algorithm (such as the exclusive $k_T$ procedure \cite{Catani:1993hr,Ellis:1993tq}), $\tau_N$ is defined as
\begin{equation}
\label{eq:tauN} 
\tau_N = \frac{\sum_k p_{T,k} \left(\min \left\{\Delta R_{1,k},\Delta R_{2,k},\ldots, \Delta R_{N,k}   \right\} \right)^\beta}{\sum_k p_{T,k} (R_0)^\beta},
\end{equation}
where the sum runs over the particles in the jet, $p_{T,k}$ is the transverse momentum of particle $k$, $\Delta R_{A,k}$ is the azimuth-rapidity distance between subjet axis $A$ and particle $k$, and $R_0$ is the characteristic jet radius defined such that $0 \leq \tau_N \leq 1$.  The constant $\beta$ is an angular weighting exponent closely related to angularities \cite{Berger:2003iw}, and 1-subjettiness roughly corresponds to jet angularities \cite{Ellis:2010rwa} with $a \equiv  2-\beta$.

To separate boosted hadronic objects from the QCD jet background, one could use the complete set of $\tau_N$ values (with different values of $\beta$) in a multivariate analysis.  However, \cite{Thaler:2010tr} showed that a simple cut on the ratio $\tau_N/\tau_{N-1}$ provides excellent discrimination power for $N$-prong hadronic objects.  In particular, $\tau_3/\tau_2$ is a successful boosted top discriminator, and $\tau_2/\tau_1$ can identify boosted $W$/$Z$ and Higgs bosons, with the angular weighting exponent $\beta = 1$ (corresponding roughly to jet broadening \cite{Catani:1992jc}) providing the best discrimination.  In subsequent work \cite{Thaler:2011gf}, Thaler and van Tilburg showed that the initial step of choosing candidate subjet axes is in fact unnecessary.  In particular, the quantity in \erefcap{tauN} can be minimised over the candidate subjet directions using a variant of the $k$-means clustering algorithm \cite{Lloyd82leastsquares}, further improving boosted object discrimination.

\subsection{Dipolarity}

A new colour flow observable, ``dipolarity'', was introduced by Hook, Jankowiak, and Wacker
to discriminate between different colour configurations of a given pair
of subjets $j_1$ and $j_2$ \cite{Hook:2011cq}. Dipolarity is given by a sum in which each constituent
of $j_1+j_2$ is weighted by its $p_T$ and its squared angular separation
$\Delta R^2$ from the line segment connecting $j_1$ and $j_2$ in the $\eta$-$\phi$
plane:
\begin{equation}
\mathcal{D} \equiv \frac{1}{\Delta R_{12}^2 }\sum_{i\in J} \frac{p_{Ti}}{p_{T_J}}\Delta R_i^2 .
\label{eq:dipdefn}
\end{equation}
For subjets $j_1$ and $j_2$ in a colour singlet configuration, the radiation pattern
is of the dipole form with most radiation clustered in the region between the two subjets.
Consequently $\mathcal{D}$ is expected to be small for colour singlet configurations
and larger for other colour configurations, in which $j_1$ and $j_2$ are colour connected
to other subjets.  

By considering the entire radiation pattern of the two
subjets at once, dipolarity is designed to be most effective in the
semi-boosted regime, where there can be considerable overlap between
the two subjets.  This is in contrast to jet pull \cite{Gallicchio:2010sw}, which was introduced
with the low boost regime in mind and which can lose discrimination
power if there is substantial overlap.  As a first application, dipolarity
has been incorporated into the HEP top tagger \cite{Plehn:2009rk}, where it was shown to improve
background rejection by probing the colour structure of the reconstructed
$W$ boson.  More work will be needed to determine whether dipolarity
can be applied effectively outside of top tagging.

\subsection{Jet substructure without trees}

Jankowiak and Larkoski developed a method for identifying
substructure within jets via angular correlations \cite{Jankowiak:2011qa}, introducing
an angular correlation function ${\cal G}(R)$:
\begin{equation}\label{corrfunc}
\mathcal{G} (R)\equiv\frac{\sum\limits_{i\ne j}p_{T i}p_{T j}
\Delta R_{ij}^2\Theta (R-\Delta R_{ij})}
{\sum\limits_{i \ne j}p_{T i}p_{T j}\Delta R_{ij}^2},
\end{equation}
where the sum runs over all pairs of jet constituents.
The angular correlation function (ACF) measures the contribution
to a jet's mass from pairs of constituents separated by an angular
scale $R$ or less.  A high-$p_T$ QCD jet has
an ACF that goes approximately like a power of $R$, since it is nearly
scale-invariant.  By contrast, a jet initiated by a heavy particle decay 
has one or more intrinsic scales, which results in an ACF with one or
more ``cliffs''.\footnote{For example, consider a jet with two well-defined narrow subjets.  Its ACF will increase steeply near $R_\text{sub}$, where $R_\text{sub}$ is the subjet separation, since pairings of constituents from each subjet begin to contribute to the sum in Eq.~\ref{corrfunc} for $R\gtrsim R_\text{sub}$.}

For a given jet, numerous infrared/collinear-safe observables can be constructed from the ACF,
and these can be used to characterise the jet's substructure.  
For example, one can look at the angular scales
at which cliffs are located as well as the corresponding cliff heights.
Cliff heights are closely related to mass drops as utilised in 
BDRS mass-drop/filtering \cite{Butterworth:2008iy} and can be used to extract mass scales that correspond to 
hard substructure in the jet.  As a first application of these ideas,
Jankowiak and Larkoski developed a top tagging algorithm
whose performance is competitive with others in the literature. Other
applications remain to be explored.  Further work on these ideas was pursued in \cite{Jankowiak:2012na}.

\subsection{Template overlap}

The energy distribution resulting from hard scatterings can be well described by energy
correlation functions in momentum space.  In QCD, these naturally describe jet cross sections in 
terms of energy flow observables, which are peaked around the states associated with the hard scattering
that subsequently initiate the jets.  Therefore, energy flow observables within the jet should be of particular
interest to substructure studies. In \cite{Almeida:2010pa}, Almeida, Lee, Perez, Sterman, and Sung developed a method based on the
quantitative comparison of the energy flow of observed jets at high-$p_T$ with the flow from selected sets (the templates) of
partonic states. 

The template overlap procedure can be summarised as follows.
Let $|j\rangle$ denote the set of particles or calorimeter towers that 
make up a jet, identified by some algorithm, and take $|f\rangle$
to represent a set of partonic momenta $p_1\dots p_n$ that represent a boosted decay, found by the same algorithm.
The functional measure ${\cal F}(j,f)\equiv \langle f | j \rangle$ quantifies how
well the energy flow $|j\rangle$ matches the (templates) $|f\rangle$.
In practice, \cite{Almeida:2010pa} found good results with a simple construction of functional overlap based on a Gaussian in energy differences within angular regions surrounding the template partons.
Any region of partonic phase space for the boosted decays, $\{f\}$, defines a template.
Knowledge of the signal and background can be used to design a
custom analysis for each resonance, to make use of differences
in energy flow between signal and background.
The template overlap of an observed jet $j$ is the defined as 
$Ov(j,f[j])={\rm max}_{\,\{f\}}\, {\cal F}(j,f)$, 
the maximum functional overlap of $j$ to a state $f[j]$ within the template region,
where  $f[j]$ stands for the state of maximum overlap, emphasising that the value of the overlap functional
depends not only on the physical state $| j \rangle$, but also on the choice for the set of template
functions $f$. 

Template overlaps provide a tool to match unequivocally arbitrary final states $j$ to partonic partners $f[j]$ at any given order.
Once a ``peak template'' $f[j]$ is found, it can be used to characterise the energy
flow of the state, which gives additional information on the likelihood that
it is signal or background. In addition, template overlaps can be combined with higher moments of
the energy distribution or jet shapes to further discriminate the event. 

\subsection{Shower deconstruction}

Shower deconstruction, proposed by Soper and Spannowsky, is a method to look for new physics in a hadronic environment \cite{Soper:2011cr}. First, one picks the part of the event that is likely to be of interest, for instance the part contained in a large-radius jet that possibly contains the decay products of a boosted heavy particle. This part of the event is divided into small radius jets called the microjets that are ideally the size of topoclusters or calorimeter towers. If there are too many microjets to analyse, one can discard the microjets with the lowest transverse momenta. Shower deconstruction uses the four-momentum and possibly $b$-tag information for each microjet.Ê

The aim of shower deconstruction is to calculate a single number $\chi$ for each event such that events with small $\chi$ are likely to be background events and events with large $\chi$ are likely to be signal events. The number $\chi$ is an approximation to the ratio $P(S)/P(B)$ of the probability $P(S)$ that a parton shower Monte Carlo that represents the sought signal process would generate the given event to the probability $P(B)$ that a parton shower Monte Carlo that represents the background process would generate the event. The function $P(S)$ is calculated as of a sum, over all possible shower histories for the signal hypothesis, of weights that are a product of splitting-kernels and Sudakov factors. $P(B)$ is calculated the same way, but for the background hypothesis.

Although shower deconstruction is not limited to boosted configurations, the computing time increases strongly with the number of microjets. Boosted configurations are known to ameliorate combinatoric problems in reconstructing a resonance that decays hadronically because all decay products can be in one wide angle jet. Thus, \cite{Soper:2011cr} presents a first application of shower deconstruction using the $HZ$ production channel, where the boosted Higgs decays to a $b\bar{b}$ pair, first discussed in \cite{Butterworth:2008iy}. The statistical significance obtained with the shower deconstruction algorithm is found to be larger than that obtained with the method of \cite{Butterworth:2008iy}.

\subsection{\hep}

Unlike other taggers, the \hep, proposed by Plehn, Salam, and Spannowsky, is not motivated by
searches for resonances decaying to two highly relativistic top
quarks \cite{Plehn:2009rk}. Instead, its first application was the notorious Higgs search
channel $pp \to t\bar{t}H$ with a hadronically decaying $H \to
b\bar{b}$ \cite{Butterworth:2008iy}. In the Standard Model, one can expect several
percent of the events to have transverse momenta in the range $p_{T,H}
\gtrsim m_H$ and $p_{T,t} \gtrsim m_t$, for the leading hadronically
decaying top quark. To extract the $t\bar{t}H$ signal from the
continuum QCD background, \cite{Plehn:2009rk} required two fat jets,
one from a boosted Higgs and one from a boosted top quark.

Another application of top tagging in a moderately boosted regime
is identifying top partners---like a supersymmetric top squark---decaying to a top quark and an invisible dark matter
agent \cite{Plehn:2010st}. Similar to the $t\bar{t}H$ channel such
searches suffer from combinatoric backgrounds. Using this tagger, one
can exploit purely hadronic top decays and extract the stop pair
signal out of backgrounds.

Algorithmically, the \hep is motivated by the BDRS Higgs
tagger.  In particular, it starts with a large, $R=1.5$, Cambridge-Aachen
jet. This size immediately translates into a minimum transverse
momentum condition of $p_{T,t} \gtrsim 200$ \gev.  This fat jet is
unclustered using an iterative mass-drop criterion, with a general
cutoff at $m_j>30$ \gev for the subjets. Next, 
filtering \cite{Butterworth:2008iy} is applied to sets of three hard subjets, using
five constituents, and a combination of three subjets is chosen with jet
mass closest to the top mass. To reconstruct the $W$ mass, notice
that for the top decay kinematics, it is surprisingly likely that more
than one of the three $m_{jj}$ combinations lies within 15\% of the
$W$ mass. Therefore, the \hep does not aim to
distinguish the $b$ jet from the two $W$ decay jets, but instead
applies a more democratic subjet mass criterion described in \cite{Plehn:2010st}. Finally, a self-consistency 
condition is applied on the reconstructed transverse momentum
$p_{T,t} > 200$ \gev.

In a recent application, Plehn, Spannowsky, and Takeuchi 
studied semi-leptonic decays of top partners
into two top quarks and missing energy \cite{Plehn:2011tf}. The
hadronic top was reconstructed with the usual tagging algorithm. For the
leptonic top, the unmeasured
neutrino three-momentum was reconstructed based on the assumption of a boosted top decay. Two of the three unknown components can be reconstructed
using the top and $W$ mass constraints. For the third component, one can
analyse the top decay in a specific rest frame and find that one of
the neutrino momentum components is strongly suppressed.  This way, one can 
approximately reconstruct the neutrino momentum and compare it to
the measured two-dimensional missing transverse momentum vector. The
results for supersymmetric top squarks are promising over a wide range
of masses with a similar reach in the hadronic and semi-leptonic
channels.

As this report was in preparation, several further extensions of the \hep were proposed \cite{Plehn:2011sj}.

\subsection{Quark vs. gluon separation}

Being able to distinguish light-quark jets from gluon jets on
an event-by-event basis could significantly enhance the reach
of many new physics searches at the LHC. The
two prongs of this effort are finding intra-jet observables
whose distributions are significantly different between the
flavours, and finding relatively pure samples of quark and gluon
jets to measure these observables in data.  Identifying quark and gluon jets was also studied by ATLAS in the context of reducing jet energy scale uncertainty \cite{ATLASqvg}.

In \cite{Gallicchio:2011xq}, Gallicchio and Schwartz
systematically examined many existing and novel jet
substructure observables to find the ones whose distributions,
for a given jet $\eta$ and $p_T$, are the most powerful single
and multi-variable discriminants. It turned out that a
combination of the charged track multiplicity and the
$p_T$-weighted linear radial moment (girth) performed almost as
well on particle-level Monte Carlo as discriminants with more
variables. Over 95\% of the gluon jets can be filtered out
while keeping more than half of the light-quark jets.

The best single observable was the number of charged particles
within the jet (which were required to have $p_T>500$~MeV).
The discrimination power improved with jet $p_T$, and the
strength relative to other observables was greatest at high signal efficiency, where mild cuts were required.

Another good single variable, and part of the best pair, was the linear radial moment --- a
measure of the ``width'' or ``girth'' of the jet --- constructed by
adding up the $p_T$ deposits within the jet, weighted by
distance from jet axis. It is defined as
\begin{equation}
    g =  \sum_{i \in \mathrm{jet}} \frac{ p_T^i }{ p_T^{\rm jet} } \, |\Delta R_i|
\end{equation}
where $\Delta R_i = \sqrt{ \Delta y_i^2 + \Delta \phi_i^2 }$ and where
the true boost-invariant rapidity $y$ should be used when
measuring with respect to the (massive) jet axis instead of the
geometric pseudorapidity $\eta$.  This is a boost-invariant
version of jet broadening, to which it reduces in the limit of
massless constituents at small angles to the jet axis.

Finding relatively pure samples was discussed by Gallicchio and
Schwartz in \cite{Gallicchio:2011xc}.  Such samples are
necessary because all intra-jet observables have distributions
with significant overlap between quark and gluon jets. Combined
distributions of an evenly mixed sample do not provide
verification, independently for quarks and gluons, of the
showering and detector simulation.

Kinematic cuts on multijet and jets+$X$ tree-level samples were
optimised to purify first quarks, then gluons. At the 7 \tev
LHC, the $pp \to \gamma+2$ jets sample can provide 98\% pure
quark jets with 200 \gev of transverse momentum and a cross
section of 5 pb. To get 10 pb of 200 \gev jets with 90\%
gluon purity, the $pp \to 3$ jets sample can be used. These
samples could provide a direct evaluation of the tagging
technique at all jet $p_T$'s, verify and help improve the Monte
Carlo generators, and provide a test of perturbative QCD.

\subsection{ISR tagging}

In \cite{Krohn:2011zp}, Krohn, Randall, and Wang studied the feasibility of identifying jets from initial state radiation (ISR) on an event-by-event basis, and considered how these jets can be used in the interpretation of new physics phenomena.  As a proof of principle, they investigated the pair production of new physics states which each decay into jets and missing energy, and suggested that ISR can be identified by looking for  jets which are distinguished in either their $p_T$, rapidity, or $m/p_T$ ratio.   Using these three criteria they report that they can identify ISR in di-squark (di-gluino) events  roughly $40\%$ $(15\%)$ of the time with a mis-tag rate of around $10\%$ $(15\%)$.

The most obvious application of the technique is in reducing the combinatoric difficulties which arise in event reconstruction.   However, the production of ISR is governed by the detailed properties of a hard scattering  event, e.g., the flavour of the initial partons and the scale of the hard interaction, and so ISR can be used to distinguish between different production mechanisms yielding events with similar visible final states.  In \cite{Krohn:2011zp}, the authors provide an example of this, showing how one can, over many samples, observe the recoil of a new-physics system against ISR and thus infer the mass scale of the system even in the presence of significant missing energy.

\subsection{Multitagging for New Physics}

The application of one or more boosted object taggers can also
be used effectively in searches for the tagged objects \emph{within}
new physics event samples themselves.  This provides the potential
to discover the Higgs boson in a way distinct from Standard Model
search strategies, as well as characterising the interactions
of the new physics by understanding the variety of boosted objects
that appear in these samples.  Kribs, Martin, Roy, and Spannowsky
demonstrated that a slightly modified BDRS algorithm
\cite{Butterworth:2008iy}
was highly effective at finding the lightest supersymmetric
Higgs boson in superpartner-enriched event samples, where a boosted
Higgs boson appeared in the cascade from heavy supersymmetric
particles decaying to light supersymmetric particles with a
gravitino \cite{Kribs:2009yh} or neutralino \cite{Kribs:2010hp}
lightest supersymmetric particle.  Vector-like fermionic top partners
also provide a rich final state amenable to the simultaneous
application of multiple boosted object taggers.
Top-partners decay $t' \rightarrow (b W, t Z, t h)$ with roughly
$(50\%, 25\%, 25\%)$ branching ratios \cite{Perelstein:2003wd}.
Kribs, Martin, and Roy showed that combining both the
HEP top tagger \cite{Plehn:2009rk}
with the BDRS algorithm \cite{Butterworth:2008iy}
could identify both boosted $t$, boosted $h$, as well as boosted
$W/Z$ (though modified filtering/subjet techniques).  This was
shown to be highly effective at identifying top partners with the
Higgs boson in LHC simulations.

\section{Jet substructure in FastJet 3}
\label{sec:software}

One of the aims of FastJet 3, available since October 2011 from
\url{http://fastjet.fr/}, is to facilitate the use and development of
jet substructure tools.
A novelty relative to the 2.X series is that jets are now
``self-aware''. So whereas previously one would access a jet's
constituents via its cluster sequence
\begin{lstlisting}
  vector<PseudoJet> constituents = cluster_sequence.constituents(jet);
\end{lstlisting}
one may now write
\begin{lstlisting}
  vector<PseudoJet> constituents = jet.constituents();
\end{lstlisting}
and similarly for other properties such as parents, subjets and, where
relevant, areas.
Aside from being more intuitive to write, this also has the advantage
that when dealing with multiple cluster sequences, one no longer needs
to remember which cluster sequence to use with which jet.

In FastJet 2.X, the only way in which one could create an object with
substructure was by creating a cluster sequence for it.
Accordingly a number of tools were written as plugins, which wasn't
necessarily the most natural way of formulating them.
In FastJet 3, an easy way of creating substructure is via the
``join'' function.
Suppose one has \texttt{\small W1}, \texttt{\small W2} and \texttt{\small b} subjets
obtained from some declustering procedure.
Then one may simply write
\begin{lstlisting}
  PseudoJet W = join(W1,W2);
  PseudoJet top = join(W, b);
\end{lstlisting}
Then, not only will the top's momentum be sensible, but \texttt{\small
  top.constituents()} will return the concatenation of the
constituents of \texttt{\small W1}, \texttt{\small W2} and
\texttt{\small b}.
The high-level top substructure can be accessed with
\begin{lstlisting}
  vector<PseudoJet> top_pieces = top.pieces()
\end{lstlisting}
where the vector contains the \texttt{\small W} and the \texttt{\small b}.
The \texttt{\small pieces()} function also works on normal jets from a
cluster sequence.

\begin{figure}
  \centering
  \includegraphics[width=\textwidth]{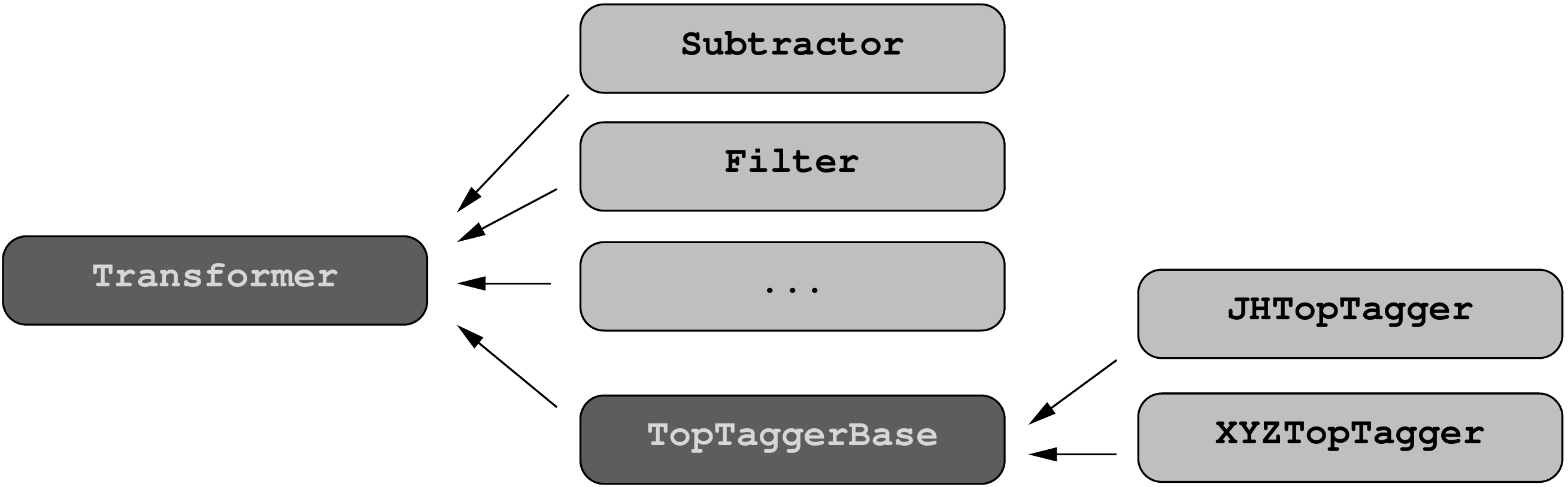}
  \caption{\small Illustration of some of the classes deriving from
    \texttt{\small Transformer} in FastJet~3. Classes shown in dark grey are
    abstract base classes.}
  \label{fig:fj3-transformers}
\end{figure}

A number of substructure tools are distributed with FastJet~3. They
derive from the \texttt{\small Transformer} class, and top taggers should
derive from \texttt{\small TopTaggerBase}, as shown in
\figrefcap{fj3-transformers}.
As an example consider the \texttt{\small Filter} class, which can be used as
follows:
\begin{lstlisting}
  #include "fastjet/tools/Filter.hh"
  #include "fastjet/Selector.hh"
  double Rfilt = 0.3;
  Selector selector_filter = SelectorNHardest(3);
  Filter filter(Rfilt, selector_filter);
  PseudoJet filtered_jet = filter(jet);
\end{lstlisting}
Note the use of the \texttt{\small Selector} (itself new to
FastJet~3), which provides an easy way of specifying the cuts on the
subjets and passing them to the \texttt{\small Filter} class.
Simply changing the selector leads to trimming~\cite{Krohn:2009th}
\begin{lstlisting}
  Selector selector_trimmer = SelectorPtFractionMin(0.05);
  Filter trimmer(Rfilt, selector_trimmer);
  PseudoJet trimmed_jet = trimmer(jet);
\end{lstlisting}
Transformed jets often have more internal structure than a normal jet.
In the case of trimmers and filters, for example, one may want to know
which subjects did not pass the selection.
For this purpose there is the
\texttt{\small structure\_of<TransformerName>()} function, which returns a
reference to a transformer-specific class with extra information about
the jet, e.g.,\ 
\begin{lstlisting}
  vector<PseudoJet> rejected_subjets = 
                       filtered_jet.structure_of<Filter>().rejected();
\end{lstlisting}
Other substructure tools that are part of FastJet~3 include a
\texttt{\small Pruner}~\cite{Ellis:2009me}, a
\texttt{\small MassDropTagger}~\cite{Butterworth:2008iy}, a
\texttt{\small JHTopTagger}~\cite{Kaplan:2008ie}, and a
\texttt{\small RestFrameNSubjettinessTagger}~\cite{Kim:2010uj}.
Given the rapid evolution of the substructure field, it is not
realistic for FastJet to always distribute a complete and up-to-date
set of substructure tools.
Accordingly, in the near future we envisage setting up a FastJet
``contrib'' area to facilitate dissemination of third-party tools.

FastJet~3 also contains changes unrelated to substructure. 
These include a new interface to pile-up and underlying background
estimation, including a new and very fast grid-based median background
estimator.
Also of use is a new facility for attaching arbitrary user information
to jets, beyond just the \texttt{\small user\_index} of FastJet 2.X. 
Illustrations of its usage together with HepMC and Pythia~8 are available
on request from the FastJet authors.
More information on these and other developments is to be found in the
FastJet manual, as well as the slides and doxygen documentation on the
FastJet web site~\cite{fastjetURL}.


\section{Benchmark samples and comparisons}
\label{sec:compare}

As proposals proliferate for LHC searches with jet substructure, the space of questions to ask about substructure techniques grows correspondingly larger.  Which techniques work best for which signals?  Does the answer depend on the $\pt$ of particles in question, or whether the analysis is a search or a measurement?  How do detector limitations and uncertainties affect each method?  How do things change as the amount of pile-up present increases?  If we study these questions with Monte Carlo event generators, do the answers depend on which generator we choose?

In this section we will address a few of these questions.  Beyond the specific results we give, we hope that the set of samples and tools we use will provide a starting point for future work.  The results here can then be seen as benchmarks for comparison.

Our starting point is the set of comparisons presented in the \boost\ 2010 report \cite{Boost2010}.  That analysis compared top taggers exclusively for several reasons.  A large number of substructure techniques are either designed for finding tops or can be used that way; the two-step decay of a top quark is an interesting example to study; and finding top jets is immediately useful at the LHC.  For the same reasons, as well as limited time and space, we will take the same approach and focus on distinguishing top jets from pure QCD background.\footnote{A thorough exploration of substructure methods for identifying $W$ jets can be found in \cite{Cui:2010km}.}  Extending last year's results, we will consider new techniques, more Monte Carlo generators --- including matched samples, and the effects of a simple detector model.

Samples new and old are publicly available at
\begin{itemize}
\item[] \url{http://tev4.phys.washington.edu/TeraScale} and
\item[] \url{http://www.hep.louisville.edu/~verm/TeraScale}.
\end{itemize}

\subsection{New samples}

This year we expand the set of benchmark samples with events generated by the \sherpa\ and \herwigpp\ Monte Carlos.  The \sherpa\ events allow us to examine the effects of higher-order matrix elements, which should provide a more accurate description of hard QCD radiation.  The \herwigpp\ events are a useful cross-check of last year's parton-shower--level results using a current-generation program.\footnote{We omit other possibilities only for reasons of time and space, and not any stance on the superiority of particular programs.}

\maketitle

\section{Description of benchmark samples}

\end{comment}

\subsubsection{Matched samples with \sherpa}

We consider LHC proton-proton collisions at a center-of-mass energy of 7 TeV. We have simulated the 
Standard Model processes $t\bar t +X$, $W/Z+X$ and pure QCD jets.  All samples were produced 
with version 1.3.1 of the \sherpa\ event generator \cite{Gleisberg:2003xi,Gleisberg:2008ta}. For 
the electroweak gauge bosons as well as the top-quarks we consider fully hadronic decays only.
For each process we generated samples of 100,000 unweighted events for \pt slices of $100$ 
GeV between $0\dots 1.6$ TeV, resulting in an almost flat \pt distribution when combining 
all samples. As slicing parameter we used the maximum of the Monte Carlo top/anti-top \pt, the gauge-boson \pt for the 
$W/Z$ channels, and the leading-jet \pt for pure jet production, where parton-level jets 
were reconstructed according to the \antikt algorithm with $R=1.5$. For all processes the 
slicing cut is applied at the matrix-element level before parton showering is invoked. 

To accurately simulate high-\pt events and to account for configurations with more than 
one hard parton inside a large jet, we employ \sherpa's matrix-element--parton-shower matching 
algorithm \cite{Hoeche:2009rj,Hoeche:2009xc}. We consider the complete sets of tree-level matrix 
elements with up to two (four) additional final-state partons for \ttbar and dijet ($W/Z$) 
production from \sherpa's matrix-element generator \comix\ \cite{Gleisberg:2008fv}. The matrix 
elements of varying multiplicity are consistently combined to form inclusive samples of fully 
exclusive events through subsequent parton shower evolution \cite{Schumann:2007mg} before being 
subjected to hadronization. The renormalization and factorization scales are dynamically 
determined on an event-by-event basis according to the matching algorithm \cite{Hoeche:2009rj}. 
For the parton separation parameter of the matching procedure we use $Q_{\rm cut} = 20$ GeV. 
The default tune of \sherpa\ 1.3.1 for the hadronization and underlying event model parameters 
was used.

Calculations of jet properties typically focus on single jet observables, also known as jet shapes.  These variables are inclusive over the jet constituents, measuring a function of the final state momenta.  By contrast, most jet substructure methods involve more complicated partitioning of the jet constituents and observables of localised structure in the jet.  While many early studies used subjet information to identify the hadronic decays of heavy particles, recent studies have explored the efficacy of jet shapes in boosted heavy object searches (e.g., \cite{Thaler:2010tr,Jankowiak:2011qa}).  Observables derived from the N-subjettiness jet shapes have been shown to be very competitive in top tagging (see also \secrefcap{compare}).  This gives hope that calculationally simpler observables may be useful in new physics searches.


In this section we will discuss the calculation of jet observables, focusing on jet mass, which provides an example of a jet observable that can be usefully calculated.  We present aspects of the calculation and discuss improvements that can be made with future work.  We then discuss other observables that are useful for measurement and point out where calculations may soon be available.

\subsection{Jet mass as an example}

Jet mass is the simplest and most phenomenologically useful jet shape.  An accurate measurement depends on a good understanding of the jet energy scale and splash-in contributions (pile-up, the underlying event, and initial state radiation).  The importance of the measurement and the accessibility of the calculation mean that a comparison between the two can be very useful.

The measurement of a single jet mass is still a complex observable.  The calculation is performed for an exclusive jet multiplicity, meaning there must be a veto on additional jets.  Consider the cross section for dijet events from the \akt algorithm, differential in both jet masses, $d^2 \sigma / dm_1 dm_2$.  We must veto on additional jets by placing a \pt cut on candidate jets found by the algorithm.  Additionally, the perturbative calculation is only valid for high-\pt jets, meaning a restriction of the phase space for the hard scattering is required.

There are two main regimes in the jet mass distribution.  The small mass regime, $m_J \muchless p_{TJ}$, contains large logs of $m_J/p_{TJ}$.  Resummation is required to tame these large logs and restore accuracy.  For larger masses, a fixed order calculation suffices to describe the distribution.  Next-to-leading order is currently state-of-the-art for a wide range of processes, and can be simulated by Monte Carlo for many processes.

The small jet mass limit is the focus of the most effort in the calculation.  Resummation must be performed in this regime, which requires proving certain properties of the cross section.  In particular, some kind of factorisation must be proven that relates the all-orders, large-log terms to the fixed order terms. 

The additional factorisation required for resummation of jet mass takes two basic forms, depending on what tools are used.  Both perturbative QCD and soft-collinear effective theory (SCET) can be used to resum the jet mass distribution, and the essential physics that each uses is the same.  However, the language each uses is distinct, and a comparison of them would be useful.  Here we give an outline of factorisation in SCET; some features of resummation in SCET and QCD are compared in the next subsection.

In the SCET formulation of the cross section, an effective theory framework is used \cite{Bauer:2000ew,Bauer:2000yr,Bauer:2001ct,Bauer:2001yt}.  Hard modes that produce perturbative emissions of energetic partons are integrated out of QCD.  The remaining soft and collinear modes have momenta parameterised by a power counting parameter $\lambda \muchless 1$.  This power counting parameter is typically determined by the kinematics of the final state; for the jet mass example, $\lambda = m_J / p_{TJ}$.  The soft and collinear fields decouple at leading power in the SCET Lagrangian \cite{Bauer:2001yt}, decoupling the $N$-jet operators used to describe the energetic jets.  Factorisation also requires showing that the measurement (such as the jet mass with the \antikt algorithm and a \pt cut on additional jets) factorizes as well.  This means the phase space constraints on soft and collinear modes must separate.  This can be shown formally by providing an operator formulation of the phase space constraints \cite{Bauer:2008dt,Bauer:2008jx}, and can also be seen at a physical level by a power counting analysis of the measurement in terms of the modes in SCET \cite{Walsh:2011fz}.  This leads to a schematic factorisation theorem of the form, for $N$ jets:\begin{equation}
\sigma = H_N \left[ B_a \times B_b \times J_1 \times \cdots \times J_N \right] \otimes S_N \,.
\end{equation}
The hard function $H_N$ is the matching coefficient onto SCET and arises from the short-distance interaction that produces the well-separated, energetic partons.  The jet functions $J_i$ and beam functions $B_{a,b}$ come from the final and initial state collinear evolution of the hard partons \cite{Stewart:2009yx}.  The soft function $S_N$ comes from the global soft gluon radiation across the event.  Each of these functions can be separately calculated in perturbation theory, and renormalisation group evolution is used to sum the large logarithms that arise in the cross section.  This framework formalises much of the resummation machinery used in perturbative QCD.

\subsubsection{Facets of the calculation}

We first discuss the perturbative contributions to the cross section, in particular the resummation.  We focus on jets in the \antikt algorithm, since jets defined in other algorithms are complicated by the non-trivial role of clustering between soft gluons.  Soft gluon clustering in these algorithms means that the soft phase space does not factorize into single particle phase space for each soft gluon, affecting the subleading logarithms in the perturbative series starting at $\mathcal{O}(\alpha_s^2 \ln^2)$ \cite{Banfi:2005gj,Delenda:2006nf}.  In the \antikt algorithm the phase space for a single soft particle is not affected by the presence of other soft particles.

The leading logs in the cross section are given by the exponentiation of single soft gluon emission in QCD, and single-log and running-coupling corrections can be included.  Equivalently in SCET, the leading logs are specified by the cusp anomalous dimensions of the one-loop jet and soft functions, with the single-log terms coming from the non-cusp anomalous dimensions.  Running-coupling effects arise in renormalisation group evolution.  In the limit of small jet radius ($R \muchless 1$), the leading logs have the form
\begin{equation}
\int_0^m \frac{\df\sigma}{\df m} \sim \sigma_0 \exp\left( - \frac{\alpha_s C_F}{2\pi} \ln^2 \frac{m}{p_T R^2} \right).
\end{equation}

In principle, subleading logarithms can be straightforwardly computed and resummed.  The extension of resummation techniques beyond next-to-leading-log (NLL) is natural using the SCET framework.  However, general jet shapes suffer from non-global logarithms that contribute at NLL \cite{Dasgupta:2001sh}.  Nearly every jet measurement made at the LHC will have non-global observables, and their resummation is more challenging and can only be performed in the leading-colour approximation \cite{Dasgupta:2001sh,Appleby:2002ke,Banfi:2002hw,Banfi:2010pa}.  The physics of the leading non-global logs at fixed order ($\alpha_s^2$) has been well understood in QCD, and their resummation calculated numerically in the large-$N_C$ limit. Recent explorations in SCET have provided additional insights into their structure and origin, providing a possible new path to renormalisation group-based resummation \cite{Kelley:2011ng,Hornig:2011iu,Hornig:2011tg}.  Non-global logs are reduced by soft gluon clustering, and so are reduced for algorithms other than anti-$\kt$~\cite{Appleby:2002ke,Hornig:2011tg}.

The small-$R$ approximation is advantageous because a simple picture of non-global logs emerges.  Each jet has its own non-global correction factor, and cross talk between jets is relegated to a power-suppressed contribution in $R$.  The non-global factor for a jet was calculated in the large-$N_c$ limit in~\cite{Dasgupta:2001sh} and used for jet mass in~\cite{Banfi:2010pa}.  The large-$N_c$ approximation will receive corrections from subleading colour terms which will modify the non-global log contribution at around the ten percent level.  Since non-global logs contribute at NLL and start at $\cO(\alpha_s^2 \ln^2)$, subleading terms in colour ought not to matter phenomenologically, as was the case for deep inelastic scattering event shapes~\cite{Dasgupta:2002dc}.  Thus in the small-$R$ approximation the cross section is described by a resummed factor for each jet dressed by a calculated non-global factor, and one can include jet mass measurements for each jet~\cite{Banfi:2010pa}.

Beyond the small-$R$ approximation, one can account for the corrections by including terms that vanish as powers of $R$.  In QCD, one computes soft gluon emission from all of the colour dipoles in the event, including the incoming partons.  Expanding the result as a power series in $R$, one finds that the terms of relative order $R^2$ make a numerical impact (at around the ten percent level even for $R=0.4$), but terms scaling as $R^4$ do not matter even for $R = 1$.  Similarly, the non-global factor must be corrected beyond the small-$R$ limit.  In SCET, the sum over soft gluon emission from colour dipoles is given in the soft function, where one expands in
\begin{equation}
t_{ij}^{-1} \equiv \frac{\tan^2 R/2}{\tan^2 \theta_{ij}/2} \,,
\end{equation}
with $\theta_{ij}$ the angle between jets $i$ and $j$ \cite{Ellis:2009wj,Ellis:2010rwa}.  The requirement $1/t_{ij} \muchless 1$  forces the jets to be well-separated, which is needed to factorize the cross section.  When jets become close together, additional modes in SCET are required to separate the scale dependence on the small dijet invariant mass~\cite{Bauer:2011uc}.  These modes parameterize soft radiation between the nearby jets, and SCET can be formally extended to SCET${}_+$, which includes these modes.

Therefore, for a wide range of jet mass measurements it should be possible to obtain a resummed result that accounts for non-global logarithms and goes beyond the small-$R$ approximation to sufficient accuracy so as to be phenomenologically useful.

A resummed calculation describes the dominant shape of the distribution.  However, many other effects must be taken into account.  The shape of the large-mass tail of the distribution is described well by fixed-order QCD.  The resummed and fixed-order calculations can be combined in a simple way:
\begin{equation}
\frac{\df\sigma}{\df m} = \frac{\df\sigma}{\df m}\bigg|_{\text{resum}} - \frac{\df\sigma}{\df m}\bigg|_{\text{resum}}^{\text{FO exp}} + \frac{\df\sigma}{\df m}\bigg|_{\text{FO}} \,.
\end{equation}
The resummed and fixed order distributions are added, and the fixed order expansion of the resummed distribution is subtracted.  In the small-mass limit, the two fixed order functions cancel up to power corrections in $m/\pt$, and the resummed distribution correctly describes the distribution.  In the large-mass tail, the resummed distributions (all orders and fixed-order expanded) cancel up to the fixed order in $\alpha_s$ of the expansion, so the fixed-order QCD distribution describes the shape up to higher-order corrections.

Hadronisation, pile-up, and underlying event contributions must also be included in a calculation of the jet mass.  These non-perturbative contributions affect the small-mass regime of the distribution, and are needed for the calculation to be compared to data.  While important, these effects are challenging to include, as no framework exists to systemically include them and they require a more detailed understanding of the data.

Basic models have been formulated to quantify contributions from the underlying event and pile-up \cite{Cacciari:2007aq,Dasgupta:2007wa,Cacciari:2008gn,Cacciari:2009dp}.  As jet shapes are measured and calculated more extensively, the comparison between the two is a good testing ground to study these models and fold them into calculations.  In the case of 2-jet event shapes in $e^+e^-$ collisions, the correction to the first moment of the event shape from non-perturbative effects has been shown to be related to a universal, measurable non-perturbative matrix element for a wide class of event shapes \cite{Lee:2006fn,Lee:2006nr}.  In the SCET framework, model soft functions that parameterise non-perturbative effects have also been introduced \cite{Hoang:2007vb}.  Further study is warranted to understand the how non-perturbative effects in a hadron collider environment can be introduced, and whether similar universality relations that exist in $e^+e^-$ collisions apply in the hadron collider case.

\subsection{Measurable and calculable observables}

A wide array of observables have been defined to study the structure of jets.  While the jet mass is phenomenologically the most important, it is important to prioritise which other observables should be measured in the near term at the LHC.  There are several key reasons to measure jet substructure.  The comparison to precision QCD calculations is obviously attractive, and understanding QCD backgrounds helps characterise observables relevant for new physics searches.  Additionally, jet substructure measurements can further understanding of detector effects and improve unfolding techniques.  This can have an impact beyond jet substructure measurements, such as a reduction in the jet energy scale uncertainty.

There are four general classes of observables used to measure jet substructure and jetty events:
\begin{itemize}\renewcommand{\labelitemi}{$\bullet$}
\item Jet mass
\item Jet shapes
\item Groomed jet observables
\item Global event shapes
\end{itemize}
Although jet mass is a jet shape, it is central enough to jet physics that special consideration should be given to it as a first measurement.  Jet grooming methods are generic substructure methods that remove isolated soft radiation in jets.  Three grooming methods have been proposed: filtering \cite{Butterworth:2008iy}, pruning \cite{Ellis:2009su}, and trimming \cite{Krohn:2009th}.  The measurement of properties of groomed jets can provide data about the structure of the jet that is difficult to capture with simpler observables such as jet shapes.  Finally, global event shapes can be measured without clustering into jets.  Several event shapes are useful in characterising soft physics that can contribute to jet substructure.

Below we discuss each class of observables, identifying key observables in each category.  These are measurements that theorists find informative, and will help in the study of jet substructure.  Many of these observables may be calculated in the near term, and measurement of them can provide motivation for a detailed comparison.

\subsubsection{Jet mass measurements}

There are several possible channels in which a jet mass distribution can be measured.  These include pure QCD events with inclusive and exclusive jet cross sections, jets produced in association with electroweak bosons, and light quark jets produced in association with top quarks.  Because the calculation of a jet mass distribution depends on the precise event topologies studied, to compare theory and experiment it is important to carefully choose the measurements.  Several key topologies will provide important jet mass measurements.  The three we focus on are $\gamma/W/Z$ + jet, dijet, and multijet events.  In each topology, there are particular jet mass observables of interest, specialised to the final state.

\begin{itemizedot}
\item \textbf{${\bf \gamma/W/Z}$ + jet events} -- These have only a single jet in the final state (assuming the $W$ or $Z$ decays leptonically) and provide the cleanest jet measurements.  This class of events is an important reference sample for jet energy scale calibration, and can be used to measure fake rates for certain substructure tagging algorithms.  They are a source of jet + lepton and jet + MET final states, and are a background to many new physics searches.  Theoretically, they can be used as precision tests of QCD and provide a more calculable environment due to the single jet multiplicity.

The observable of interest is the differential jet mass distribution, $\df\sigma/\df m_J$, in bins of jet \pt.  A jet veto must be imposed to obtain an exclusive sample of single jet events.  A hard veto on additional jets, such as $p_{T_J} < 0.1\, p_T^{\gamma}$, can be useful for jet energy scale calibration.  The jet mass distribution in this case is of interest because of the restriction on radiation outside the singe jet.  For any value of the jet veto, if the jet mass distribution is finely binned in \pt (several bins instead of one or two), the impact of various components of the calculation can be compared across \pt.

\item \textbf{dijet events} -- Dijet events with high-\pt jets have a large cross section and make up a significant fraction of the pure QCD events containing high-\pt jets at the LHC.  These events are backgrounds to many new physics searches, and their large cross section means that they must be precisely understood.

Because there are multiple jets in the event, different jet mass measurements can be made.  The doubly differential distribution,
\begin{equation}
\frac{\df^2 \sigma}{\df m_1 \df m_2} ,
\end{equation}
with $p_{T1} > p_{T2}$, is interesting because correlations between the jet masses can be studied.  This was explored in the CDF analysis of jet mass \cite{Aaltonen:2011pg}, where the ratio of event numbers in different bins in the double differential distribution was measured.  Define two single-jet mass bins $b_1$ and $b_2$ (in the CDF study $b_1 = (30,50)$ GeV and $b_2 = (130,210)$ GeV).  These bins define four regions:
\begin{align}
A&: \, m_1 \in b_1 , \quad m_2 \in b_1 \,, \nn \\
B&: \, m_1 \in b_2 , \quad m_2 \in b_1 \,, \nn \\
C&: \, m_1 \in b_1 , \quad m_2 \in b_2 \,, \nn \\
D&: \, m_1 \in b_2 , \quad m_2 \in b_2 \,. \\
\end{align}
Then define the ratio $R_{\text{mass}}$ to be
\begin{equation}
R_{\text{mass}} \equiv \frac{N_B N_C}{N_A N_D} \,.
\end{equation}
If the jet mass distributions were independent, then $R_{\text{mass}} = 1$.  Correlations between these distribution exist because of soft gluon exchange, although the expected value of $R$ is not known.  This observable is of interest due to the discrepancy between Monte Carlo simulation and data, and a theory calculation could help resolve the differences.

In addition to the doubly differential mass distribution, the projection onto a single mass is theoretically accessible.  Common choices are the heavy jet mass and the sum of jet masses, which have been studied at $e^+ e^-$ colliders.  The hardest jet's mass is harder to predict theoretically because the relative \pt in dijet events is affected primarily by soft physics.

\item \textbf{multijet events} -- Events with more than two jets provide a more complex background.  The colour correlations between jets are enhanced as the detector becomes more crowded, and these events typically have a hierarchy of jet {\pt}'s.  Multijet events are more important at the LHC due to the increased phase space for high-\pt parton production, and significant attention has been paid to understanding them better theoretically and experimentally.  Multijet events present a theoretical challenge to describe, but also offer the opportunity to learn about effects such as colour correlations that figure less prominently in simpler events.

In multijet events, placing a hard \pt cut on the leading three jets will give a multijet event sample that allows one to study the structure of the hardest jet as the kinematics of the events change.  In particular, the jet mass distribution of the hardest jet can be measured in bins of the $\Delta R$ separation to the nearest jet.  As jets become closer together, inter-jet soft radiation can have a large effect on the mass distribution.  This can be predicted theoretically, and the result compared with the Monte Carlo.
\end{itemizedot}

\noindent Because the jet mass is sensitive to the environment of the event, it is important to examine the effect of event-wide characteristics on the mass.  As pile-up increases, the jet mass can serve as a probe of the size of the pile-up and the effectiveness of subtraction methods (see \secrefcap{pileupatlas}).

The rapidities of the hard jets will affect the jet mass distribution, for a number of reasons.  First, large-rapidity jets imply a large momentum asymmetry between the initial partons, meaning that one carries a large fraction of the proton momentum ($x\sim 0.1$).  Since the quark parton distribution functions dominate in this regime, the fraction of quark jets will be enhanced relative to central events.  Second, the shape of the parton distribution functions implies there is likely to be more initial state radiation, since they are less steep in the large $x$ region.  Coupled with the fact that the large-rapidity jets are closer to the beams, it is interesting to see the dependence of jet mass on the jet rapidity.

Finally, the total transverse momentum in the event, \HT, is a measure of the activity of the event.  Soft jets that are vetoed contribute to \HT and will affect the jet mass, but still fit into the same event selection criteria.  The contribution of underlying event and pile-up will similarly affect the jet mass distribution, and \HT is a simple proxy for the size of these observables.

\subsubsection{Jet shape measurements}

Jet shapes measure the radiation pattern within a jet by calculating a function of the momenta in the jet.  The jet mass is a special jet shape, but there are other important jet shapes whose measurement would be very informative.  We discuss several.  Both $N$-subjettiness and planar flow are calculable, as they take moments of the momenta in the jet.

\begin{itemizedot}
\item $N$-subjettiness is a useful variable for jet substructure~\cite{Thaler:2010tr,Thaler:2011gf}.  $N$-subjettiness is defined with respect to a set of subjet axes $q_i$:
\begin{equation}
\tau_N^{(\beta)} = \frac{\sum_k p_{T,k} \left(\min \left\{\Delta R_{1,k},\Delta R_{2,k},\ldots, \Delta R_{N,k}   \right\} \right)^{\beta}}{\sum_k p_{T,k} (R_0)^{\beta}},
\end{equation}
where $\beta > 0$ is a real parameter, $R_0$ is the radius of the jet, and $\Delta R_{i,k}$ is the angular separation between particle $k$ and subjet axis $q_i$.  The shape $\tau_N^{(\beta)}$ probes how ``$N$-subjet-like'' a jet is, and provides an effective veto on additional subjets.  Because adding more axes only lowers the value of $\tau_N^{(\beta)}$, $N$-subjettiness cannot provide a veto against fewer subjets.

The case $N=1, \beta = 1$ is also known as girth, broadening, or width; $\beta = 2$ is related to the jet mass for small masses.  These are the most common choices for $\beta$, and it would be interesting to compare them.  Additionally, other choices of $\beta$ can be informative, as they will probe the structure of the jet as a function of the angle to the 1-subjettiness axis.  $\tau_1^{(\beta)}$ is similar to jet angularities, a jet shaped defined for $e^+e^-$ events~\cite{Berger:2003iw,Hornig:2009vb,Ellis:2009wj,Ellis:2010rwa}.  The distribution in jet angularity $\tau_a$ was calculated to next-to-leading log accuracy in $\tau_a$~\cite{Ellis:2010rwa}, and the theoretical tools exist to extend this calculation to the distribution of $\tau_1^{(\beta)}$ at the LHC.

2- and 3-subjettiness are also of interest.  These can probe jets with multiple subjets, and the substructure of these jets will more closely resemble a boosted heavy object decay.  (Using $\tau_3/\tau_2$ to identify top jets was studied in \secrefcap{compare}.)

\item Planar flow is an interesting shape variable that was proposed to distinguish top and QCD jets~\cite{Almeida:2008yp,Thaler:2008ju}.  Planar flow is defined in terms of the eigenvalues of a 2-by-2 matrix, $I_w$, constructed from the components of momenta perpendicular to the jet axis:
\begin{equation}
I_w^{kl} = \frac{1}{m_J} \sum_{i\in J} \frac{p_i^{\perp k} p_i^{\perp l}}{E_i} \,,
\end{equation}
where $p^{\perp k}$ is the $k^{\rm th}$ component of $p$ perpendicular to the jet axes (with respect to a pair of basis axes in the perpendicular plane).  From $I_w$, planar flow is defined
\begin{equation}
Pf = \frac{4\, {\rm det}\, {I_w}}{{\rm tr}(I_w)^2} \,.
\end{equation}
If the radiation in the jet is clustered along a line in the perpendicular plane going through the jet axis, then $Pf$ will be close to 0.  This will be the case for QCD jets and jets with 2-body kinematics, such as Higgs or $W$ jets.  For jets with 3-body (or more) kinematics, such as the top, $Pf$ will be $\mathcal{O}(1)$.

\item Subjet and track multiplicities have been shown to be effective discriminators of quark and gluon jets~\cite{Gallicchio:2011xq,Gallicchio:2011xc}.  Subjets can be found by reclustering the jet with a small radius, and finer resolution of the subjet size can improve discrimination between jets from different sources.  The study of subjet multiplicity offers the chance to study the sensitivity to \pt or energy cutoffs placed on jet constituents or subjets.  Does the energy resolution for subjets significantly improve with a higher energy cutoff on jet constituents?  Does the multiplicity become sensitive to pile-up and the underlying event if the cutoff is taken too low?  Similarly, the charged track multiplicity is a useful observable whose power depends on the angular resolution for track identification as well as the performance over the range of charged particle momenta in the jet.

\end{itemizedot}

\subsubsection{Groomed jet measurements}

Many new physics searches can utilise jet grooming algorithms.  Filtering, pruning, and trimming are jet grooming algorithms that modify the substructure of the jet to remove contributions from the underlying event and pile-up~\cite{Butterworth:2008iy,Ellis:2009su,Ellis:2009me,Krohn:2009th}.  These algorithms can be applied to a wide range of channels, and comparing the performance of the three algorithms can instructive (cf. Section 6 of Ref.~\cite{Boost2010}).  Because each algorithm operates on the substructure differently, it is useful to compare all of them.  Experimental studies can quantify the performance of each algorithm and determine which is best for a given application.

Each grooming algorithm operates on a single jet and produces a groomed jet.  Therefore any observable can be measured on the original jet and the groomed jet.  The distributions before and after grooming are always of interest to understand the effects of grooming, and grooming algorithms can be used to study a variety of jet properties.  The most interesting observables and objects to study with jet grooming are
\begin{itemizedot}
\item Jet mass.
\item Pile-up and the underlying event are a large motivation for jet grooming methods.  The performance with respect to pile-up for some jet grooming algorithms is currently being studied, and is quantified by studying tracks from secondary vertices.  Quantifying the performance with respect to the underlying event is more challenging, but measures exist \cite{Cacciari:2009dp}.
\item If $N$-subjettiness is measured, then the effect of the grooming algorithms will be interesting to study.  Grooming algorithms are designed to remove isolated soft radiation that can come from the underlying event or pile-up and contribute to poor mass resolution in the jet.  This radiation will also have a large effect on $N$-subjettiness.  Studying the $N$-subjettiness distribution (for $N = 1,2,3$) will give insight to grooming methods using an observable where it is simple to visualise jets with multiple energetic subjets.
\item Given the good tracking and angular resolution at the LHC, the experiments can study the performance of grooming methods using only charged tracks.
\end{itemizedot}

\subsubsection{Global event shapes}

Finally, event shapes are also of interest.  While these are not direct jet substructure observables, they have a strong connection to jet substructure.  Initial studies of jet substructure were driven by Monte Carlo studies, but more recently the theoretical foundations of jet substructure have been explored.  Event shapes have traditionally been a testing ground for theoretical tools, and jet shapes are a by-product of these studies.  $N$-subjettiness was built in analogy to $N$-jettiness, an event shape used to veto against additional jets in an event~\cite{Stewart:2010tn,Berger:2010xi,Jouttenus:2011wh}.  Like $N$-subjettiness, $N$-jettiness measures how $\le \! N$-jet-like an event is.  Measurements of event shapes can be compared directly to calculations, and the comparison is useful in building theoretical tools to calculate jet substructure properties.  These observables are especially amenable to calculation, and in fact distributions of these observables have already been calculated in a variety of applications.

Observables of interest include:
\begin{itemizedot}
\item 0-jettiness, also known as beam thrust, is a measure of total hadronic activity away from the beams.  In events without energetic jets, this can be used to study the underlying event and assist in tuning the Monte Carlo.  In new physics searches with a veto on any jets (e.g., $h\to WW$), the beam thrust distribution can be used to understand the theory uncertainties on the jet veto.
\item 1-jettiness can be similarly implemented on events with a single jet, and can be used to veto against additional jets for an exclusive 1-jet sample.  An ample background of $\gamma/W/Z +$ jet exists in which to study the 1-jettiness distribution.
\item A number of event shapes have been proposed for study in dijet type events in \cite{Banfi:2004nk} and examined in detail in \cite{Banfi:2010xy}, including observables related to thrust, broadening and jet rates ($Y_{23}$). Thrust-like observables show enhanced sensitivity to the underlying event, possibly allowing for powerful constraints on its properties, while $Y_{23}$ provides a more direct sensitivity to perturbative effects. The distribution of $Y_{23}$ has also been studied in \cite{brooijmans}. The above observables have all been defined with particles as inputs. One can also study event shapes with jets as inputs and in one such study \cite{cmsshapes}, Monte Carlos with matched matrix elements covering a range of jet multiplicities gave a worse description of event shape data than parton shower Monte Carlos (which only include $2\to2$ matrix elements matched directly onto the parton shower).  This is troubling, and comparison to more event shapes can provide a way to differentiate different effects in the Monte Carlo simulations that may be contributing to this issue.

\end{itemizedot}

\section{Concluding remarks}
\label{sec:conclusions}


The field of jet substructure has made remarkable progress since its inception more than twenty years ago \cite{Seymour:1991cb}.  In this we report we have endeavoured to present a contemporary overview of jet substructure theory and practice.  The experimental results published so far are encouraging and promising.  They suggest that in the coming years jet substructure will play an important role in both Standard Model measurements and beyond-the-Standard-Model searches.  New theoretical ideas for using and understanding jets continue to be developed.  As substructure techniques proliferate, a unified understanding will be vital.  The theoretical progress and goals discussed above are important steps along this path.  Meanwhile, software tools like FastJet and SpartyJet provide a complementary method for understanding and exploring jet substructure techniques.  We have given some examples of such exploration in this report, and we hope that the benchmark samples, tools, and comparisons we are making available will be useful tools in advancing jet substructure in 2012 and beyond.

\section*{Acknowledgements}

We thank the Princeton Center for Theoretical Science for hosting and
providing financial support for \boost\ 2011.

SH, CKV, and JRW were supported in part by the US National Science Foundation, grants NSF-PHY-0969510 and NSF-PHY-0705682, the LHC Theory Initiative, Jon Bagger, PI.  SH's work was supported by the US Department of Energy under contract DE-AC02/-76SF00515.  JRW was supported in part by the US Department of Energy under contract DE-AC02-05CH11231.  IWS was supported in part by the Office of Nuclear Physics of the US Department of Energy under the grant DE-FG02-94ER40818.
GPS was supported in part by grants ANR-09-BLAN-0060 from the French Agence
Nationale de la Recherche and PITN-GA-2010-264564 from the European
Commission.
Computing resources were provided by the Universities of Washington and Louisville, supported by US National Science Foundation contract ARRA-NSF-0959141 and US Department of Energy contract DE-FG02-98ER41089.

\section*{References}
\bibliographystyle{utphys}
\bibliography{boost2011}{}

\end{document}